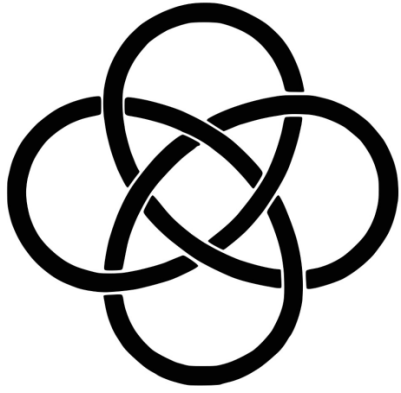

**Inter-University Centre for Astronomy and Astrophysics**

**Post-Bag 4, Ganeshkhind, Pune 411007 MH, India**

# Energetics of the solar atmosphere

**Abhishek Rajhans**

**Supervisor: Prof. Durgesh Tripathi**

**Co-Supervisor: Dr. Vinay Kashyap**

Thesis submitted for the degree of

Doctor of Philosophy to

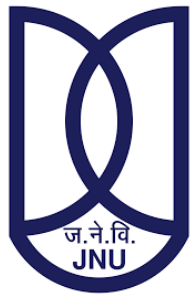

**Jawaharlal Nehru University, New Delhi, India**

# Decalaration from Candidate

I hereby declare that the material present in the thesis is based on work done at the Inter-University Centre for Astronomy and Astrophysics, Pune. This thesis is original and has not been submitted previously for any academic cause. Work of other researchers utilized by us have been properly cited.

**Abhishek Rajhans**

Signature:

Date:

# Declaration from Supervisors

This is to certify that this thesis titled 'Energetics of the solar atmosphere' is based on work done by Mr. Abhishek Rajhans at the Inter-University Centre for Astronomy and Astrophysics, Pune under our supervision. To the best of our knowledge this thesis is original and has not been submitted previously for any academic cause.

**Supervisor: Prof. Durgesh Tripathi**

Affiliation: Inter-University Centre for Astronomy and Astrophysics, Pune, MH, India

Signature:

Date:

**Co-Supervisor: Dr. Vinay Kashyap**

Affiliation: Center for Astrophysics | Harvard and Smithsonian, Cambridge, MA, United States of America

Signature:

Date:

# Declaration from the Head of the Institute

This is to certify that this thesis titled 'Energetics of the solar atmosphere' is based on work done by Mr. Abhishek Rajhans at the Inter-University Centre for Astronomy and Astrophysics, Pune. To the best of my knowledge this thesis is original and has not been submitted previously for any academic cause.

**Prof. Somak Raychaudhury**

Director, Inter-University Centre for Astronomy and Astrophysics, Pune, MH, India

Signature:

Date:

# Acknowledgements

I would first like to thank Durgesh Tripathi and Vinay Kashyap for supervising this thesis and providing their valuable suggestions and comments at all stages of my graduate school. They patiently helped me in learning new techniques and develop a better understanding of some of the topics. The calm working atmosphere facilitated fruitful discussions.

I thank James Klimchuk (GSFC, NASA) and Stephen Bradshaw (Rice University) for multiple valuable discussions and suggestions. I also thank P.S. Athiray (University of Alabama), Aveek Sarkar (Physical Research Laboratory), and IUCAA solar physics colleagues Nishant Singh, Megha Anand, Vishal Upendran, Soumya Roy, and Biswanath Malekar. I also thank former IUCAA colleagues Girjesh Gupta, Sargam Mulay, C.R. Sangeetha, Avyarthana Ghosh, Aishawnnya Sharma, and Nived V. N.

I have used data from Atmospheric Imaging Assembly (AIA) and Helioseismic and Magnetic Imager (HMI) onboard Solar Dynamics Observatory (SDO). SDO is a NASA mission. Interface Region Imaging Spectrograph is a small explorer mission of NASA developed and operated by LMSAL. FOXSI is a sounding rocket mission funded by NASA. The analysis was done in Interactive Data Language (IDL) using the Solarsoft package. The computational facilities were provided by the SUIT server of IUCAA. I have also used facilities provided by the IUCAA library. I thank all the members associated with all the above-listed facilities.

IUCAA gave me the company of excellent batchmates, Anshuman Borgohain, Shreejit Jadhav, Anusree K.G., Mousumi Mahato, Debasish Mondal, Ashish Mhaske, Uday Nakade, and Prakash Tripathi. In particular, I ap-plaud Shreejit's patience in successfully tolerating me as his neighbor. I have enjoyed the company of IUCAA colleagues, including Rajendra Bhatt, Sourabh Chabra, Sunil Choudhary, Prateek Dabhade, Sayak Dutta, Suraj Dhiwar, Akash Garg, Manish Kataria, Bikram Pradhan, Pushpak Pandey, Shishir Sankhyayan, Divya Rana, Priyanka Rani, Divya Rawat, Shilpa Sarkar, and Mayur Shende.

This work would not have been possible without immense emotional support from my family, especially my parents and my sister. My niece, the youngest member of my family, played a unique role in cheering me up. My special thanks are to my friend Sumedh Bhivsanie. I also thank Avinash Kumar, Anshika Chugh, Raj Jaiswal, and Amit Negi, my friends from Delhi University. This list of acknowledgments would be incomplete without thanking the people working at the tea and vada pav stall at Ambedkar Chowk. Tea and vada pav were important motivations for finishing work on time in the evening. Last but not least, the playful animals around me, in particular, Gundu and Sheru (cats of IUCAA), acted as very efficient stress busters.

*Dedicated to my parents*

# Abstract

The excess temperature of the solar corona over the photosphere poses a challenge. Multiple energetic events contribute to maintaining the corona at such high temperatures. The energy released in different events can vary across several orders of magnitude. Large energy events of geomagnetic importance like flares and coronal mass ejections (CMEs) contribute little to the global energetics of the solar corona. Therefore, events with several (9-10) orders of magnitudes of lower energy, with much higher frequency of occurrence, need to be studied in great detail. Observations suggest that these impulsive events with different energies follow a power-law distribution, indicating a common underlying mechanism. We perform observation-motivated modeling of coronal loops (magnetic flux tubes) to understand the energetics of these small transient events and their similarity with impulsive events like flares. This thesis uses the EBTEL code based on the 0D hydrodynamical description of coronal loops. This approach is appropriate for getting quick estimates of the energetics of the system over a wide range of parameters. We then discuss the improvement of EBTEL to make it suitable over a broader range of parameters. This is followed by using improved EBTEL to explore the possibility of simulating impulsive events of different energy generated using a single power-law distribution. Comparison between observed emissions from various components of multi-thermal plasma and hydrodynamical models suggest the heating to be impulsive. Since field-aligned flows induced due to impulsive events are a crucial part of our modeling of coronal loops, we discuss the implications of such flows in the context of transition region heating.

# Contents

**7    Center to limb variation of Doppler shifts in active regions of solar transition region    135**

**8    Summary, conclusions, & outlook    153**

**Bibliography    157**

# List of Figures

# List of Tables

# Abbreviations

**AIA** Atmospheric Imaging Assembly

**AR** Active Region

**CLV** Center to Limb Variation

**DEM** Differential Emission Measure

**EBTEL** Enthalpy-Based Thermal Evolution of Loops

**EUV** Extreme UltraViolet

**FOXSI** Focusing Optics X-ray Solar Imager

**HMI** Helioseismic and Magnetic Imager

**HYDRAD** Hydrodynamics and Radiation code

**IRIS** Interface Region Imaging Spectrograph

**PLD** Power-Law Distribution

**SDO** Solar Dynamics Observatory

**TR** Transition Region

# 1

---

# Introduction

The Sun is a G-type main-sequence star, with a surface temperature of about $6 \times 10^3$ K. It is at an average distance of $1.5 \times 10^{13}$ cm from the Earth, and has a mass ($M_\odot$) and radius ($R_\odot$) of about $2 \times 10^{33}$ g and $7 \times 10^{10}$ cm, respectively. The Sun has a luminosity of about $4 \times 10^{33}$ ergs s$^{-1}$, which is due to energy generated by nuclear fusion at its core, where more than 99 percent of the energy is released by the proton-proton chain reaction. It refers to the fusion of hydrogen atoms leading to the formation of helium atoms (Broggini, 2003). Less than 1 percent of energy is released by the CNO cycle. Theories predict that the importance of the CNO cycle will eventually increase with the aging of the Sun (Goupil et al., 2011). At the present epoch, the Sun is composed of about 74.9 % of Hydrogen, 23.8 % of Helium, and traces of other heavier elements. It has been suggested that the initial composition could have been slightly different (Lodders, 2003).

The Sun may have been formed about 4.6 billion years ago from the gravitational collapse of material within a molecular cloud (Connelly et al., 2012; Bonanno et al., 2002). Since the Sun shows a higher abundance of heavy elements like gold and uranium relative to other stars, Falk et al. (1977) have suggested that shock waves from a supernova(e) triggered its formation. With passing time, the hydrogen content in the core will decrease eventually, making the fusion rate insufficient to keep the Sun in hydrostatic equilibrium. Consequently, its core will contract and undergo an increase in temperature and density. The outer layers will continue expanding, and the Sun will gradually become a red giant (Boothroyd and Sackmann, 1999). Once the outer layers get shredded, it will eventually become and remain a white dwarf after nuclear fusion has ceased.

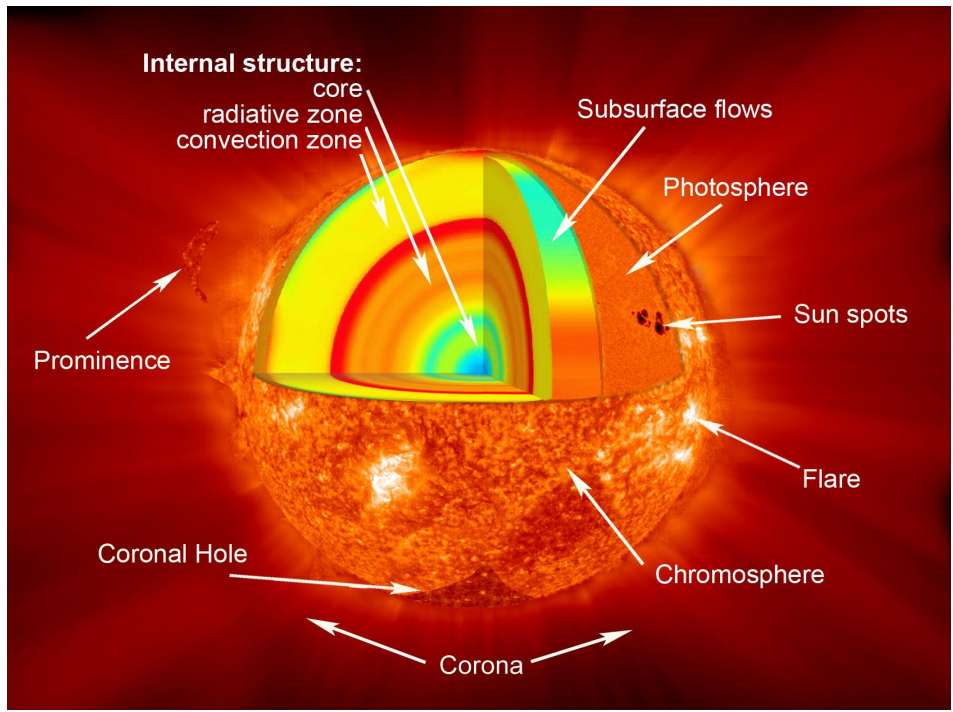

Figure 1.1: An artistic impression of the Sun including its interior and atmosphere. Image credit: NASA (https://www.nasa.gov/sites/default/files/images/462977main_Sun_layers_full.jpg)

Being the nearest star, the Sun provides a unique opportunity to study the various physical processes in detail. The study of the Sun can be broadly divided into two parts: the solar interior and the solar atmosphere. An artistic picture showing the Sun shown in Figure 1.1 depicts the different layers of the Sun.

## 1.1  Solar interior

Based on the physical characteristics, the solar interior can be divided into the following sub categories.

### 1.1.1  Solar core

It is the innermost region of the Sun where nuclear fusion takes place. Mathematical models suggest its extent from the center to about 0.2-0.25 $R_\odot$ (García et al., 2007) with a mass density of about 150 g cm$^{-3}$ (Basu et al., 2009) and temperature of about $1.6 \times 10^7$ K (Ricci and Fiorentini, 2003). Nuclear fusion produces high-energy gamma-ray photons, which are absorbed within a length of a few mm due to large densities in the core. Lower energy photons are then re-emitted in random directions. A sequence of absorption and re-emission of photons in the solar interior thermalizes the photons when they reach the solar surface. Consequently, the solar surface emits mainly in visible light waveband.

### 1.1.2  Radiative zone

The radiative zone extends above the core from about 0.25–0.7 $R_\odot$ and witnesses a temperature variation from $7 \times 10^6$ K to $2 \times 10^6$ K. The transfer of energy across this region happens through radiation. Densities drop by two orders of magnitudes from about 20 g cm$^{-3}$ near 0.25 $R_\odot$ to about 0.2 g cm$^{-3}$ near 0.7 $R_\odot$ (Christensen-Dalsgaard et al., 1996).

### 1.1.3  Convective zone

The radiative zone is surrounded by a convective zone that extends from about 0.7 R$_\odot$ to near the surface, where R$_\odot$ denotes solar radius. The temperature varies from about $2 \times 10^6$ K to $6 \times 10^3$ K over 0.3 R$_\odot$. Consequently, the gradient is more than the adiabatic lapse rate, i.e., the temperature gradient, which can allow the adiabatic motion of the gas. This leads to convection. Low temperature and density make radiation a much lesser efficient mode of energy transfer than convection, which is further facilitated by low density.

At the interface of the uniformly rotating radiative and differentially rotating convective zone, there is believed to be a tachocline (Spiegel and Zahn, 1992) that may play an important role in the solar dynamo (Charbonneau, 2010), which is the proposed mechanism for the generation of solar magnetic fields. These magnetic fields rise into different layers of the solar atmosphere.

## 1.2  Solar atmosphere

The Solar atmosphere consists of the following layers :

### 1.2.1  Photosphere

Photosphere extends roughly from the visible surface of the Sun[*] to about 4 $\times 10^7$ cm above it. The photosphere has a temperature of about 5778 K. The solar atmosphere above this layer is almost transparent to the visible light photons. Figure 1.2 shows the image of the Sun taken in the visible light continuum, 1700 Å, and 4500 Å. These images correspond to photosphere observations in different wavelengths bands.

---

[*]The surface is defined as the position below which optical depth ($\tau$) in the Fe I line at 5500 Å becomes greater than unity. The surface is highly anisotropic and dynamic and depends on the wavelength in which optical depth is measured.

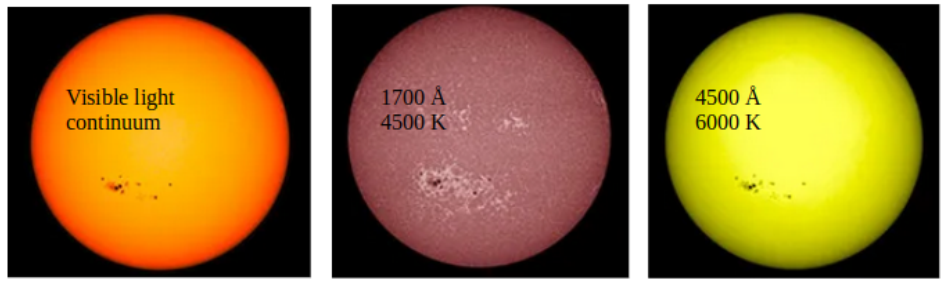

Figure 1.2: The Sun imaged in the visible light continuum (left panel), 1700 Å/ Temperature ≈ 4500 K (middle panel) and 4500 Å/ Temperature ≈ 6000 K (right panel), using the AIA and HMI telescopes on board SDO. The bandwidths of 1700 Å and 4500 Å are roughly 200 Å and 500 Å respectively.
Image credit: C. Alex Young (https://www.theSuntoday.org/missions/sdo/)

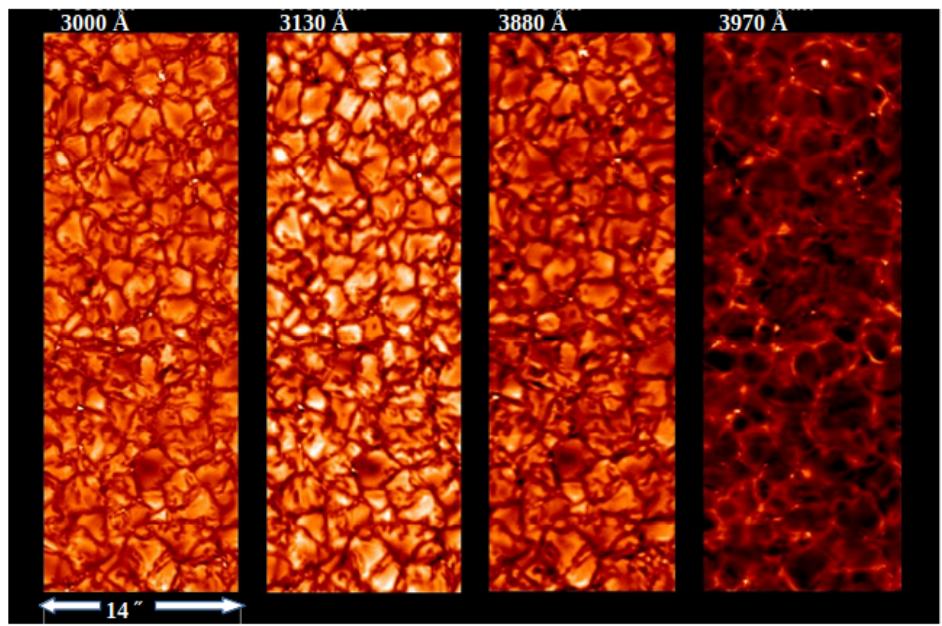

Figure 1.3: Granulation patterns on Sun observed in four different wavelengths in near ultraviolet light using Sunrise Filter Imager (SuFI) on board Sunrise I. The smallest granular structures have a length scale of $10^7$ cm. Image credit: Max Planck Institute for Solar System Research (https://www2.mps.mpg.de/en/aktuelles/pressenotizen/)

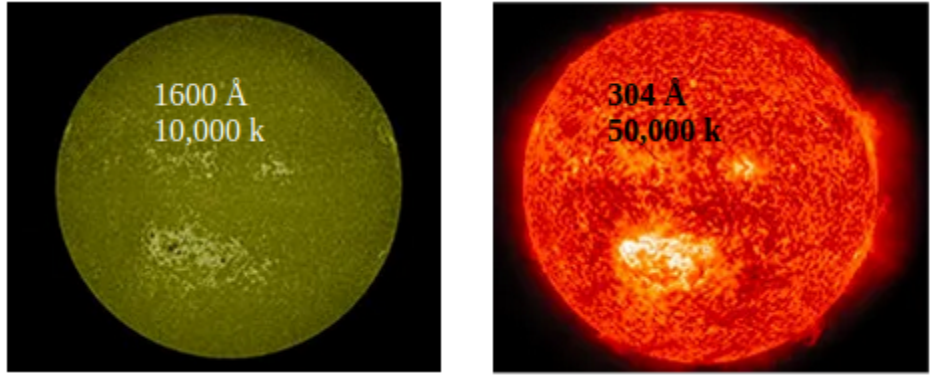

Figure 1.4: The upper photosphere imaged in 1600 Å/ Temperature $\approx 10^4$ K (left panel), and the upper chromosphere observed in 304 Å/ Temperature $\approx 5 \times 10^4$ K (right panel) using AIA onboard SDO. Image credit: C. Alex Young (https://www.theSuntoday.org/missions/sdo/)

The photosphere shows regions called Sunspots, which appear darker in visible light waveband. These regions have higher magnetic fields of the order of $10^{2-3}$ G compared to other areas with the magnetic field of the order of 10 G. Figure 1.3 shows the granulation patterns arising in the photosphere due to the motion in the convective zone (see for e.g. Schrijver et al., 1997). These granular patterns have a typical length scale of $10^7$ cm. These granules appear bright in white light and are surrounded by dark lanes. Supergranulation patterns with a length scale 100 times larger than those of granulation patterns have also been observed in the photosphere (Rieutord and Rincon, 2010).

### 1.2.2 Chromosphere

The photosphere is surrounded by chromosphere, which derives its name from the Greek word *chromos*, meaning color because of its red flash-like appearance at the beginning and end of the total solar eclipse. This reddish appearance is due to strong emission in $H_\alpha$ lines. Observations using spectroscopic methods show its extension from around $4 \times 10^7$ cm to about $2 \times 10^8$ cm above the solar surface. Its temperature varies from around $4 \times 10^3$ K to $10^4$ K (Hall, 2008). Figure 1.4 shows the upper photosphere in 1600 Å and the corresponding image of the chromosphere in 304 Å in the left and right panels, respectively. These images show regions of enhanced emission above strong photospheric magnetic features. In addition, the left panel also shows Sunspots as observed in the chromosphere.

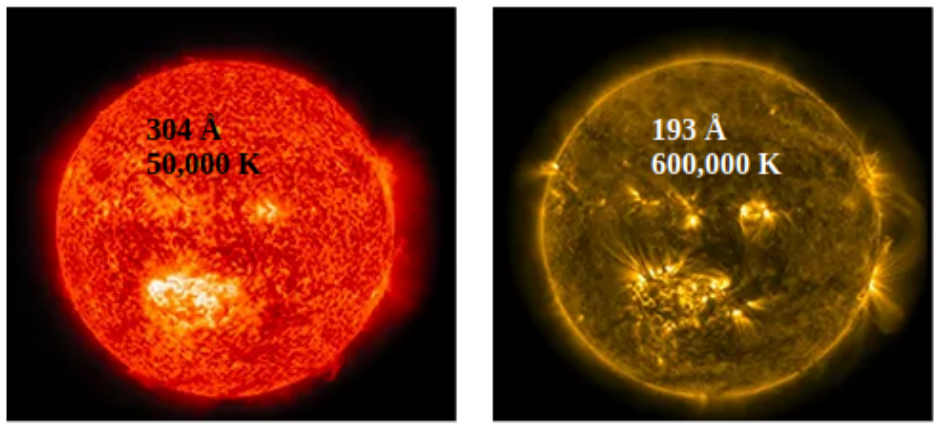

Figure 1.5: The Sun imaged in 304 Å / Temperature $\approx 5 \times 10^4$ K (left panel), and 171 Å/ Temperature $\approx 6 \times 10^5$ K (right panel) using AIA onboard SDO. Image credit: C. Alex Young (https://www.theSuntoday.org/missions/sdo/)

A striking feature in the chromosphere is the ubiquitous presence of spicules. These correspond to spikes in gases which rise above the surface and fall back immediately (for a review see e.g. Sterling, 2000). The other interesting feature present in the chromosphere is reverse granulation. The intergranular lanes, which are dark in the photosphere, become bright network regions in the chromosphere observed in Ca II line, and the region above bright photospheric granules look darker in the chromosphere (see for eg. Cheung et al., 2007).

### 1.2.3 Transition region

The transition region is narrowly above the chromosphere and below the corona. The temperature rises from about $10^4$ K at the top of the chromosphere to about 1 MK at the coronal base over a height of about $10^7$ cm. It can be observed in UV and EUV band (Mariska, 1986). Figure 1.5 show observations of the transition region in 304 and 193 Å in the left and right panels, respectively (Note that observations in 304 Å have a contribution from both upper chromosphere and the transition region).

The transition region separates the partially ionized chromosphere and fully ionized corona. It also marks the separation of the chromosphere dominated by gas pressure and the corona dominated by magnetic pressure. In active regions, the transition region shows features like spicules and loops. In the active regions, spicules and loops are observed in the transition region.

### 1.2.4 Corona

The corona lies above the transition region and has an average temperature of about 1-2 $\times$ $10^6$ K. In the case of transients, the temperature can become as high as 2 $\times 10^7$ K (Benz, 2017). The small density of the solar corona ($10^{9-11}$ cm$^{-3}$) lying above the photosphere ($10^{15-17}$ cm$^{-3}$) as compared to the surface makes it extremely faint. The corona predominantly emits in EUV and X-rays. Figure 1.6 shows observations of the solar corona in four wavelengths, namely 211, 335, 94, and 131 Å.

The solar corona can be broadly divided into three regions. These are active regions, coronal holes, and quiet Sun. The active regions contain bright loop-like structures and show pronounced heating. The coronal holes are identified as dark regions in EUV and X-ray. The latter facilitates plasma to escape with relative ease. This escaping plasma is referred to as solar wind. The quiet Sun in the corona is dominated by diffused emissions. Since they are most widespread in terms of surface area, they offer the largest contribution to the global energetics of the solar corona.

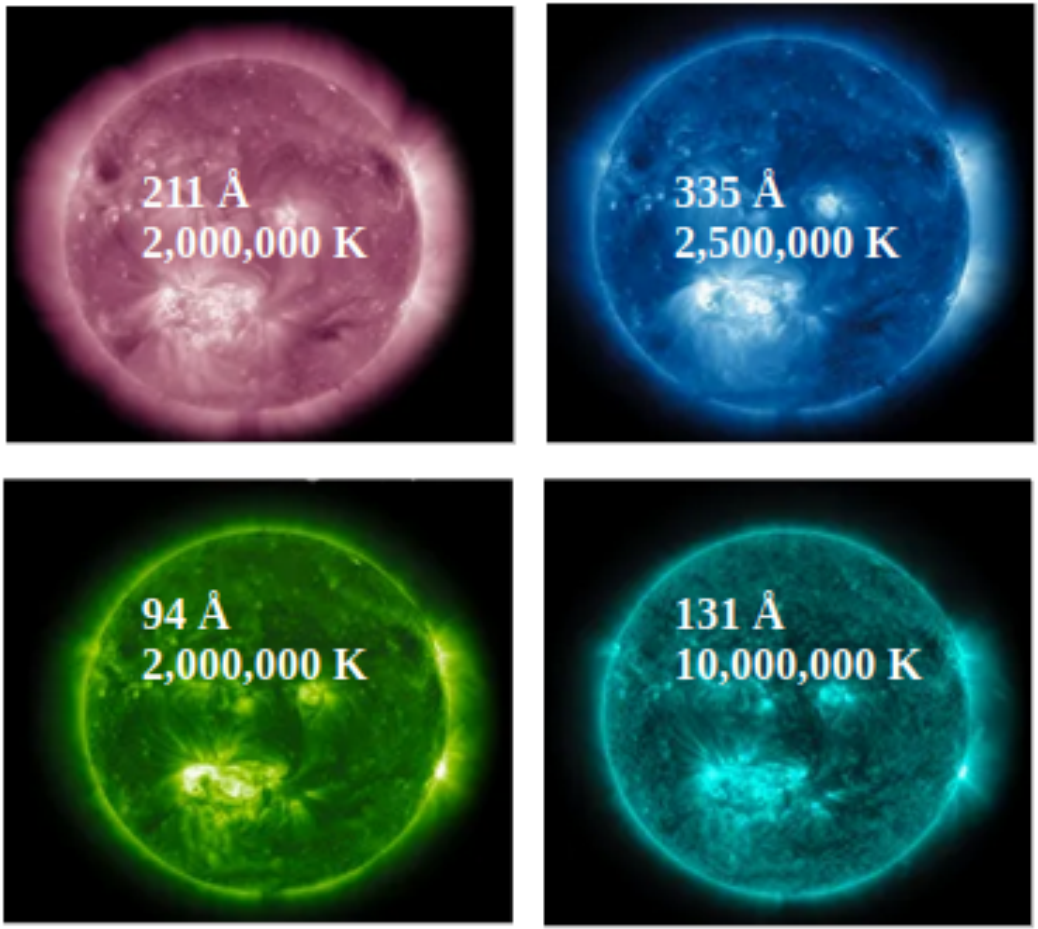

Figure 1.6: The solar corona imaged in 211 Å / Temperature $\approx$ 2 $\times$ $10^6$ K (top left), 335 Å/ Temperature $\approx$ 2.5 $\times$ $10^6$ K (top right), 94 Å/ Temperature $\approx$ 2 $\times$ $10^6$ K (bottom left), and 131 Å/ Temperature $\approx$ $10^7$ K (bottom right) recorded by AIA on board SDO. Image credit: C. Alex Young (https://www.theSuntoday.org/missions/sdo/)

## 1.2.5 Heliosphere

The corona is surrounded by the heliosphere, beginning at distances where solar winds and plasma flows attain super Alfvénicic speeds. In this region, magnetic field strengths are low compared to the corona, and magnetic pressure becomes smaller than thermal pressure, unlike the corona. Solar winds in this layer travel outwards in outer space until they hit the heliopause, where stellar winds from other stars start becoming important Dessler (1967).

## 1.3 Problem of Solar atmospheric heating

In 1887 Young and Harkness observed an emission line at 5303 Å in the Solar corona during a total Solar eclipse. At that time, this line could not be explained using known emission lines. Noting the success of spectroscopy in predicting unknown elements in celestial objects, Gruenwald proposed that these emission lines were due to an unknown element and gave it the name *coronium*. After seven decades, around 1940, Grotrian and Edlén identified this emission line as from Fe XIV, i.e., iron ions formed after stripping neutral atoms of 13 electrons (Grotrian, 1939; Edlén, 1943). By 1941 all coronal emission lines could be explained using emissions from highly ionized states of heavy elements like iron, nickel, magnesium, etc. Under the condition of local thermodynamic equilibrium presence of strong emission lines from highly ionized states of these elements require temperatures of the order of $10^6$ K. This established that hotter plasma was present over a much cooler photosphere.

After the discovery of the coronal heating problem, chromospheric heating was established in 1942 by Redman (Redman, 1942). Redman observed thermal broadening of $H_\alpha$ lines and deduced a temperature of about $3 \times 10^4$ K. These widths could not be explained by mass motion because of the presence of a group of strong metal lines which had only one-tenth of the thermal broadening.

Using a semi-empirical 1D hydrodynamical model, Vernazza et al. (1981) found that the transition from chromosphere with a temperature of about $3 \times 10^4$ K and corona with a temperature of about $1 \times 10^6$ K happens in a small layer of thickness of about $10^7$ cm. While plasma is partially ionized in the chromosphere it becomes fully ionized in the corona. Energy injected in partially ionized plasma can increase temperature or cause ionization. The fraction of ionized plasma increases very rapidly in the transition region. This, in turn, results in a rapid increase in energy available for increasing temperature. Consequently, the temperature shoots up in the transition region. In the corona, the energy dissipated goes mainly into increasing the temperature, and hence temperature starts saturating.

While the corona is highly inhomogeneous and dynamic, simplified 1D models shed valuable light on the solar atmosphere's overall temperature and density profile. Figure 1.7 shows the average variation of temperature and density across various layers of the solar atmosphere based on a 1D model. Positive height represents the solar atmosphere above the photosphere, while negative height represents the interior of the Sun. On moving above toward the chromosphere, the temperature starts increasing and goes up to the order of $10^6$ K in the corona.

## 1.4   Estimates of required energy budget

Here we use energy conservation equations to find a rough estimate of the energy budget for heating and maintaining the chromosphere and the corona. This section is based on the discussion in Mullan (2009). The equations in this section have been taken from Mullan (2009)

In the chromosphere, radiation is the predominant mechanism of cooling.

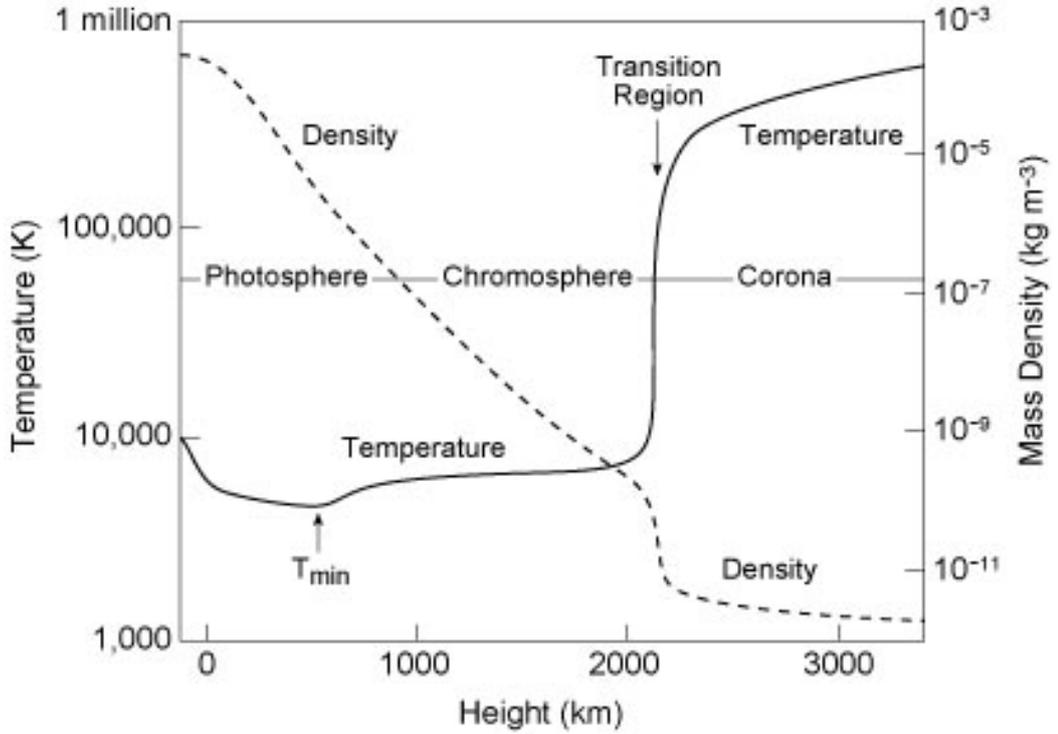

Figure 1.7: Variation of temperature and density of plasma with altitude across the solar atmosphere. Image Credit: Kenneth R. Lang, 2010

Contribution from continuum emission and line emission can be assumed to be rough of the same order. Hence for simplicity, we consider only continuum emission. If the chromosphere were a black body (which it is not), the power radiated by it would be

$$P_{rad} = 4\sigma_b(T^4 - T_0^4) \quad (1.1)$$

where $P_{rad}$ is radiation loss flux in ergs cm$^{-2}$ s$^{-1}$, $\sigma_b$ is Stefan Boltzmann constant in cgs units, T is the local temperature of the chromosphere and $T_0$ is photospheric temperature.

Since the chromosphere is not a perfect blackbody, the actual radiated energy is lesser and reduced by a dimensionless factor of $\tau$ ($< 1$). It can be related to density ($\rho$) and opacity ($\tau$) of the system by the relation $\tau = \kappa\rho ds$. Taking this into account, radiated power becomes

$$P_{rad} = 4\sigma_b \kappa \rho (T^4 - T_0^4) ds \qquad (1.2)$$

For conditions present in the chromosphere, the opacity can be written as

$$\kappa = 10^{-32} \rho^{0.3} T^9 \qquad (1.3)$$

Consequently one gets

$$P_{rad} = 4\sigma_b 10^{-32} \rho^{1.3} T^9 (T^4 - T_0^4) \qquad (1.4)$$

For maintaining the chromosphere at a temperature of about $10^4$ K, the rate at which energy is radiated should be equal to the rate at which energy is deposited. Since we are interested in rough estimates, we assume that this required rate is constant throughout the chromosphere. Taking typical thickness of chromosphere as $\sim 10^8$ cm, typical temperature of $10^4$ K and typical density of $10^{-12}$ g cm$^{-3}$, the required energy flux is $\approx 10^8$ ergs cm$^{-2}$ s$^{-1}$.

Unlike the chromosphere, the corona radiates predominantly by emission lines. Furthermore, conduction losses play a vital role in its energetics because the coronal plasma is completely ionized. The magnitude of conduction flux from the corona can be estimated using Spitzer conductivity as

$$F_C = \kappa_0 T^{\frac{5}{2}} \frac{dT}{ds} \sim \kappa_0 \frac{T^{\frac{7}{2}}}{L} \qquad (1.5)$$

where $T$ is the typical temperature of active region corona, i.e., $2 \times 10^6$ K, $L$ is the typical length scale of corona which is $\approx 10^9$ cm in active regions and $\kappa_0 = 10^{-6}$ in cgs units. Therefore, the typical conduction flux from the corona is about $10^7$ ergs cm$^{-2}$ s$^{-1}$.

The corona is optically thin, and hence radiation passes almost freely. The radiated power per unit volume for a collisionally excited and de-excited plasma is

$$E_{rad} = n_i \; n_e \; \Lambda(T) = n_e^2 \; \Lambda(T) \tag{1.6}$$

where $n_i$, and $n_e$ are ion and electron number densities, respectively, which are equal due to quasi-neutrality. $\Lambda(T)$ is the temperature-dependent optically thin radiative loss function calculated empirically by Rosner et al. (1978). For typical coronal temperature and electron number density of $10^6$ K and $10^9$ cm$^{-3}$, respectively (see for eg. Tripathi et al., 2006, 2009, 2010; O'Dwyer et al., 2010a; Subramanian et al., 2014; Del Zanna et al., 2015b) the power radiated per unit volume is about $10^{-3.5}$ ergs cm$^{-3}$ s$^{-1}$ (Note that $\Lambda(T) = 10^{-21.5}$ ergs cm$^3$ s$^{-1}$ for T = $10^6$ K). Using a typical coronal length scale of $10^9$ cm in active regions, the total power radiated per unit area is $10^{5.5}$ ergs cm$^{-2}$ s$^{-1}$. Conduction losses are more than an order of magnitude larger than radiative losses in the corona.

Hence, for maintaining the active regions of the corona at a typical temperature of $10^6$ K, the combined conduction and radiation power loss per unit area should be equal to the total energy dissipated in the corona per unit area, i.e., of the order of $10^7$ ergs cm$^{-2}$ s$^{-1}$. Similar calculations for quiet Sun and coronal holes (open loops) estimate an energy budget of around $10^5$ ergs cm$^{-2}$ s$^{-1}$ (Athay, 1976; Withbroe and Noyes, 1977).

## 1.5 Role of magnetic fields and available magnetic energy

The excess temperature of the solar atmosphere than the surface ($\approx 6 \times 10^3$ K) poses a challenge. This is because the second law of thermodynamics prohibits a physical process whose sole outcome is heat transfer from a cooler surface to a hotter atmosphere. Hence an agency is needed to perform mechanical work. Many observations suggest that solar magnetic fields play a very important role in atmospheric heating (Klimchuk (2006) and references therein). It is now accepted that the agency performing mechanical work is photospheric motions which stress and store energy in the magnetic field. Such motions also produce waves which add to energy transported into the atmosphere.

Figure 1.8 shows an image of the solar atmosphere in different filters sensitive to different temperatures and, therefore, different heights in the atmosphere. A remarkable feature is that strong magnetic field regions in the photosphere are spatially co-aligned with bright EUV features in the upper solar atmosphere.

Below we perform a rough estimate of energy stored in the stressed magnetic fields that is available for dissipation (Parker, 1988; Klimchuk, 2006). The Poynting flux ($F$) in ergs cm$^{-2}$ s$^{-1}$ associated with the work done by photospheric flows in stressing of coronal magnetic fields is given by

$$\mathbf{F} \sim -B_v \mathbf{B}_h . \mathbf{V}_h \qquad (1.7)$$

where B$_v$ and B$_h$ are the vertical and horizontal components of magnetic fields in Gauss and v$_h$ is the horizontal component of velocity of the photospheric foot points due to convective motion Choudhuri et al. (1993).

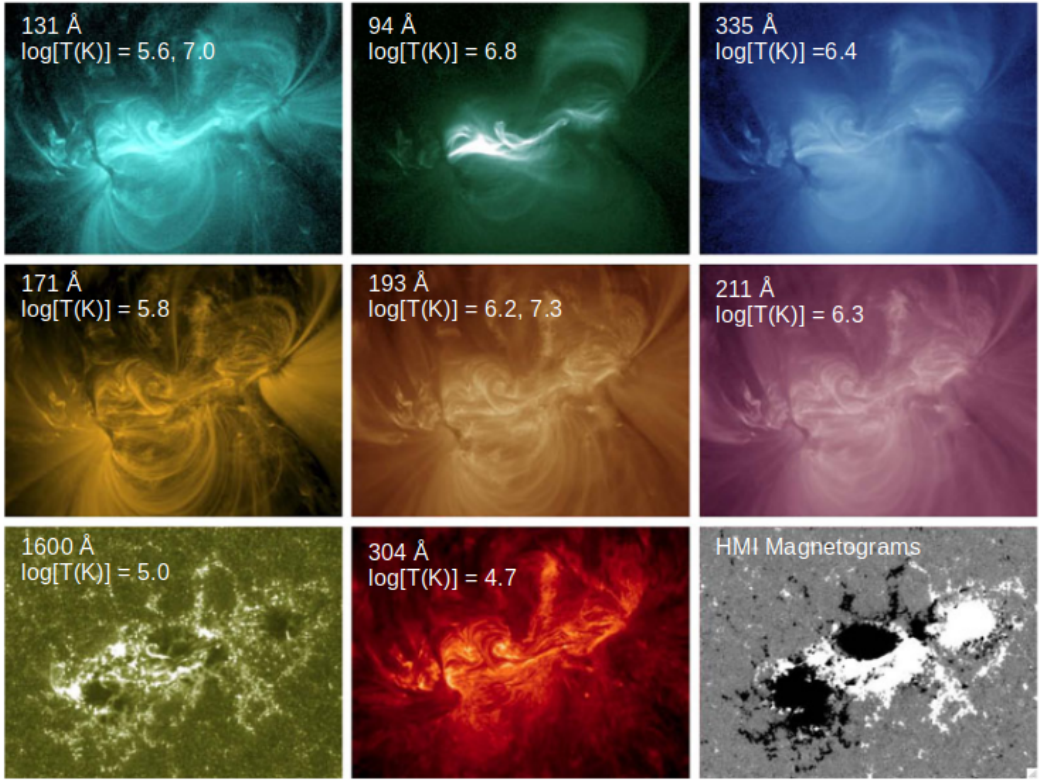

Figure 1.8: An active region observed through different ultraviolet filters of AIA and the corresponding magnetograms. The plasma temperature, which provides maximum emission in different AIA wavebands, is also listed.
Image courtesy: Lemen et al. (2012a)

$$F \sim B_v^2 V_h \tan(\theta) \qquad (1.8)$$

where $\theta$ is the angle subtended between the field and the vertical in a direction opposite the motion. A typical active region measurement of magnetic field estimates about 100 G Parker (1988). The typical velocities of magnetic flux tubes due to turbulent convection in the photosphere are about $10^5$ cm s$^{-1}$ (Muller and Roudier, 1984; Berger and Title, 1996). If $\theta$ is in the range of 10–20 degrees, then Poynting flux ($10^8$ ergs cm$^{-2}$ s$^{-1}$) is more than sufficient to account for the heating requirement of corona and chromosphere ($10^7$ ergs cm$^{-2}$ s$^{-1}$).

These are the order of magnitude estimates, and the precise values can be different for systems with different magnetic field configurations present in the photosphere. However, these estimates show that Poynting flux from the photosphere is sufficient to maintain the chromosphere and corona at temperatures of the order of $10^4$ K and $10^6$ K, respectively. However, energy simply cannot be transferred radiatively, lest it will violate the second law of thermodynamics. Energy can be transferred from a system at a lower temperature to a higher temperature only by mechanical work.

## 1.6 Mechanisms of energy dissipation

In the late 1940s, Biermann and Schwarzschild proposed the idea of the dissipation of energy in waves generated due to the motion of matter in the solar atmosphere. However, with the possibility of space-based missions providing a strong correlation between magnetic fields and X-ray/EUV brightenings in the corona, mechanisms related to magnetic fields started getting attention. Furthermore, observational studies of acoustic waves revealed that the measured acoustic flux is insufficient to explain the heating of the solar atmosphere (Athay and White, 1978; Mein and Schmieder, 1981; Fossum and Carlsson, 2005; Rajaguru et al., 2019). While acoustic flux can account for a small fraction of the required energy budget, the major contribution should come from other mechanisms. Such mechanisms can be broadly divided into categories: AC and DC mechanisms.

AC mechanism refers to processes that involve energy dissipation from MHD waves. The three promising candidates for the dissipation of MHD waves are (i) resonant absorption, (ii) phase mixing, and (iii) Alfvén wave turbulence.

Resonant absorption was first proposed as a coronal heating mechanism by Ionson (1978). According to this theory, MHD waves are generated on the surface of coronal loops. The surface of the coronal loop marks the separation between regions of different local Alfvénic speeds (higher inside the loop and lower outside the loop). In addition, the loop field lines support shear Alfvén waves. When there is resonance due to the phase velocity of the surface wave matching with the local Alfvén speed, energy can be dissipated efficiently. Such dissipative processes are expected to take place in a small region enveloping the coronal loops called resonant absorption sheath.

Phase mixing was first proposed as a viable coronal heating mechanism by Heyvaerts and Priest (1983). This mechanism exploits the spatial variation of local Alfvénic speeds. Due to this, different magnetic field lines support shear Alfvénic waves of varying frequency and wavelengths. With the upward propagation of these waves towards the loop top, these start becoming out of phase, and this gradient in wave fronts can dissipate energy.

In addition to resonant absorption and phase mixing, turbulence in the propagating Alfvén waves has also been proposed as a mechanism for explaining coronal heating (Cranmer et al., 2007; van Ballegooijen et al., 2017).

DC heating requires mechanisms involving the release of magnetic stress. Since the foot points of magnetic field lines are located in the photosphere, these are in constant random motion due to strong convection. This leads to magnetic stress, which may be released by magnetic reconnection (Parker, 1988). In this process, current sheets develop where magnetic resistivity becomes important and ultimately leads to energy dissipation.

The observational signatures and theoretical understanding of these mechanisms are not totally conclusive and still have open questions to be answered. Since the convective motion of plasma near the photosphere can result in both generation of MHD waves and magnetic reconnection, it seems highly improbable that a single mechanism will be able to explain all aspects of heating of the Solar atmosphere. Hence different mechanisms might be at work at different locations during different times (Klimchuk, 2006).

## 1.7 Power-law distribution of energetic events

Irrespective of the mechanism of heating, the following features may hold in general, i.e., the heating occurs at small length scales, and it is impulsive. While big events like solar flares are important from a space weather perspective, they contribute little to the global energetics of the solar corona. It is assumed that events with energy less than $10^{25}$ ergs occur at a frequency much higher than the observed transients. This scenario was proposed by Parker (1988) as a "swarm of nanoflares."

Hudson (1991) observed that the relationship between the number of events and their energies (E) obey a power-law distribution (see Figure 1.9)

A power-law distribution can be expressed as

$$\frac{dN}{dE} \propto E^{-\alpha} \qquad (1.9)$$

where $dN$ is the rate of occurrence of events having energy in the range $[E, E+dE]$ and $\alpha$ is a positive number (see Hannah et al. (2011) for a review). Hudson (1991) also conjectured that to maintain the corona at a temperature greater than 1 MK with the help of impulsive events, there must be a large frequency of such events with smaller energy. The power-law distribution of flares of different energies may be a sign of the underlying phenomenon of self-organized criticality (Lu and Hamilton (1991)). For the heating to be dominated by nanoflares, the power-law index ($\alpha$) should be greater than 2. This has led to many observational studies (Shimizu, 1995; Berghmans et al., 1998, 2001; Krucker and Benz, 1998, 2000; Parnell and Jupp, 2000;

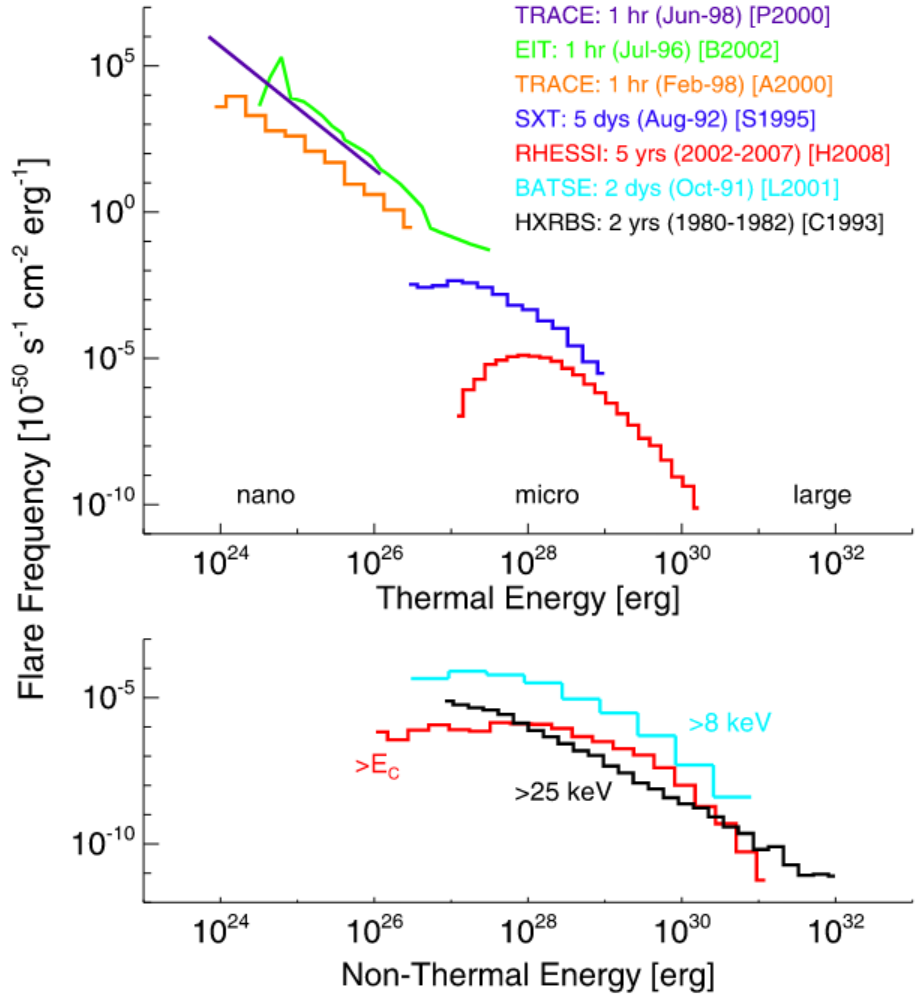

Figure 1.9: Power-law distribution of energetic events in the solar atmosphere. The upper part shows the energy radiated from thermal plasma. The lower part shows the non-thermal energy distribution for large flares. Image courtesy: Hannah et al. (2011)

Aschwanden et al., 2000a,b; Christe et al., 2008; Hannah et al., 2008) based on the counting of different types of transient events that result in varied negative slopes of power-law ranging from $1.6 \leq \alpha \leq 2.2$. However, there are limitations to such studies due to constraints on the cadence, passbands, and resolutions of instruments. Moreover, there remains a chance that flares of different energy, particularly at lower energies, are undercounted (see, e.g., Pauluhn and Solanki, 2007; Upendran and Tripathi, 2021). Additionally, it is possible that different events may be generated due to different mechanisms, and hence they would not necessarily follow the same power-law distribution.

## 1.8 Outline of the thesis

In this section, we briefly discuss the structuring of the remaining thesis.

### 1.8.1 Chapter 2: Instruments

This chapter discusses the instruments whose data we have used in our projects for performing observational studies or motivating numerical modeling. We first discuss the Atmospheric Imaging Assembly (AIA) and the Helioseismic and Magnetic Imager (HMI) onboard the Solar Dynamics Observatory (SDO). AIA provides UV/EUV images in 8 passbands centered around 94 (Fe XVIII), 131 (Fe VIII, XXI), 171 (Fe IX), 193 (Fe XII, XXIV), 211 (Fe XIV), 304 (He II), 335 (Fe XVI), 1600 (C IV and nearby continuum), and 1700 (continuum) Å. It provides full-disk images of the solar atmosphere extending up to 0.5 solar radii above the limb. HMI observes the Fe I absorption line at 6173 Å to measure intensities, Doppler shifts, and vector magnetic fields in the solar photosphere. We then discuss the Interface Region Imaging Spectrograph (IRIS), which provides UV spectra in windows centered around 1334.5 (C II), 1335.7 (C II), 1349.4 (Fe XII), 1354.1 (Fe XXI), 1355.6 (O I), 1393.8 (Si IV), 1399.8 (O IV), 1401.2 (O IV), 1402.8 (Si IV), 2796.4 (Mg II k), 2803.5 (Mg II h), and 2820 (Mg II wing) Å. The spectra obtained has a velocity resolution of 1 km s$^{-1}$ in these windows. This is followed by a description of the Focusing Optics X-ray Imager (FOXSI). It is a sounding rocket designed to observe X-ray photons in the range of 4–20 keV.

### 1.8.2 Chapter 3: Numerical modeling

This chapter discusses the software used for the numerical modeling of coronal loops. We begin with a brief discussion of the conditions under which the fluid description can be used to study magnetized plasma. This description is called magnetohydrodynamics (MHD). We then discuss the MHD equations under conditions suitable for the solar corona and the transition region. This is followed by the derivation of the 1D field-aligned hydrodynamical (HD) description of coronal loops from MHD. HD is a reliable framework for studying the plasma response in coronal loops to a heating event. We then discuss the 0D hydrodynamical description of coronal loops, which studies the evolution of instantaneous coronal averages of physical quantities. We describe the Enthalpy-Based Thermal Evolution of Loops (EBTEL), a 0D code used prominently in our modeling efforts. It computes instantaneous coronal averages of pressure, temperature, and density. It also computes velocity at the coronal base and instantaneous differential emission measures in the corona and the transition region.

### 1.8.3 Chapter 4: Hydrodynamics of small transient brightenings in the solar corona

This chapter is based on our first publication Abhishek Rajhans et al. 2021 ApJ 917 29, which models weak transient bright events discovered in Hi-C observations. As discussed, the dominant contribution to global energetics of the solar corona should come from events with several orders of magnitudes lower energy ($< 10^{25}$ ergs) and much higher frequency. Hence, such events need to be studied in great detail. An important question is whether different energetic events of varying magnitudes come from the same physical processes. Some of the smallest brightenings were discovered using images obtained by Hi-C (High-Resolution Coronal Imager) in the 193 Å passband. The energetics of these events was previously studied using thermal diagnostics estimated with the help of images obtained from the six extreme ultraviolet filters of AIA. Using stationary loop approximations, conduction was found to be the dominant cooling mechanism in the corona. This is a feature shared by large flares, microflares, and nanoflares.

The objective of this chapter is to perform numerical simulations to check whether these brightenings can be modeled using the same physics as that involved in larger flares. We use EBTEL for hydrodynamical simulations and produce synthetic light curves to compare with AIA observations. We first identify a set of input parameters (loop half-length, energy budget, and duration of heating) that produce synthetic light curves similar to those obtained by AIA. For simplicity, we only study the brightenings with a single prominent peak. Simulation results obtained for these input parameters were used for studying the time evolution of conduction, radiation, and enthalpy.

We model these transients as loops of $\sim$1.0 Mm length depositing energies $\sim 10^{23}$ ergs in $\sim$50 seconds. The simulated synthetic light curves show reasonable agreement with the observed light curves. During the initial phase, conduction flux from the corona dominates over the radiation, similar to those observed in flares. Our results show that the time-integrated net enthalpy flux is positive, hence into the corona. The fact that we can reasonably model the observed light curves of these transients by using the same physics as those for nanoflares, microflares, and large flares suggests that these transients may also be caused by impulsive heating.

### 1.8.4 Chapter 5: Flows in the enthalpy-based thermal evolution of loops

This chapter is based on Abhishek Rajhans et al. 2022 ApJ 924 13, which includes kinetic energy term in EBTEL. 0D hydrodynamics codes like EBTEL are very useful when a large parameter space of heating functions and loop lengths needs to be explored. They are also well suited when light curves of several hundreds of thousands of seconds are needed for comparison for studying statistical properties of coronal heating. While performing the study discussed in Chapter 4, we used the version of the code named EBTEL2. We realized that one of its main limitations was neglecting kinetic energy in the energy evolution equation at all stages. This approximation is valid if the flows remain subsonic throughout the system's evolution. This was the case for the parameters used in the simulations of brightenings studied in Chapter 4. However, this condition does not always hold. Computation of Mach numbers by EBTEL2 code returned transonic and supersonic velocities in many cases where 1D field-aligned simulations computed subsonic flows. This was a limitation of ignoring kinetic energy in the energy evolution equation in EBTEL2. This chapter discusses the work done for including the kinetic energy term in the EBTEL2 code and develops the updated version of the code EBTEL3. The upgraded code EBTEL3 has an adaptive time grid making it roughly ten times faster than EBTEL2.

We compare the solutions from EBTEL3 with those obtained using EBTEL2 and the state-of-the-art field-aligned 1D hydrodynamics code HYDRAD. We find that the match in pressure between EBTEL3 and HYDRAD is better than that between EBTEL2 and HYDRAD. We notice that the density computed by HYDRAD matches slightly better with EBTEL2 than EBTEL3. The reason behind this is discussed. We note the velocities predicted by EBTEL3 are in close agreement with those obtained with HYDRAD, especially when the flows are subsonic.

There are cases where 1D field-aligned simulations also compute supersonic flows. However, due to spatial information, 1D codes can potentially treat the consequent shock dissipation. EBTEL computes the time evolution of coronal averages of physical quantities like pressure, density, and temperature, along with the velocity at base and differential emission measures of the corona and transition region. Since there is no spatial information, EBTEL cannot tackle shocks. Hence, it is necessary to predict the reliability of results produced by EBTEL, which forms the second part of this project. Using the mismatches in the solution, we also propose a criterion to determine the conditions under which EBTEL may be used to study flows in the system.

## 1.8.5 Chapter 6: Simulations of AR studied by FOXSI and AIA: single PLD of events

This chapter describes multistranded 0D simulations for studying active region events. Previous simulations performed for studying transients provide a constant uniform heating in addition to the event of interest. This is done to ensure coronal temperature always remains above $0.5 \times 10^6$ K. Here, we consider the possibility of background coming from events of different energies lower than that of the transient. We assume that all the events come from the same power-law distribution. This would be expected for events with identical mechanisms but different energy ranges. Hence, this project allows us to test the hypothesis that different energy events are similar and come from the same power-law distribution. We use the upgraded 0D code EBTEL3 since it no longer has the limitation of neglecting kinetic energy. The aim of these simulations is to model X-ray observations in FOXSI and EUV observations from AIA.

We approximate the loop length under a semicircular loop approximation. The radius of an individual strand and the total energy budget of the event have been constrained by Hi-C observations and FOXSI luminosity, respectively. Using these observation-motivated constraints, we vary the minimum and maximum energy $[E_{min}, E_{max}]$ that can be dissipated in a single event along with the slope of power-law distribution ($\alpha$). We simulate multiple representations of the same case, each lasting for the duration of observed light curves. Different cases i.e., different combinations of $E_{min}, E_{max}$, and $\alpha$, have been studied. Our analysis of results suggests that power-law slopes larger than two cannot explain FOXSI observations. We discuss the implications of these results.

## 1.8.6  Chapter 7: CLV of Doppler shifts in ARs of solar transition region

One of the features related to impulsively heating the corona is field-aligned flows, i.e., flows along the magnetic field lines. If these are the only flows in the corona and the transition region, then it is natural to expect center-to-limb variation (CLV) of measured Doppler shifts in the corona and the transition region. These should vanish as one approaches the limb. However, observation of Doppler Shifts in the transition region defies this expectation. This has been a long-standing puzzle in earlier observational studies. A limitation of earlier studies was the absence of cool neutral lines for performing wavelength calibration. IRIS can observe several such lines and use them for wavelength calibration.

A previous study used IRIS data in the Si IV line and performed wavelength calibration using the O I line. It measured the center-to-limb variation of Doppler shifts in a Si IV line for a single active region using IRIS observations, showing that although there are hints of CLV in the data, there are significant Doppler shifts observed near the limb. Moreover, velocities associated with these Doppler shifts are an order of magnitude larger than those predicted by field-aligned simulation of coronal loops heated by impulsive events. It has been suggested that the flows observed in transition regions measured using Si IV are primarily due to type-II spicules in a chromospheric well (associated with classical type-I spicules) that diminishes the CLV of the Doppler shifts and may produce non-zero Doppler shifts at the limb. However, this work lacked substantial coverage of solar longitudes. This work supports these results by studying 50 active regions on the solar disk at different locations.

### 1.8.7 Chapters 8 : Summary, conclusions, & Outlook

We present a summary of the results obtained in this thesis and their influences on enhancing our understanding of the energetics of the solar atmosphere in Chapter 8. The thesis ends with possible future directions.

# Instruments

*In this chapter we discuss briefly the instruments whose observations we have used in this thesis for motivating simulations or data analysis. We first discuss AIA and HMI onboard SDO. We have used UV images provided by AIA and magnetic field information provided by HMI. We next discuss IRIS which we have used for obtaining spectra in UV wavelength windows. This is followed by a discussion of FOXSI which images the solar corona in hard X-rays.*

The solar corona provides emissions in wavelengths ranging from X-ray to radio. However, the dominant contribution is from ultraviolet and X-ray wavelengths. Due to the absorption of these photons by the earth's atmosphere, the advent of space-based observatories provided a turning point in our understanding of the solar atmosphere.

The earliest observations in these wavelengths were recorded using rocket missions in the early 1960s (see for review Evans and Pounds, 1968; Davis et al., 1975)). The corresponding data were available for a limited duration depending on the flight time of the rocket. These rocket missions were followed by small satellites, namely Orbiting Solar Observatories (OSO; Neupert et al., 1967; Bruner, 1977b). Skylab was the first space-based observatory dedicated to study the solar atmosphere (Poland et al., 1973).

Since then, many missions dedicated to such studies have been accomplished.

These include P78-1 (Doschek, 1983), High-Resolution Telescope and Spectrometer (HRTS; Brueckner and Bartoe, 1983), Coronal Helium Abundance Spacelab Experiment (CHASE; Breeveld et al., 1988), Solar Maximum Mission (SMM; Acton et al., 1980), Hinotori (Tanaka et al., 1982), Yohkoh (Culhane et al., 1991), Solar and Heliospheric Observatory (SoHO; Domingo et al., 1995b), Reuven Ramaty High-Energy Solar Spectroscopic Imager (RHESSI; Lin et al., 2002), Hinode (Kosugi et al., 2007a), Solar Dynamics Observatory (SDO; Pesnell et al., 2012b)), Interface Region Imaging Spectrograph (IRIS; De Pontieu et al., 2014b)), Parker Solar Probe (PSP; Fox et al., 2016), and Solar Orbiter (SO; Müller et al., 2020). Additionally, these have been complemented by several imagers on sounding rocket missions, e.g., High-Resolution Coronal Imager (Hi-C; Kobayashi et al., 2014a; Rachmeler et al., 2019), Focusing Optics X-ray Solar Imager (FOXSI; Krucker et al., 2014), Christe et al. (2016), and Marshall Grazing Incidence X-ray Spectrometer (MAGIX; Kobayashi et al., 2010), Champey et al. (2016). This chapter will discuss the missions and the instruments whose observations have been used in the studies constituting this thesis, *viz.* SDO, IRIS, and FOXSI.

## 2.1 Solar Dynamics Observatory (SDO)

The Solar Dynamics Observatory is a NASA mission to study the energetics of the solar atmosphere and its coupling with the earth's atmosphere (Pesnell et al., 2012b). The mission was launched on $11^{th}$ February 2010. SDO carries three payloads, namely Atmospheric Imaging Assembly (AIA), Extreme Ultra-violet Variability Experiment (EVE), and Helioseismic and Magnetic Imager (HMI) (see Figure 2.1).

It was designed to have a geosynchronous orbit with an inclination of 28

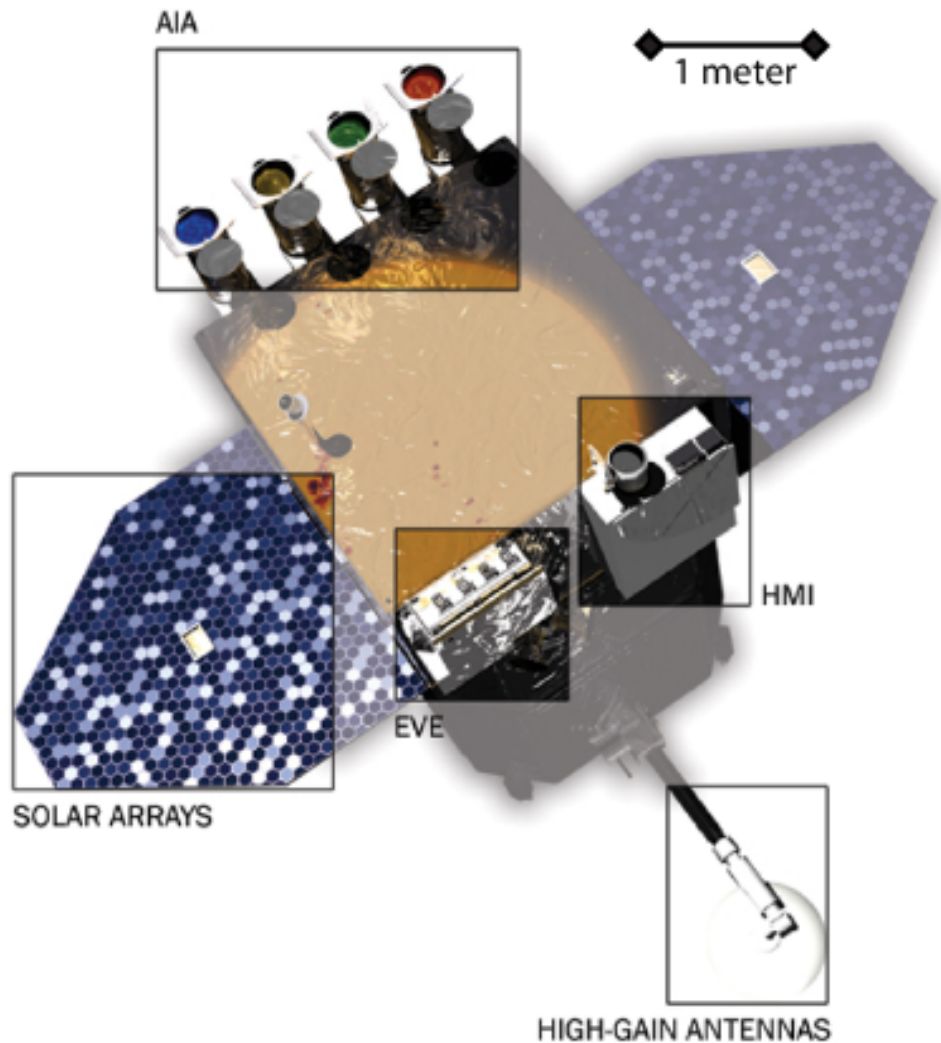

Figure 2.1: A layout of SDO illustrating the three scientific payloads (AIA, EVE, and HMI). Image courtesy: Pesnell et al. (2012b)

degrees about the longitudinal plane of its ground station in New Mexico (USA). Though this mission was originally planned for five years, it has been successfully providing data for over a decade. For this thesis, we have used observations from AIA and HMI. In this section, we provide a few salient features of these instruments.

### 2.1.1 Atmospheric Imaging Assembly (AIA)

The AIA consists of 4 telescopes that observe the solar atmosphere in seven extreme ultraviolet (EUV) bands and two continua (Lemen et al. (2012a)). It provides full-disk images of the solar atmosphere extending up to 0.5 $R_\odot$ above the limb. The images have a pixel size of $\approx$0.6″ and temporal resolution (cadence) of 12 seconds for EUV filters and 24 seconds for filters observing UV lines and nearby continuum. Table 2.1 lists all the filters, the dominant component of the plasma observed by these filters, and their characteristic temperature (Lemen et al., 2012a). Table 2.2 shows the line/continuum emission dominating in different EUV filters for features like (i) flares (FL), (ii) active regions (AR), (iii) quiet Sun (QS), (iv) coronal holes (CH) (see for e.g. O'Dwyer et al., 2010b; Boerner et al., 2012). In this thesis, we have used observations from 94, 131, 171, 193, 211, 335, and 1600 Å wavebands of AIA.

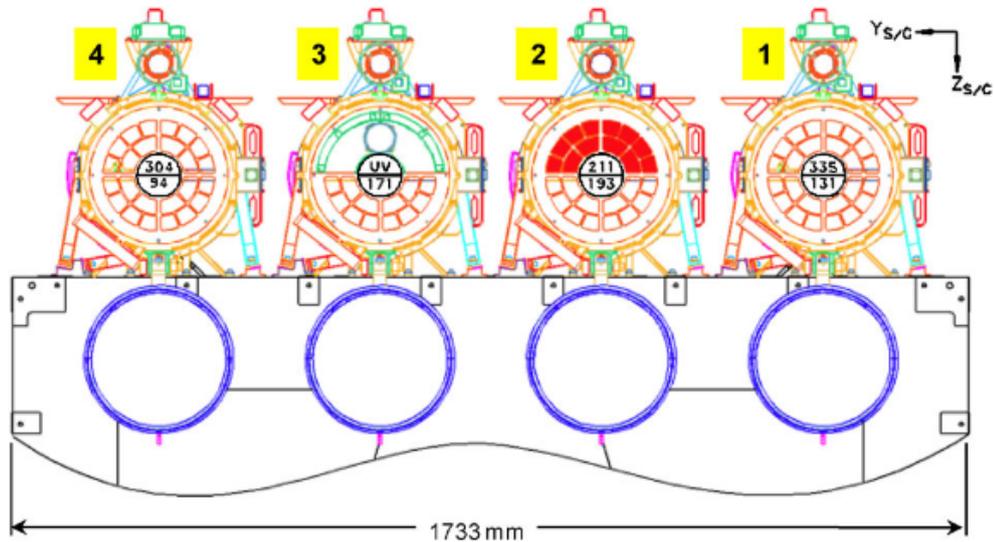

Figure 2.2: A layout of AIA illustrating the four telescopes and eight filters. Image courtesy: Lemen et al. (2012a)

The response function can be computed using (i) theoretical modeling based on atomic physics[†], and (ii) responsiveness of filter to radiation at different wavelengths. In Figure 2.3 the response function of six EUV filters, which mainly observe iron emission lines, are shown. The response function of the 304 Å filter is not shown since atomic physics-based modeling of the emission line at 304 Å corresponding to He II is problematic. This filter is consequently not included while performing thermal diagnostics.

Table 2.1: The primary ions observed by different channels of AIA. Column 1 shows the ion(s), Column 2 and 3 show their characteristic temperature ($\log[\mathrm{T(K)}]$) and the layer of atmosphere in which they are formed. (Source Lemen et al. (2012b))

| Channel | Ions observed | ($\log[\mathrm{T(K)}]$) | Layer of atmosphere |
|---------|---------------|-------------------------|---------------------|
| 4500 Å | continuum | 3.7 | photosphere |
| 1700 Å | continuum | 3.7 | photosphere |
| 304 Å | He II | 4.7 | chromosphere, transition region |
| 1600 Å | C IV and nearby continuum | 5.0 | transition region, upper photosphere |
| 171 Å | Fe IX | 5.8 | quiet corona, upper transition region |
| 193 Å | Fe XII, XXIV | 6.2, 7.3 | corona and hot flare plasma |
| 211 Å | Fe XIV | 6.3 | active region corona |
| 335 Å | Fe XVI | 6.4 | active region corona |
| 94 Å | Fe XVIII | 6.8 | flaring corona |
| 131 Å | Fe VIII, XXI | 5.6, 7.0 | transition-region, flaring corona |

Table 2.2: Emission lines predominantly observed by different AIA filters in different coronal features. Source O'Dwyer et al. (2010a)

| Channel | FL ($\log[\mathrm{T(K)}]$) | AR ($\log[\mathrm{T(K)}]$) | QS ($\log[\mathrm{T(K)}]$) | CH ($\log[\mathrm{T(K)}]$) |
|---------|------|------|------|------|
| 94 Å | Fe XVIII (6.85) | Fe XVIII (6.85) | Fe X (6.05) | Fe X (6.05) |
| 131 Å | Fe XXI (7.05) | Cont.* | Fe VIII (5.6) | Fe VIII (5.6) |
| 171 Å | Fe IX (5.85) | Fe IX (5.85) | Fe IX (5.85) | Fe IX (5.85) |
| 193 Å | Fe XXIV (7.25) | Fe XII (6.2) | Fe XII (6.2) | Fe XI, XII (6.15, 6.2) |
| 211 Å | Cont. | Fe XIV (6.3) | Fe XI, XIV (6.15, 6.3) | Fe XI (6.15), Cont. |
| 304 Å | He II (4.7) | He II (4.7) | He II (4.7) | He II (4.7) |
| 335 Å | Fe XVI (6.45) | Fe XVI (6.45) | Fe X (6.05), Mg VIII (4.7) | Mg VIII (4.7) |

---

[†]CHIANTI database is an atomic physics database which contains information about most of the emission lines observed in the solar atmosphere Dere et al. (1997a); Del Zanna et al. (2015a); Young et al. (2016). A complete list of different versions can be found at https://chiantidatabase.org

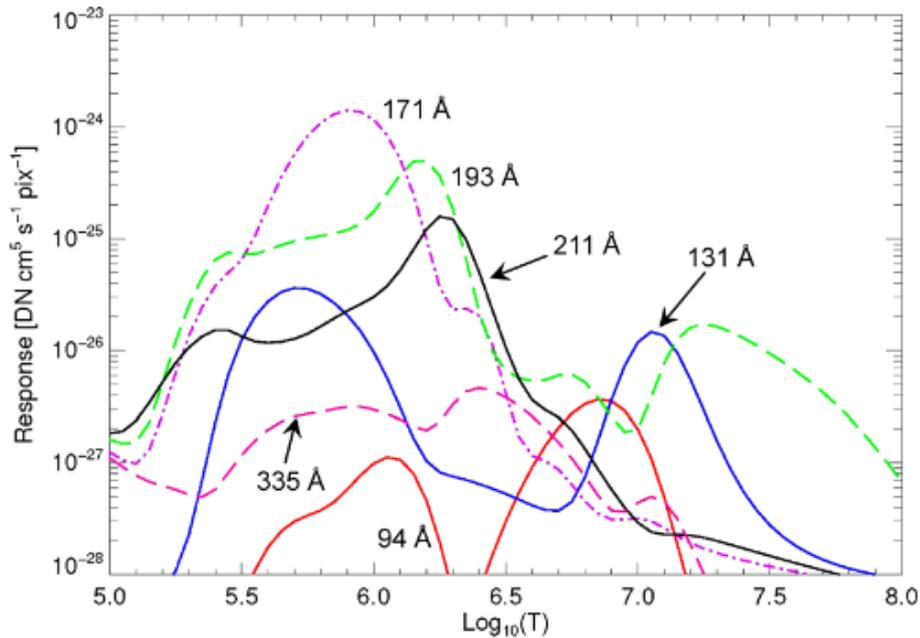

Figure 2.3: Response function of six AIA filters which chiefly observe different iron emission lines. Image courtesy: Pesnell et al. (2012a)

## 2.1.2 Helioseismic and Magnetic Imager (HMI)

The HMI on board the SDO (Scherrer et al. (2012); Schou et al. (2012b)) consists of three components, viz. optics package, electronics box, and a harness to connect them (see Figure 2.4). It observes the Fe I absorption line at 6173 Å and uses spectral and polarimetric techniques for providing a measurement of Doppler shifts, vector magnetic fields, and intensities at the photosphere. The full disk measurements of (i) Doppler shifts, (ii) magnetic flux along the line of sight, and (iii) continuum intensities are made at a cadence of 45 seconds and pixel size of 0.5". These are recorded on 4096×4096 pixels CCD camera. Two sets of full disk vector magnetic fields are recorded at a cadence of 90 and 135 seconds.

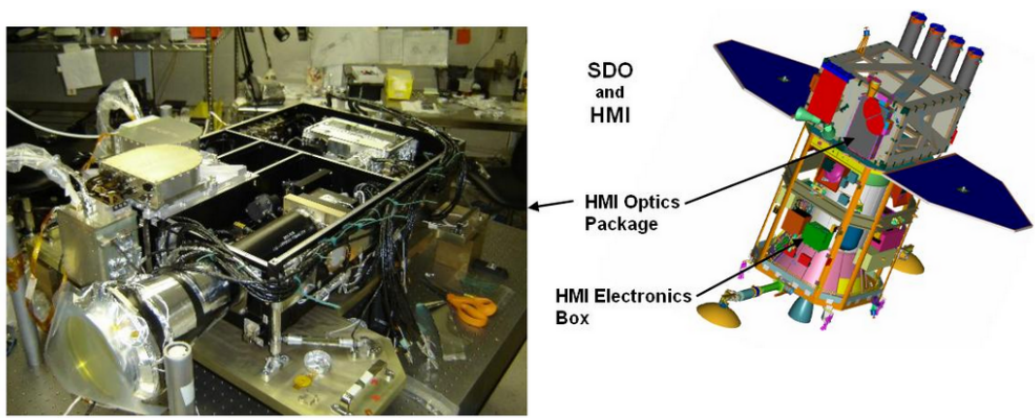

Figure 2.4: A view of HMI onboard SDO and its two main constituents: optics package and electronics box. Image courtesy: HMI homepage Stanford Solar group, Stanford University (http://hmi.stanford.edu).

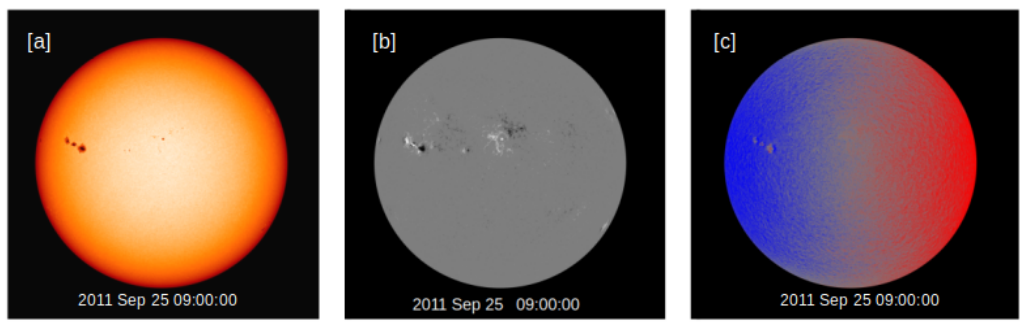

Figure 2.5: Full disk intensity maps in Fe-I absorption line [a] recorded on $25^{th}$ of September, 2011 at 09:00:00 UT. Corresponding full disk images showing derived line of sight magnetic fields [b] and Doppler shifts [c]. Image courtesy: Science Visualization Studio, NASA (https://svs.gsfc.nasa.gov)

## 2.2 Interface Region Imaging Spectrograph (IRIS)

The Interface Region Imaging Spectrograph (IRIS) was launched in 2013 (De Pontieu et al. (2014b)) into a Sun-synchronous orbit. It borrows its name from its ability to observe the chromosphere and the transition region, which act as an interface between the photosphere and the corona. It provides spectra and images with spatial resolutions varying between 0.33 (FUV) and 0.4" (NUV) and a cadence of up to 2 s. The spectra obtained allow us to resolve velocities of $10^5$ cm s$^{-1}$. The effective spectral resolution of IRIS varies from 26 mÅ

h]

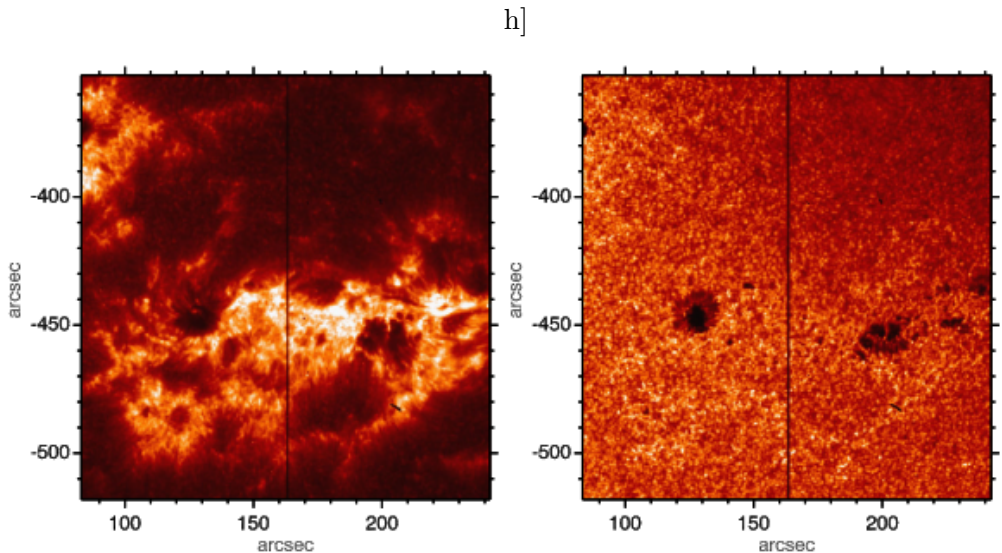

Figure 2.6: IRIS slit-jaw images at 2796 Å (left) and 2830 Å (right) of AR 11817. These wavelength windows view plasma roughly from the upper chromosphere (2796 Å) and upper photosphere (2830 Å). The dark line in the center of the image shows the slit position. Image courtesy: De Pontieu et al. (2014b)

(FUV) to 53 mÅ(NUV). The field of view can extend up to 175" × 175". Figure 2.6 shows images taken by IRIS in two wavelength windows centered on 2796 Å and 2830 Å, respectively. These roughly correspond to the upper chromosphere and upper photosphere, respectively.

The temperature of the plasma, which IRIS can observe, covers around four orders of magnitude extending from $5 \times 10^3$ K to $10^7$ K. This helps in studying different layers and their dynamic coupling. The wavelength windows observed by IRIS along with the ionized spectral lines are observed and the peak temperature of formation are listed in Table 2.3 (De Pontieu et al., 2014b). IRIS can provide spectral rasters in the following basic modes (i) dense rasters (the difference between consecutive raster locations is equal to the slit width), (ii) sparse/coarse rasters (the difference between consecutive raster locations is larger than the slit width), (iii) sit and stare (no rastering), and (iv) multi-point dense/sparse rasters (taken at specific locations).

Table 2.3: Wavelength windows in IRIS and some properties. Source De Pontieu et al. (2014a)

| Central Wavelength | Ion | Temperature ($\log[T]$) |
|---|---|---|
| 2820 Å | Mg II wing | 3.7-3.9 |
| 1355.6 Å | O I | 3.8 |
| 2803.5 Å | Mg II h | 4.0 |
| 2796.4 Å | Mg II k | 4.0 |
| 1334.5 Å | C II | 4.3 |
| 1335.7 Å | C II | 4.3 |
| 1402.8 Å | Si IV | 4.8 |
| 1393.8 Å | Si IV | 4.8 |
| 1399.8 Å | O IV | 5.2 |
| 1401.2 Å | O IV | 5.2 |
| 1349.4 Å | Fe XII | 6.2 |
| 1354.1 Å | Fe XXI | 7.0 |

## 2.3 Focusing Optics X-ray Solar Imager (FOXSI)

Focusing Optics X-ray Solar Imager (FOXSI) was launched using sounding rockets on three occasions in 2012, 2014, and 2018. FOXSI is designed to observe hard X-rays emitted due to the presence of non-thermal electrons, which could be signatures of the theorized concept of nano-flares that occur even in the quiet Sun (Krucker et al. (2014)). In this thesis, we have used observations from FOXSI 2 (2014). It had detectors made from Si and CdTe. The angular resolution of Si and CdTe detectors was 7.7″ and 6.2″, respectively. It was designed to observe X-ray photons in the 4–20 keV range. However, the spectral response of the detectors to photons below 5 keV energy is not well understood (Christe et al. (2016)). FOXSI can be used to study the emission from plasma at temperatures larger than $\log[T(K)] \approx 6.5$. These observations can be used in conjunction with AIA observations of relatively cooler plasma to cover emissions from plasma over a wide range of temperatures. Figure 2.7 shows the temperature response of FOXSI-2 in three energy bands (5–6, 6–7, and 7–8 keV).

Athiray et al. (2020) constructed the temperature response for FOXSI-2 from multiple isothermal emission models in the range of $1$–$30\times10^6$ K in steps of $\delta\log(\text{T}) = 0.05$. Using coronal abundances, they created a synthetic X-ray photon spectrum using the CHIANTI database (Dere et al., 1997b; Landi et al., 2013). Each synthetic photon spectrum was then used along with the FOXSI-2 instrument response obtained from ground calibration data to obtain synthetic counts as a function of photon energy. They obtained average counts integrated over one keV energy bin in the 4–10 keV range. This provides predicted counted rates as a function of plasma temperature and photon energy.

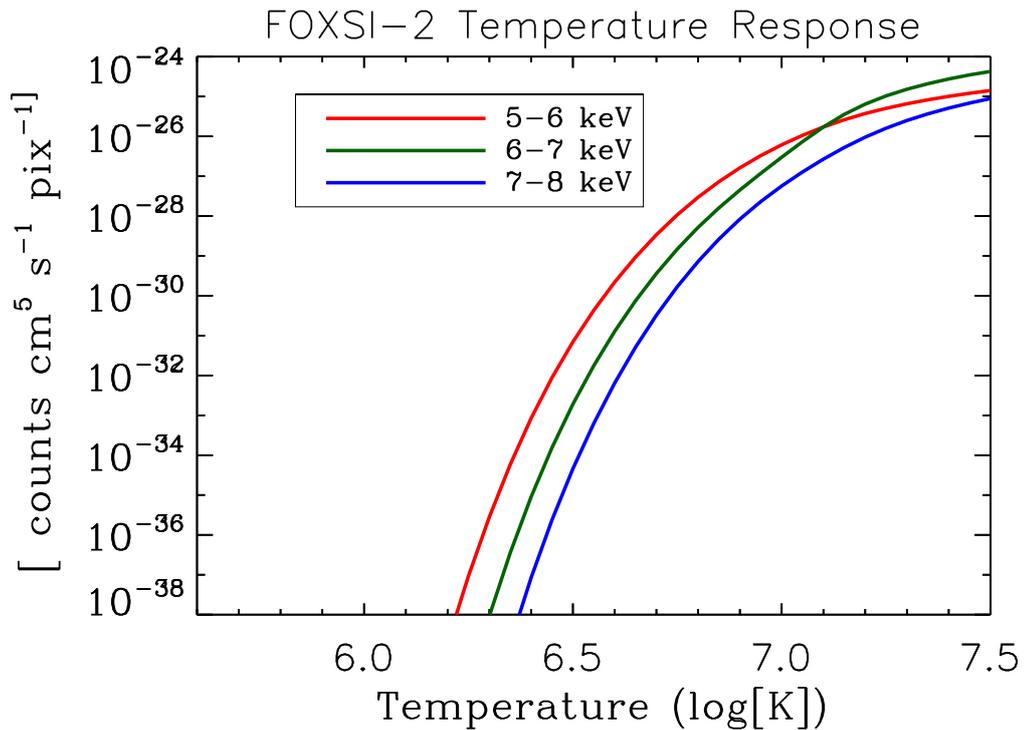

Figure 2.7: The temperature response of three energy bands of FOXSI-2 was used in this work. The red, green, and blue curves show the temperature response of FOXSI-2 in 5-6 keV, 6-7 keV, and 7-8 keV energy bands, respectively.

# Numerical modeling

*In this chapter we discuss the description of coronal loops used for performing observation-motivated modeling. We first discuss the fluid description of magnetized plasma in corona (MHD) and the regime of its application along with justification of the important approximations. We then use conditions in solar corona to derive simplified magnetic field-aligned hydrodynamics equations. This is followed by a further simplified 0D description of coronal loops. This thesis uses the EBTEL code for performing simulations based on this 0D description. Parts of Section 3.2 discussing EBTEL formalism developed by Klimchuk et al. (2008) and Cargill et al. (2012a) have been reproduced with permission from Rajhans et al. 2022, "Flows in enthalpy-based thermal evolution of loops", ApJ, 924, 13 (DOI: 10.3847/1538-4357/ac3009). Figures 3.2 and 3.3 have been taken with permission from Rajhans et al. (2021).*

The solar corona has a very high magnetic Reynold's number ($> 10^{8-12}$) (Priest, 2014). Therefore, it is reasonable to consider magnetic flux to be frozen in the plasma. The additional condition of magnetic pressure dominating gas pressure, i.e., low $\beta$, means that plasma flows in directions perpendicular to the magnetic field lines are heavily suppressed. Due to these properties, the corona contains multiple loop-like structures. The plasma confined in magnetically closed corona in the active regions is brighter than the background and can be observed easily. This makes them an important system to study using simulations. Figure 3.1 shows an extreme ultraviolet image of the solar atmosphere taken from AIA onboard SDO in 171 Å channel. Bright loops are readily visible in the active regions.

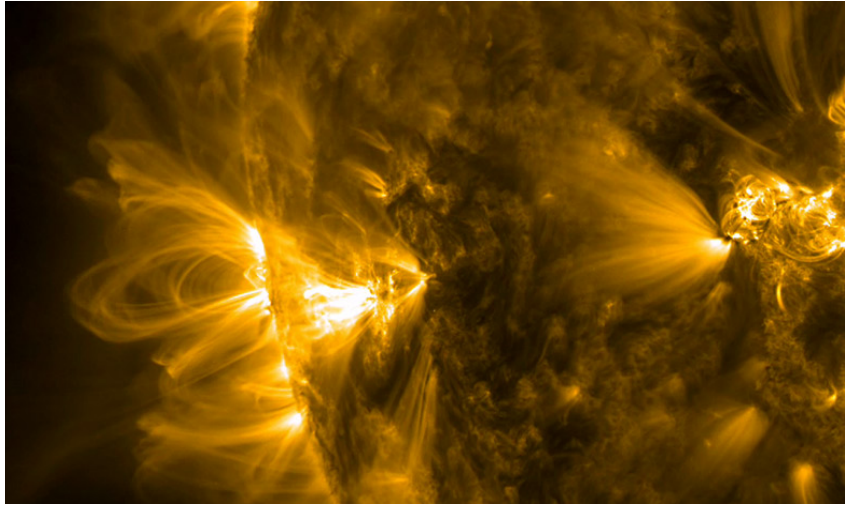

Figure 3.1: An extreme ultraviolet image of the solar atmosphere recorded by AIA using 171Å channel. Image credit: NASA SDO (https://www.nasa.gov/sites/default/files/festooningloops.jpg)

It is believed that the corona is heated by the dissipation of energy at length scales much smaller than the characteristic length scales of the system (Klimchuk, 2006). The identification of such length scales is much beyond the current computational power. Hence, a cut-off at length scales larger than dissipation lengths is provided using artificial resistivity and viscosity.

When plasma is in local thermodynamical equilibrium, the fluid (continuum) picture is sufficient to understand its several aspects. The theoretical formulation of the fluid picture of plasma is known as Magneto-Hydro-Dynamics (MHD). The fluid picture is valid at (i) length scales much larger than the mean free path length and ion gyro-radius, and (ii) time scales much larger than the mean free path time and ion gyro-period. Ion gyro-radius and gyro-period are larger than those of electrons because of the much higher mass of the ions.

# 3.1 Magnetohydrodynamic and field-aligned hydrodynamic equations

MHD equations need to be solved simultaneously taking into account radiation loss and energy transport. These are subject to suitable initial and boundary conditions. The typical velocities in the solar corona are of the order of $10^{6-7}$ cm s$^{-1}$. These are non-relativistic and an order of magnitude higher than the velocities associated with solar rotation ($10^5$ cm s$^{-1}$) (see for e.g. Priest, 2014). If the plasma is sufficiently collisional, which is the case in corona, electrons and ions can be assumed to be at the same temperature (Priest, 2014). Under these conditions, the set of equations for single fluid plasma includes the mass conservation equation

$$\frac{\partial \rho}{\partial t} + \nabla.(\rho \mathbf{v}) = 0 \tag{3.1}$$

equation of motion

$$\rho \left( \frac{\partial \mathbf{v}}{\partial t} + \mathbf{v}.\nabla \mathbf{v} \right) = -\nabla P + \frac{1}{\mu_0}(\mathbf{j} \times \mathbf{B}) + \nu \nabla^2 \mathbf{v} + \nu \nabla(\nabla.\mathbf{v}) + \rho \mathbf{g} \tag{3.2}$$

induction equation

$$\frac{\partial \mathbf{B}}{\partial t} = \nabla \times (\mathbf{v} \times \mathbf{B}) + \eta \nabla^2 \mathbf{B} \tag{3.3}$$

and energy conservation equation

$$\frac{\partial}{\partial t} \left( \frac{P}{\gamma - 1} + \frac{\rho}{2}v^2 + \frac{B^2}{2\mu_0} \right) + \nabla. \left( \frac{\gamma P \mathbf{v}}{\gamma - 1} + \frac{\rho \mathbf{v} v^2}{2} + \mathbf{F} + \frac{\mathbf{E} \times \mathbf{B}}{\mu_0} \right) = X + \rho \mathbf{g}.\mathbf{v} - R \tag{3.4}$$

where $\rho$, $\mathbf{v}$, $P$, $\mathbf{j}$, $\mathbf{B}$, $\mathbf{E}$, and $g$ represent density, velocity, gas pressure, current density, magnetic field, electric field, and acceleration due to gravity, respectively. $\gamma$ is the polytropic index, $\nu$ is the coefficient of viscosity, $\eta$ is the magnetic diffusivity, $\mu_0$ is permeability of free space, and $e$ is charge on electron. $\mathbf{F}$ and R represent the conduction flux and radiation loss respectively. X is a collection of terms involving viscosity and magnetic diffusivity

(Galtier, 2016) given by

$$X = \nabla.\left(\frac{4}{3}\nu(\nabla.\mathbf{v})\mathbf{v} - \nu\mathbf{v} \times \nabla \times \mathbf{v}\right) - \nu|\nabla \times \mathbf{v}|^2 - \mu_0\eta|\mathbf{j}|^2 - \frac{4}{3}\nu(\nabla.\mathbf{v})^2 \quad (3.5)$$

The current density is related to the magnetic field by Maxwell's equation

$$\mu_0\mathbf{j} = \nabla \times \mathbf{B} \quad (3.6)$$

The electric field is related to the magnetic field and current density using Ohm's law given by

$$\eta\mathbf{j} = \mathbf{E} + \mathbf{v} \times \mathbf{B} \quad (3.7)$$

The magnetic field is subject to the constraint

$$\nabla.\mathbf{B} = 0 \quad (3.8)$$

Finally, the above systems are closed by an equation of state given by

$$P = 2nk_BT \quad (3.9)$$

where $k_B$ is the Boltzmann's constant. Note that $P$ is the total pressure of electrons and ions, while $n$ is the electron number density. The plasma can be approximated to be a fully ionized hydrogen plasma Klimchuk et al. (2008). The factor of 2 shows that in such a plasma where the main contribution is protons and electrons, half of the contribution $(nk_BT)$ comes from electrons and the other half from ions. However, Klimchuk et al. (2008) show that the presence of the second most abundant element i.e. Helium (in completely ionized form) can be incorporated by modifying $k_B$. The presence of other species is neglected while calculating the pressure. The electron number density $n$ is related to $\rho$ and effective atomic mass $\mu$ by the relation: $\rho = n\mu$.

The major drawback is that performing such MHD simulations requires a lot of computational power. Additionally, it is challenging to interpret and reconcile the results with observations. A simpler description is possible if the study's objective is to model the plasma response in coronal loops to a heating event. This is the magnetic field-aligned description of plasma, which we discuss below.

On length scales larger than those at which viscous and diffusive dissipation occur, all terms with coefficients of $\eta$ and $\nu$ can be neglected in MHD equations. In this case equation 3.7 (Ohm's law) reduces to

$$\mathbf{E} + \mathbf{v} \times \mathbf{B} = 0 \tag{3.10}$$

and equation 3.3 (induction equation) simplifies to

$$\frac{\partial \mathbf{B}}{\partial t} = \nabla \times (\mathbf{v} \times \mathbf{B}) \tag{3.11}$$

However, dissipation effects cannot be neglected completely because they will result in local heating and heat flux across the loop. Hence we introduce an ad hoc heating function $Q$ in our energy conservation equation (Priest, 2014). After further simplification and using equations 3.10 and 3.11 we can write equation 3.4 (energy conservation equation) as

$$\frac{\partial}{\partial t}\left(\frac{P}{\gamma - 1} + \frac{\rho}{2}v^2\right) + \nabla.\left(\frac{\gamma P}{\gamma - 1}\mathbf{v} + \frac{\rho}{2}v^2\mathbf{v} + \mathbf{F}\right) = \rho\mathbf{g}.\mathbf{v} - R - \mathbf{j}.(\mathbf{v} \times \mathbf{B}) + Q \tag{3.12}$$

We now consider equation 3.2 (equation of motion) and decompose it into motion parallel and perpendicular to magnetic field. We get

$$\rho\frac{d\mathbf{v}_{||}}{\partial t} = -\nabla_{||}P + \rho\mathbf{g}_{||} \tag{3.13}$$

and

$$\rho\frac{d\mathbf{v}_{\perp}}{\partial t} = -\nabla_{\perp}P + \mathbf{j} \times \mathbf{B} + \rho\mathbf{g}_{\perp} \sim \mathbf{j} \times \mathbf{B} \tag{3.14}$$

where the subscripts $||$ and $\perp$ denote the components parallel and perpendicular to the magnetic fields, respectively. We have used the fact that $\mathbf{j} \times \mathbf{B}$ has no component along $\mathbf{B}$. Also, since magnetic forces dominate in the solar corona due to low plasma $\beta$, the contribution of all terms in the equation for $\mathbf{v}_\perp$ can be neglected. The solar corona can be assumed to be force-free for static magnetic field configurations (see for e.g. Wiegelmann and Sakurai, 2012). As $\mathbf{j} \times \mathbf{B}$ affects only $v_\perp$ and dominates other forces, the force-free condition implies

$$\mathbf{j} \times \mathbf{B} = 0 \tag{3.15}$$

Since there is no force perpendicular to $\mathbf{B}$ we neglect any motion perpendicular to it, i.e., $v_\perp = 0$. This means that $\mathbf{v} \times \mathbf{B}$ vanishes. Using this condition, equation 3.11 (induction equation) becomes

$$\frac{\partial \mathbf{B}}{\partial t} = 0 \tag{3.16}$$

and equation 3.12 (energy conservation equation) becomes

$$\frac{\partial}{\partial t}\left(\frac{P}{\gamma - 1} + \frac{\rho}{2}v^2\right) + \nabla.\left(\frac{\gamma P}{\gamma - 1}\mathbf{v} + \frac{\rho}{2}v^2\mathbf{v} + \mathbf{F}\right) = \rho\mathbf{g}.\mathbf{v} - R \tag{3.17}$$

In low $\beta$ coronal plasma, $|\mathbf{F}_\perp| \ll |\mathbf{F}_{||}|$ (Braginskii, 1965). Hence the only spatial degree of freedom in these equations is the field-aligned coordinate denoted by $s$.

The field-aligned equations can be described in terms of $n$ by

mass conservation equation

$$\frac{\partial n}{\partial t} + \frac{\partial J}{\partial s} = 0 \tag{3.18}$$

where J is the electron flux given by $nv$,

equation of motion

$$\left(\frac{\partial v}{\partial t} + v\frac{\partial v}{\partial s}\right) = -\frac{1}{n\mu}\frac{\partial P}{\partial s} + g_{||} \tag{3.19}$$

and energy conservation equation

$$\frac{\partial}{\partial t}\left(\frac{P}{\gamma - 1} + \frac{n\mu}{2}v^2\right) + \frac{\partial}{\partial s}\left(\frac{\gamma P}{\gamma - 1}v + \frac{n\mu}{2}v^3 + F\right) = n\mu g_{||}v + Q - R \tag{3.20}$$

where $F$, and $v$ have been used instead of $F_\parallel$ and $v_\parallel$ for brevity.

Emission from the solar corona is primarily a result of the excitation of ions when they collide with ions. Since the densities of both are equal in a quasi-neutral fully ionized hydrogen plasma we can write

$$R = n^2 \Lambda(T) \tag{3.21}$$

where $\Lambda(T)$ is the temperature dependent optically thin radiative loss function (Klimchuk et al., 2008).

## 3.2 Enthalpy-Based Thermal Evolution of Loops (EBTEL)

Field-aligned 1D hydrodynamic simulations serve as an efficient tool for studying the response of plasma in coronal loops to a generic time-dependent heating event (see, e.g. Klimchuk, 2006, 2015b; Reale, 2014). However, in situations where one needs to study the effects of variation of parameters like loop length, energy budget, and heating function profile, many runs are required. Additionally, more realistic scenarios involve multi-stranded loops. Performing field-aligned simulations for such complicated but realistic systems is also relatively computationally expensive. To overcome such issues, 0D codes have been developed.

1D description of coronal loops suggests that the instantaneous variation in physical quantities like pressure, density, and temperature across the corona loop is not drastic. The variation is such that average values over the coronal loop are a fair representation of the system at a given time (see for e.g. Bradshaw and Mason, 2003; Bradshaw and Cargill, 2006; Cargill et al., 2012b). The study of the instantaneous coronal averages is referred to as a 0D description of coronal loops.

Enthalpy-based thermal evolution of loops (EBTEL) is a code based on this description (Klimchuk et al., 2008; Cargill et al., 2012a; Barnes et al., 2016). It solves for $\bar{P}$, $\bar{n}$ and $\bar{T}$ (the bar indicates coronal averages). It also solves for the velocity at the coronal base $v_0$, the instantaneous differential emission measures (DEM) from the corona, and the transition region, which can be used for generating synthetic light curves. We now describe briefly the EBTEL framework outlined in Klimchuk et al. (2008) and Cargill et al. (2012a).

Figure 3.2 summarizes the important features of the EBTEL framework. The loop is assumed to be symmetric and has a constant cross-section. Due to symmetry, quantities like $F$, $J$, and $v$ vanish at the loop's apex ($s = L$), and we need to simulate only half of the loop. The coronal base is the position at which thermal conduction changes from a cooling term to a heating term. Mathematically, this is the point where the second spatial derivative of temperature changes its sign. Field-aligned simulations show that such change in the sign occurs at a position along the loop where the temperature is approximately 0.6 times the temperature at the loop apex (Cargill et al., 2012a).

The kinetic energy and gravitational energy are neglected from the 1D field-aligned energy equation in the EBTEL framework. Due to coronal temperatures of the order of $10^6$ K and consequent large sound speeds ($1.5 \times 10^4\ T^{\frac{1}{2}}$ cm s$^{-1}$), it seems reasonable to assume the flows to be subsonic. In cgs units, the corona has a characteristic electron number density of $10^9$, the velocity of $10^{6-7}$, $\Lambda$ of $10^{-21}$ (for $T = 10^6 K$). Note that $g$ is of the order of $10^5$ in cgs units. Using these values, the gravitational energy term can also be neglected in comparison to the radiative loss term. The field-aligned equation for energy conservation (equation 3.20) reduces to

$$\frac{\partial}{\partial t}\left(\frac{P}{\gamma - 1}\right) + \frac{\partial}{\partial s}\left(\frac{\gamma P}{\gamma - 1}v + F\right) = Q - n^2\Lambda(T) \qquad (3.22)$$

On integrating equation 3.22 from $s = 0$ (coronal base) to $s = L$ (apex of loop), we get

$$L \frac{d}{dt} \left( \frac{\bar{P}}{\gamma - 1} \right) - \frac{\gamma P_0 v_0}{\gamma - 1} - F_0 = \bar{Q} L - R_{cor} \qquad (3.23)$$

where $R_{cor} = \bar{n}^2 \Lambda(\bar{T}) L$ is the total radiative loss from corona. We have assumed that the average of products is the same as products of averages. This is justified for quantities involving $P$, $n$, and $T$, which differ by only a few factors along the loop at a given time.

$F_0$ is the heat flux across the coronal base. The flux due to particles in the thermal pool is a combination of Spitzer flux and saturation flux. The former accounts for Coulomb collisions and the latter is correct for over-prediction by classical expression. Non-thermal electrons might carry away a significant fraction of energy. We can also readily include the effect of non-thermal electrons. For this purpose, we need the flux of non-thermal electrons across the coronal base ($J_{nt0}$) and average energy per non-thermal electron ($E_{nt}$) as additional inputs. Under such a scenario, $F_0$ is the sum of the thermal conduction flux ($F_{t0}$) and the energy flux carried by non-thermal electrons ($E_{nt}J_{nt0}$), across the coronal base. Similarly, $J_0$ becomes the sum of the flux of electrons in the thermal pool ($n_0 v_0$) and non-thermal electrons ($J_{nt0}$) across the coronal base.

The transition region can be assumed to be in a steady state and any energy flux across the base of the transition region ($s = -l$) is assumed to be negligible. Consequently on integrating equation 3.22 from $s = -l$ to $s = 0$ we get

$$\frac{\gamma P_0 v_0}{\gamma - 1} + F_0 = \bar{Q} l - R_{tr} \qquad (3.24)$$

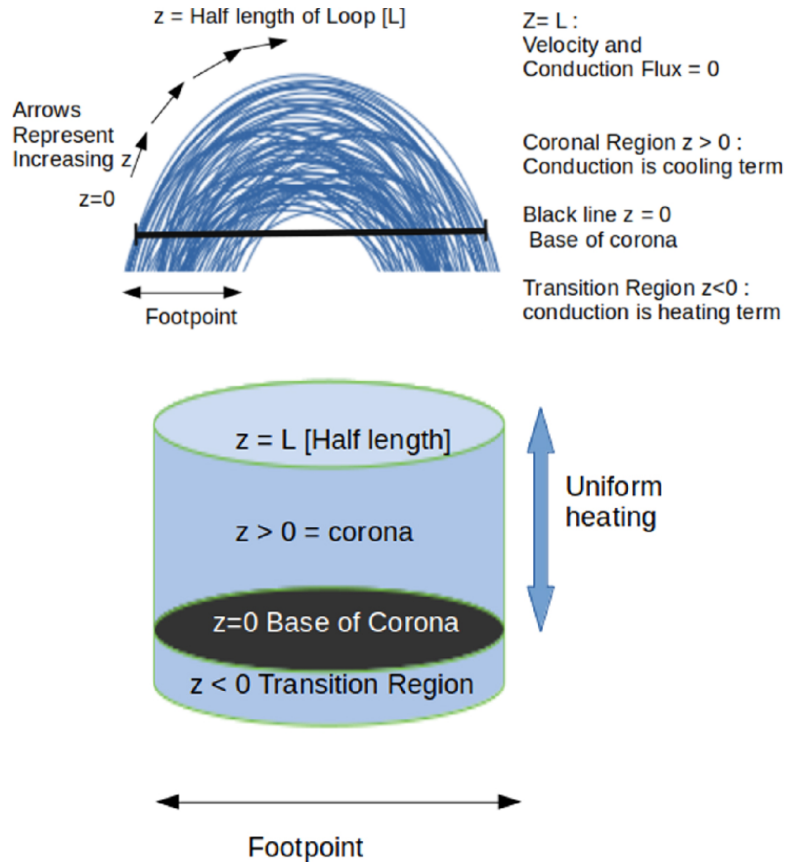

Figure 3.2: Sketch of the loop set-up. (Top) A flux tube filled with multiple strands rises up from the chromosphere. The thick horizontal line represents the base of the corona, and the loop is assumed to be symmetrical around the apex. (Bottom) The representation of half of the loop in EBTEL. The plasma aligned along the field lines is shown straightened out, with the various regions corresponding to the physical system labeled. (Image adapted from Rajhans et al. (2021).)

where $R_{tr}$ is the total radiative loss from the transition region given by

$$R_{tr} = c_1 R_{cor} \qquad (3.25)$$

Klimchuk et al. (2008) used results from 1D simulations to show that $R_{cor}$ and $R_{tr}$ maintain a constant ratio at all times and assumed $c_1$ to be 4. However, Cargill et al. (2012a) computed it dynamically within the code.

On adding energy conservation equations integrated over the corona and transition region (equations 3.23 and 3.24) one gets

$$L\frac{d}{dt}\left(\frac{\bar{P}}{\gamma-1}\right) = \bar{Q}L - (1+c_1)\bar{n}^2\Lambda(\bar{T})L \qquad (3.26)$$

where $\bar{Q}(l+L) \approx \bar{Q}L$ because $l \ll L$.

The equation for mass conservation (equation 3.18) can be integrated from $s = 0$ to $s = L$ to give

$$L\frac{d\bar{n}}{dt} = J_0 \qquad (3.27)$$

where $J_0$ is the flux of electrons across $s = 0$ and is given by $n_0 v_0 = \frac{P_0}{2k_B T_0}v_0$. If there is an additional flux of non-thermal electrons, $J_0 = \frac{P_0}{2k_B T_0}v_0 + J_{nt0}$. In either case for expressing $P_0$ in terms of $\bar{P}$, it is assumed that the loop is in hydrostatic equilibrium and is isothermal, with a temperature equal to the average coronal temperature, i.e. $\left[\frac{P_0}{\bar{P}}\right] = \left[\frac{P_0}{\bar{P}}\right]_{hse}$. The subscript *hse* denotes hydrostatic equilibrium at a uniform temperature ($\bar{T}$).

The temperature at loop top ($T_a$) and coronal base ($T_0$) are related to $\bar{T}$ by a constants $c_2$, and $c_3$ such that

$$c_2 = \frac{\bar{T}}{T_A} = \quad \& \quad c_3 = \frac{T_0}{T_A} \qquad (3.28)$$

Results from 1D simulations performed for different loop lengths and heating functions show the best overall agreement for $c_2 = 0.9$ and $c_3 = 0.6$ (Klimchuk et al., 2008; Cargill et al., 2012a).

The system of equations (3.23, 3.26, and 3.27) is closed by equation of state

$$\bar{P} = 2\bar{n}k_B\bar{T} \qquad (3.29)$$

The presence of multi-thermal plasma can be expressed by the differential emission measure (DEM).

$$DEM(T) = n^2 \left(\frac{\partial T}{\partial s}\right)^{-1} \qquad (3.30)$$

where $n$ is the electron number density and $s$ is the coordinate along the line of sight.

The instantaneous temperature across the loop in the corona varies at most by a factor of 0.6. Hence, coronal contribution to DEM ( see equation 3.30) is computed assuming that the entire emission ($\bar{n}^2 L$) is uniformly distributed over the temperature range $[T_0 < T < T_A]$. We cannot make this assumption in the transition region. This is because density varies across the coronal loops by a few factors, while it varies by three-four orders of magnitudes in the transition region. However, under the assumption of the transition region being in a steady state, DEM from this region can be easily computed. In the absence of any direct heating in the transition region, equation 3.22 becomes

$$\frac{\partial}{\partial s}\left(\frac{\gamma P}{\gamma - 1}v + F\right) = -n^2 \Lambda(T) \tag{3.31}$$

Classical expression for conduction flux is suitable for temperatures in transition regions that are lower than the corona, which is given by

$$F = -\frac{2}{7}\kappa_0 \frac{\partial}{\partial s}\left(T^{\frac{7}{2}}\right) \tag{3.32}$$

where $\kappa_0$ is $10^{-6}$ in cgs units. If the scale lengths of temperature and conduction flux are same in the transition region (say $L_{sTR}$) we can write

$$\frac{\partial F}{\partial s} \approx -\frac{2}{7}\kappa_0 \frac{T^{\frac{7}{2}}}{L_{sTR}^2} = -\frac{2}{7}\kappa_0 T^{\frac{3}{2}}\left(\frac{T}{L_{sTR}}\right)^2 \approx\approx -\kappa_0 T^{\frac{3}{2}}\left(\frac{\partial T}{\partial s}\right)^2 \tag{3.33}$$

Combining equation 3.33 along with uniform electron flux ($J_0 = n_0 v_0$) and pressure across the transition region is given by

$$\kappa_0^{\frac{3}{2}}\left(\frac{\partial T}{\partial s}\right)^2 - 5k_B J_0\left(\frac{\partial T}{\partial s}\right) - \left(\frac{\bar{P}}{2k_B T}\right)^2 = 0 \tag{3.34}$$

This is a quadratic equation and can be readily solved. The sum of DEM from the corona and transition region is then used to generate synthetic light curves

Throughout this thesis, we have used symmetric triangular profiles for modeling transient events. Since we do not have theoretical constraints on the heating profile from the first principle, we select a triangular heating profile which is a fair representation of an impulsive event. It also avoids numerical issues which can be caused by abrupt changes in ratios of ambient heating rate and that associated with the transient heating event. To simulate an impulsive event, we need to characterize it by its energy budget and impulsiveness. These are fairly captured by the peak heating rate and the time in which energy is dissipated. The precise shape of the heating profile with the same peak heating rate and dissipation time is not necessary for the purpose of our study, which is to get rough estimates of the thermal evolution of coronal loops. Since the decay phase of the plasma response proceeds slower than the impulsive phase, the rise time of the heating profile is more important. For simplicity, we take the decay time of the heating function to be the same as the rise time.

An illustrative example of how EBTEL is used is shown in Figure 3.3. The upper panel shows the heating function applied to a loop of half-length $10^8$ cm. The energy deposited begins to rise linearly, starting at 60 s from an ambient value to a peak at 105 s and then declines linearly back to the ambient value at 150 s. The baseline ambient level is necessary to establish the presence of a corona, in this case, at a temperature of 0.92 $\times 10^6$ K and an electron number density of $7.8 \times 10^9$ cm$^{-3}$. This triangular heating pulse leads to an increase in the temperature of the plasma in the loop (middle panel), which reaches a maximum at a time close to the maximum heating. In contrast, the plasma density (bottom panel) rises more shallowly. It reaches a peak after the plasma fills the loop in response to the enthalpy flux from the transition region before it declines. Notice that the loop temperature drops below the ambient value after the heating pulse ends. This is due to the heated plasma becoming over-dense and cooling rapidly (due to the $T \propto n^2$ scaling law ) until the excess plasma is drained.

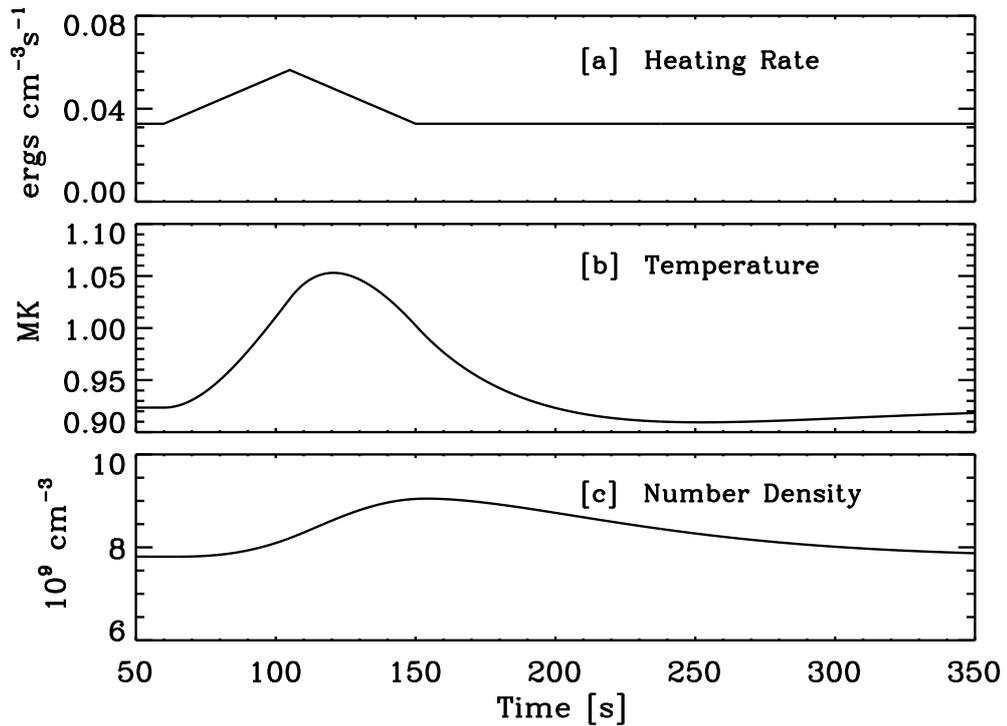

Figure 3.3: Plasma response of a loop of half length of $10^6$ cm to a triangular heating function computed using EBTEL. (a) The heat is given per unit volume to the loop in units of ergs cm$^{-3}$ s$^{-1}$. (b) Time evolution of temperature in $10^6$ K (c) Time evolution of electron number density in $10^9$ cm$^{-3}$. Image source: Rajhans et al. (2021)

# Hydrodynamics of small transient brightenings in the solar corona

*Small-scale transients occur in the Solar corona at much higher frequencies than flares and play a significant role in coronal dynamics. Here we study three well-identified transients discovered by Hi-C and also detected by AIA/SDO. We perform hydrodynamical simulations and produce synthetic light curves to compare with AIA observations. We have modeled these transients as loops of $\sim$1.0 Mm length depositing energies $\sim 10^{23}$ ergs in $\sim$50 seconds. During the initial phase, conduction flux from the corona dominates over the radiation, like impulsive flaring events. Our results further show that the time-integrated net enthalpy flux is positive, hence into the corona. The fact that we can model the observed light curves of these transients reasonably well by using the same physics as those for nanoflares, microflares, and large flares, suggests that these transients may have a common origin. This chapter, including all figures and tables, has been reproduced from Rajhans et al. 2021, "Hydrodynamics of small transient brightenings in the solar corona", ApJ, 917, 29 (DOI:10.3847/1538-4357/ac03bb).*

## 4.1 Introduction

The presence of high-temperature (>1 MK) plasma in the solar corona was discovered in the 1940s. How this plasma, above the much cooler photosphere, is heated to such high temperatures has been one of the most challenging questions in astrophysics. Though our understanding of the energy dissipation in the corona has improved substantially, the full solution to the problem of coronal heating remains elusive and the transfer of mass and energy between layers of the solar atmosphere is not completely understood (see, e.g., Klimchuk, 2006; Reale, 2014, for a review). Multiwavelength observations of the Sun show that different layers couple through magnetic fields. By and large, theories related to omnipresent coronal heating fall into two groups: AC heating and DC heating (see, e.g., Klimchuk, 2015a; Walsh and Ireland, 2003). Depending on the frequency of occurrence, heating events can be classified into high and low-frequency heating, with the former may contribute to steady and the latter to transient events (see, e.g., Tripathi et al., 2011; Winebarger, 2012).

Solar flares provide the best-observed examples of impulsive events taking place in the solar atmosphere.Hudson (1991) conjectured that to maintain the corona at a temperature greater than 1 MK with the help of impulsive events, there must be a large frequency of such events with smaller energy, and found that the relationship between a number of events and their energies obeys a power-law distribution $\frac{dN}{dE} \propto E^{-\alpha}$, where $dN$ is the rate of occurrence of events having energy in the range $[E, E + dE]$ and $\alpha$ is a positive number (see, e.g., Hannah et al., 2011, for a review). The power law distribution of flares of different energies is a sign of the underlying phenomenon of self-organized criticality (Lu and Hamilton, 1991). For the heating to be domin-ated by nanoflares, the power-law index ($\alpha$) should be greater than 2. This has led to many observational studies (see, e.g., Shimizu, 1995; Berghmans et al., 1998, 2001; Krucker and Benz, 1998, 2000; Parnell and Jupp, 2000; Aschwanden et al., 2000a,b; Christe et al., 2008; Hannah et al., 2008) based on the counting of different types of transient events that result in varied negative slopes of power-law ranging from $1.6 \lesssim \alpha \lesssim 2.2$. However, there are limitations to such studies due to constraints on the cadence, passbands, and resolutions of instruments. Also, there remains a chance that flares of different energy, particularly at lower energies, are undercounted (see, e.g., Pauluhn and Solanki, 2007; Upendran and Tripathi, 2021) because of multiple events of lower energies as being counted as a single event of larger energy. Additionally, it is possible that different events may be generated due to different mechanisms and hence they would not necessarily follow the same power-law distribution.

Habbal and Withbroe (1981) studied coronal bright points and used Ly$\alpha$ emission as a proxy for conduction losses from the corona into the chromosphere. Preś and Phillips (1999) were then able to establish that conduction losses are at least an order of magnitude larger than radiation losses, implying that radiation loss from the corona is a small fraction of total energy dissipated. The smallest brightenings detected thus far (Régnier et al., 2014; Subramanian et al., 2018) are due to the observations recorded by Hi-C (Kobayashi et al., 2014b). Subramanian et al. (2018) identified 27 such events in Hi-C images and performed a detailed study to understand their energetics using simultaneous observations obtained with Atmospheric Imaging Assembly (AIA; Lemen et al., 2012a) on board the Solar Dynamics Observatory (SDO). The study found conduction to be the dominant cooling mechanism in the corona. This is a feature shared by impulsive events like flares, microflares, and nano-flares, suggesting that the same physical mechanism is shared by these small transient brightenings. We note, however, that a number of simplifying assumptions, such as detailed thermal balance and stationary loop structures, were made in this study.

In this work, we carry out hydrodynamic simulations to gain a theoretical understanding of the energetics of small brightenings identified by (Régnier et al., 2014; Subramanian et al., 2018). We have selected three of the brightenings (BR-00, 07, and 26) from Subramanian et al. (2018) for a detailed study, as they show the simplest profiles, with a single peak, and a clearly visible decay phase (see Section 4.5.1-4.5.3). We have used a 0D numerical code called Enthalpy Based Thermal Evolution of Loops (EBTEL; Klimchuk et al., 2008; Cargill et al., 2012a). The results obtained from EBTEL simulations were used along with AIA response functions to produce synthetic light curves, mimicking observations of these small brightenings.

## 4.2 Data

Hi-C (High-Resolution Coronal Imager) is a sounding rocket mission that observed the Sun in the 193 Å channel with a pixel size of 0.1" (Kobayashi et al., 2014b). It was launched on July 11, 2012 and recorded observations of active region AR-11520 for $\approx$ 5 minutes. One of several interesting phenomena observed in detail * were multiple tiny brightenings within a system of fan loops rooted in the active region (Régnier et al., 2014; Subramanian et al., 2018).

Subramanian et al. (2018) identified 27 such point-like brightenings using the automatic detection algorithm of Subramanian et al. (2010), and performed a detailed study of energetics involved in these events. For this purpose, the Hi-

---

*https://hic.msfc.nasa.gov/publications.html#hic1_pubs

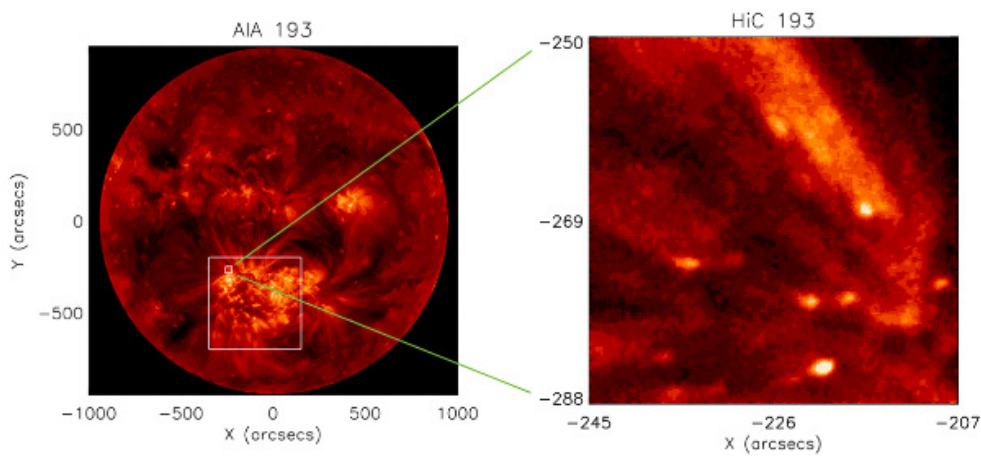

Figure 4.1: The site of the bright points considered here. (Left) The full disk AIA/193 Å filter image of the Sun, with an inset box showing the Hi-C (193) field of view (large square box), as well as the location of the fan-loop structure where bright points were detected (small square box). (Right) The Hi-C (193) image corresponding to the small square box showing the location of the fan loop region where the transient brightenings were observed. Image credit: Subramanian et al. (2018).

C observations were cross-calibrated with simultaneously obtained SDO/AIA (Solar Dynamics Observatory/Atmospheric Imaging Assembly; Lemen et al. (2012a); O'Dwyer et al. (2010a)) data.

AIA provides full disk images of the solar atmosphere in 7 separate EUV passbands with a cadence of 12 s and pixel size of 0.6″. Fig. 4.1 displays the AIA full disk image recorded in the 193 Å bandpass filter, the Hi-C field of view (larger inset white box), and the region where brightenings were found (smaller inset white box). The latter is zoomed-in in the right panel of Fig. 4.1.

To determine the thermal structure of the brightenings, Subramanian et al. (2018) obtained the *Differential Emission Measure* (DEM) using six optically thin filters of AIA. For this purpose, the `PINTofALE` package (Kashyap and Drake, 1998) was used. The DEMs were used for computing emission measure weighted temperatures and electron densities. It was found that all these events have temperatures $\approx 10^{6.2-6.6}$ K, electron number densities of the order of $10^9$ cm$^{-3}$ and estimated radiative energy losses of $10^{24-25}$ ergs.

Assuming that a magnetic flux density of $\sim$50 G is involved in each of these brightenings, Subramanian et al. (2018) estimated the total magnetic energy to be of the order of $10^{26}$ ergs and hence should be sufficient to power these brightenings with radiative energies of the order of $10^{24}$ ergs. Based on cooling timescales in static equilibrium, they suggested that thermal conduction was the dominant cooling mechanism in the corona for these brightenings.

It is important to note that the estimates of Subramanian et al. (2018) are crude, as they rely on stationary equilibrium estimates of conduction and radiative cooling timescales of corona and are integrated estimates over the lifetime of the brightenings. Furthermore, these neglect flows and density variations. Hence it is necessary to, first, validate the results using more realistic dynamical loop construction, and second, to identify the region of the parameter space, where loops that can mimic the observed light curves exist.

## 4.3 Hydrodynamic modeling

The reduced computational cost and improved speed of computing 0D solutions using EBTEL make studying the temporal evolution of length-averaged physical quantities a fruitful alternative to carrying out detailed numerical hydrodynamical simulations. Cargill et al. (2012b) demonstrated that the results obtained with the EBTEL differ from 1D hydrodynamic simulations by at most 15-20% despite being much faster.

## 4.3.1 Input parameters

To run the EBTEL simulations, we must specify some fundamental input parameters. EBEL requires two inputs: the half-length of the loop and the heating function. The heating function is defined as the rate at which the heat is deposited in the loop per unit volume (in units of ergs cm$^{-3}$ s$^{-1}$).

## 4.3.2 Estimation of Conduction and Radiation Losses and Enthalpy

For studying the various cooling processes and the relative importance of different energy terms, we need to compute the energy fluxes associated with conduction, radiation, and enthalpy from outputs provided by EBTEL.

The conduction flux,

$$F_c = \frac{F_{sp} F_{sat}}{\sqrt{F_{sp}^2 + F_{sat}^2}} \tag{4.1}$$

with $F_{sp}$ and $F_{sat}$ being the Spitzer and saturation fluxes respectively and

$$F_{sp} = -\frac{2}{7} \kappa_0 \frac{T_A^{7/2} - T_0^{7/2}}{L} \approx -\frac{2}{7} \kappa_0 \frac{T_A^{7/2}}{L} \quad \& \quad F_{sat} = -0.25 \sqrt{\frac{k_B^3}{m_e}} \bar{n} \bar{T}^{1.5} \tag{4.2}$$

where $T_A$ is the temperature at the apex of loop, $\kappa_0 = 8.12 \times 10^{-7}$ [cgs], $k_B$ is Boltzmann's constant, $m_e$ is the electron mass, and $\bar{n}$ and $\bar{T}$ are length-averaged electron number density and temperature. The radiation flux,

$$F_R = (1 + c_1) \bar{n}^2 L \Lambda(\bar{T}) \tag{4.3}$$

where $c_1$ is the ratio of the total radiation losses from the transition region and the corona as computed within EBTEL, and $\Lambda(\bar{T})$ is the optically thin power loss function of temperature and is computed using the routine rad_loss.pro from Solarsoft software, using the CHIANTI database (Dere et al., 1997b; Landi et al., 2013) and coronal abundances from Grevesse et al. (2007).

The enthalpy flux through the base of the loop is

$$F_P = \frac{5}{2}\bar{P}v_0 \qquad (4.4)$$

where $\bar{P}$ is the average pressure in the coronal part of the loop and $v_0$, is the velocity at the base of the loop.

### 4.3.3 Synthetic Light Curves

The intensity($I$) recorded by a filter indexed by subscript $i$ is

$$I_i = \int_0^\infty R_i(T)DEM(T)dT \qquad (4.5)$$

where $R_i(T)$ is the response function. $I_i$ has units of in DN s$^{-1}$ pix$^{-1}$, while $R_i(T)$ is measured in units of DN cm$^{-5}$ s$^{-1}$ pix$^{-1}$ (see for e.g. Boerner et al. (2012)).

## 4.4 Analysis

We probe the range of input parameters for generating synthetic light curves consistent with observations. Since these events were detected and studied using joint observations from AIA and Hi-C, with the latter observing the solar atmosphere for five minutes only in 193 Å, Subramanian et al. (2018) obtained the lifetime of these brightenings using 193 Å filter of AIA. Consequently, we have used AIA-193 Å observations as our reference. To determine the input parameters that best describe the transients, we employ two methods: In the first, we assume that the transients are part of the same dynamical system as the ambient corona and use the pre-transient intensity in AIA 193 to constrain the inputs (Section 4.4.1); in the second, we assume that the transients and the ambient corona are dynamically distinct, we match the characteristics of the background-subtracted light curves (Section 4.4.2).

## 4.4.1 Method 1: Modeling the transients with background

Here we seek to find a suitable combination of heat input and loop half-length that replicates the observed background intensities, the rise time of the transient, and the average background-subtracted intensities of the transient as observed in AIA 193 Å.

### 4.4.1.1 Range of parameters used in simulations

Subramanian et al. (2018) estimated the spatial extent of the bright points to be $\approx 2 \times 2$ AIA pixels. The pixel size of AIA is 0.6" which is roughly $8.4 \times 10^7$ cm. Assuming that the loop would be semicircular, the expected half-length is $\approx 0.65 \times 10^8$ cm. To account for uncertainties, we simulate loops with half lengths ranging between $0.1$–$1.5 \times 10^8$ cm. The heating function provided in EBTEL consists of two parts, viz. steady heating that creates the background and time-dependent heating that causes the transient. For each loop in this range, we set up a background heating ($Q_{bkg}$) in [ergs cm$^{-3}$ s$^{-1}$] following the scaling law used in EBTEL,

$$Q_{bkg} \approx \frac{2}{7} \kappa_0 \frac{T_A^{\frac{7}{2}}}{L^2} = \frac{2}{7} \left(\frac{10}{9}\right)^{\frac{7}{2}} \kappa_0 \frac{\bar{T}^{\frac{7}{2}}}{L^2} \tag{4.6}$$

where $\kappa_0 = 8.12 \times 10^{-7}$ in cgs units, $T_A$ is the temperature at loop summit, $\bar{T}$ is average temperature of the coronal part of loop and $\bar{T} \approx 0.9 T_a$ (Cargill et al. (2012a)). We seek solutions where $\bar{T} \approx 1$ MK.

To simulate the transient, we use a triangular heating profile characterized by a maximum heating rate $H_m$ ergs s$^{-1}$ cm$^{-3}$, and a total duration $t_{dur}$ seconds (Cargill et al. (2012a); Klimchuk et al. (2008)). Then the total energy $E$ dissipated in the semi-loop,

$$E = \frac{1}{2} \, A_{foot} \, L \, t_{dur} \, H_m \; [ergs].$$ (4.7)

For the exemplar case shown in Figure 3.3, we have $H_m = 0.2$ ergs s$^{-1}$ cm$^{-3}$ and $t_{dur} = 90$ s with cross-sectional area $A_{foot} = 1.76 \times 10^{15}$ cm$^2$, corresponding to one AIA pixel. To fix the time-dependent heating function, we vary the total energy budget instead of the volumetric heating rate. We run the simulations by varying the amount of total energy in the range $10^{22}$ to $10^{25}$ ergs, in steps of 0.1 dex. This range was chosen as it brackets the radiative losses for these transients as determined in Subramanian et al. (2018). The typical lifetime of these transients of the order $\approx 100$ s and therefore we consider $t_{dur}$ ranging from 10 to 200 s in steps of 10 s.

### 4.4.1.2 Parameters for specific brightening

Our goal is to identify the set of parameter values (viz., loop half-length, heating duration, and heating rate, $[L, t_{dur}, E]$) which mimic the characteristics of the observed brightenings. To find the best parameter set, we seek to compare and match the following characteristics of the observed and simulated light curves:

1. The average background level of observed light curves and the background-subtracted intensities averaged over the lifetime of the events as observed in AIA 193 Å images

2. The rise times of the events in the observed light curves obtained from AIA 193 Å filters. We require to match the rise times instead of total duration because the decay times are subject to large systematic errors due to the difficulty of identifying precisely when the model intensities become indistinguishable from the background.

We find that for all the three brightenings considered in this study, there exist physically plausible loop lengths, heating rates, and duration that capture the behavior of the observed AIA light curves. We emphasize, however, that the best parameters sets were not obtained by performing fits to the data; such a process would be unrealistic given the simplicity of the models we consider. To identify the input parameters for EBTEL, we follow the following procedure:

1. First, we identify the most suitable loop half-length. We note from equation 4.6 that for a given temperature, the background heating is a function of loop length. Therefore, we compare the background intensities in the light curves of AIA 193 Å with those obtained using equation 4.6 for all values of $L$ within the range 0.1–1.5 $\times 10^8$ cm (see Section 4.4.1.1), while fixing the temperature at 1 MK. We select that value of $L$ that provides the background intensity closest to the observed values in the 193 Å light curves.

2. Second, we identify the total duration of the heating events. For this, we deposit an heating event with total energy $E = 10^{22}$ ergs by varying the time duration $t_{dur}$ within the range of 10 to 200 s (see Section 4.4.1.1), for the values of $L$ previously obtained. Given $(L, E)$, the rise times are primarily dependent on $t_{dur}$. The values of $t_{dur}$, which give the closest agreement with the rise times of the observed events, are selected.

3. Finally, for the given $L$ and $t_{dur}$, the average intensity (DN pix$^{-1}$ s$^{-1}$) of the brightening in the 193 Å filter is estimated for values of $E$ ranging from $10^{22} - 10^{25}$ ergs. The value of $E$, which generates an intensity closest to that observed, is then selected.

## 4.4.2 Method 2: Modeling the transients without background

In contrast to the method outlined above, we detail an alternate method where background levels are ignored and only the rise time and average intensities of the transients are matched with model predictions to obtain a suitable set of EBTEL parameters. This effectively treats the transients as dynamically distinct events concerning the ambient corona.

### 4.4.2.1 Parameter range used in simulations

Using the assumption of semicircular loops confined within an area equivalent to 2×2 AIA pixels (1 AIA pixel corresponds to $8.4 \times 10^7$ cm), we fix the loop half-length to be $0.65 \times 10^8$ cm. The background heating rate is generally taken to be two to three orders of magnitude smaller than the main heating event (Klimchuk et al., 2008; Cargill et al., 2012a). For a loop length of the order of $1 \times 10^8$ cm with a cross-sectional area of 1 AIA pixel, we require a volumetric heating rate of the order of 0.01 to 0.1 ergs cm$^{-3}$ s$^{-1}$, such that $\sim 10^{23}$ ergs is deposited in $\sim 50$ s. Hence we have set a uniform background heating rate of $10^{-4}$ ergs cm$^{-3}$ s$^{-1}$, such that the background temperature and densities are an order of magnitude lower than the peak values.

The remaining task is to find the heating duration and total energy budget for the event. They were determined in the same way as mentioned in Section 4.4.1.1 once loop length and background heating rates are set. After fixing the half length of the loop and background heating rate to $0.65 \times 10^8$ cm and $10^{-4}$ ergs cm$^{-3}$ s$^{-1}$, respectively, we follow steps 2 & 3 enumerated in §4.4.1.1 to obtain the doublet $[t_{dur}, E]$ (see Table 4.1).

Table 4.1: Input parameters used in simulations for studying transient brightenings

| Index | Energy (E [log(ergs)]) | | Half Length (L [$10^8$ cm]) | | Duration of heating ($t_{dur}$ [s]) | |
| --- | --- | --- | --- | --- | --- | --- |
| | Method 1 bkg. modeled | Method 2 bkg. not modeled | Method 1 bkg. modeled | Method 2 bkg. not modeled | Method 1 bkg. modeled | Method 2 bkg. not modeled |
| 00 | 23.0 | 23.2 | 1.00 | 0.65 | 60 | 40 |
| 07 | 23.1 | 23.3 | 0.90 | 0.65 | 70 | 60 |
| 26 | 22.8 | 23.1 | 1.1 | 0.65 | 60 | 70 |

## 4.5 Results

Using the two methods described above, we identify the triplet $[L, t_{dur}, E]$ of inputs required for EBTEL that best describes the transient under investigation (see Table 4.1). We plot the observed and simulated light curves for AIA 193 Å (panels a & d), simulated plasma temperature (panels b & e) and density (panels c & f) obtained from both methods; explicitly modeling background levels (method 1; Section 4.4.1) and excluding background from the modeling (method 2; Section 4.4.2) for BR00, BR07 and BR26 in Figures 4.2, 4.5 and 4.8, respectively. We have investigated the robustness of both models by bracketing the nominal duration by half and twice the selected $t_{dur}$. Moreover, we have also obtained the synthetic light curves using both methods for other AIA filters, viz., 94, 131, 335, 211, and 171, and compared them with the observed light curves in Figures 4.3, 4.6, & 4.9.

The average model intensities of the events in each AIA filter, corresponding to input parameters selected using both methods, are reported in Table 4.2 and are compared with the measured intensities from Subramanian et al. (2018). We also show the range in the calculated intensities that arise due to possible systematic uncertainty in the event duration. This is done by computing average intensities when the duration of heating is $\frac{1}{2}\times$ and $2\times$ the most suitable value. These results demonstrate that the observed intensities in AIA 193 Å filter are robustly modeled, and where they differ from other filters, point to limitations in the plasma temperature reconstructions with EBTEL.

We note that EBTEL also allows the inclusion of non-thermal particle flux within the simulation. However, it assumes that the entire energy of the non-thermal particles goes into enthalpy and neglects any chromospheric/transition region radiation in response to non-thermal particles. Hence EBTEL cannot incorporate non-thermal emission, which is reasonable for the small energy events we are interested in. We find that even dissipating as much as half of the total energy budget to be dissipated by non-thermal particles (for loop parameters relevant to our study, as in Section 4.4), the light curves change by <5%. Additionally, the presence of non-thermal particles worsens the agreement between simulated and observed EM-weighted temperatures. Therefore, we have switched off this option in the code.

To study the energetics of brightenings of this class, we now look into the EBTEL simulations to understand how the energy transfer processes operate. We show in Figs. 4.4, 4.7, & 4.10, the conduction loss (blue), radiation loss (green), enthalpy (red), and the heating rate (black) for each brightening simulated using input parameters obtained from method 1 (panels [a-b]) and method 2 (panels [c-d]). We note that for the sake of visibility, loss curves are shown only for the most suitable $t_{dur}$. To demonstrate both the absolute and relative magnitudes of the energy losses, the contribution of background heating is included in the upper panels of these figures and is excluded in the lower panels. Note that the values are as computed for the full loop, not just the semi-loop as computed by EBTEL. We next discuss the results for each event in sequence.

## 4.5.1 Modeling the Dynamics of BR-00

### 4.5.1.1 Method 1: Background modelled

We find that a loop of half-length $1 \times 10^8$ cm, with $10^{23}$ ergs deposited within 60 s mimics the observed rise times, peak intensity (DN pix$^{-1}$ s$^{-1}$), and approximate duration. The pre-event ambient background heating, designed to match the base intensity (DN pix$^{-1}$ s$^{-1}$) level in AIA 193, maintains plasma at a temperature of 0.92 MK and density 7.8×10$^9$ cm$^{-3}$, which is similar to the cooler loops found in active regions (Ghosh et al. (2017)). The EBTEL simulation is run for 60 s before a triangular heat pulse is applied for a duration $t_{dur}$=60 s. We follow the plasma evolution during and after this pulse and use the DEMs obtained from the simulation to predict the light curves for all AIA channels.

Simulation light curves of AIA 193 intensity (DN pix$^{-1}$ s$^{-1}$), plasma temperature, and plasma density are shown in the panels a, b, and c of Figure 4.2, for the nominal heating duration of $t_{dur}$=60 s (black curve), as well as for 30 s (red dashed curves) and 120 s (red dot-dashed curves). As expected, the simulated light curves match the achieved intensity (DN pix$^{-1}$ s$^{-1}$) and duration of the observed light curve. Note that the peak intensity (DN pix$^{-1}$ s$^{-1}$), temperature, and density correlate with the heating duration; this is due to more impulsive events heating the plasma on smaller timescales, leading to a sharper rise in temperature and consequently a higher density due to stronger evaporation.

We also show the simulated light curves for the other AIA filters in panels a-e of Figure 4.3. In these plots, the simulated lightcurves have been offset by a case-specific value to match the observed pre-event background intensity level. The background level has been offset from that predicted based on matching to AIA 193 (see Table 4.3) to match the model light curve intensities to the pre-event observed intensities for each filter. As seen above for the 193 Å filter, the anti-correlation between peak intensity and heating duration persists for all filters. It is seen that observed intensities in all AIA filters except 94 Å peak before the synthetic light curves for the most suitable heating duration $t_{dur} = 60$ s. We now average the observed over the lifetime of the brightenings. The simulated light curves are integrated over a lifetime as determined from when the synthetic AIA 193 Å light curve drops to 5% of the peak intensity and is divided by the lifetime to obtain a comparison with the observed average intensities (see columns 2 and 3 of Table 4.2). The sim-ulated average intensities match observations well for AIA 193, though they differ by factors ranging from 0.8× (AIA 335) and 1.2× (AIA 211) to 3.1× (AIA 171) and 3.4× (AIA 131). Hence, the match is best in AIA 211 and worst in AIA 131 and AIA 171. This is a remarkable correspondence with observations considering the limitations of the 0-D monolithic loop system we consider.

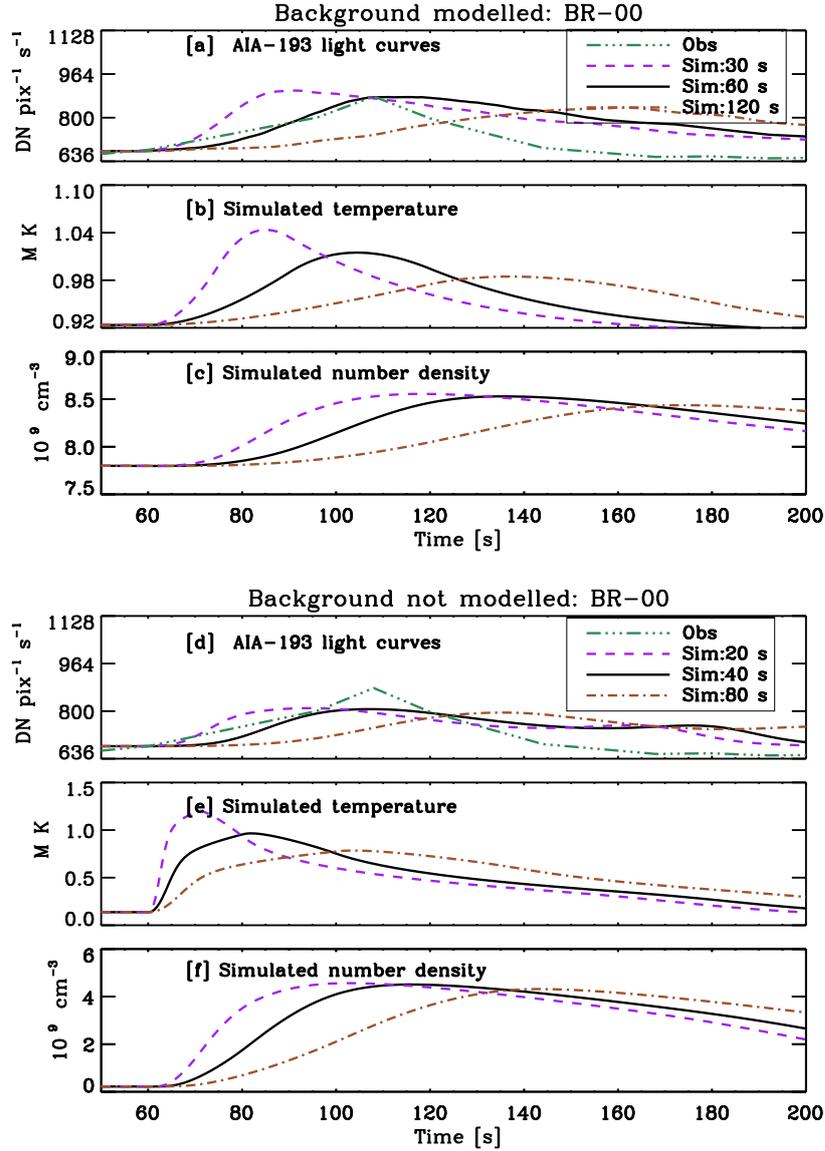

Figure 4.2: Data and EBTEL simulations of brightening BR00 using input parameters set using method 1, which uses transient + background for determining input parameters (top) and method 2 which uses only transient for determining input parameters (bottom). Panels a & d show observed and simulated light curves as labeled. Panels (b, e) and (c, f) show the simulated plasma temperature and density respectively. The solid curves represent model curves obtained using the parameters in Table 4.1, the pink dashed and brown dot-dashed curves represent model curves made for heating durations $\frac{1}{2}\times$ and $2\times$ the nominal, and the green triple-dot-dashed curves represent observed light curves.

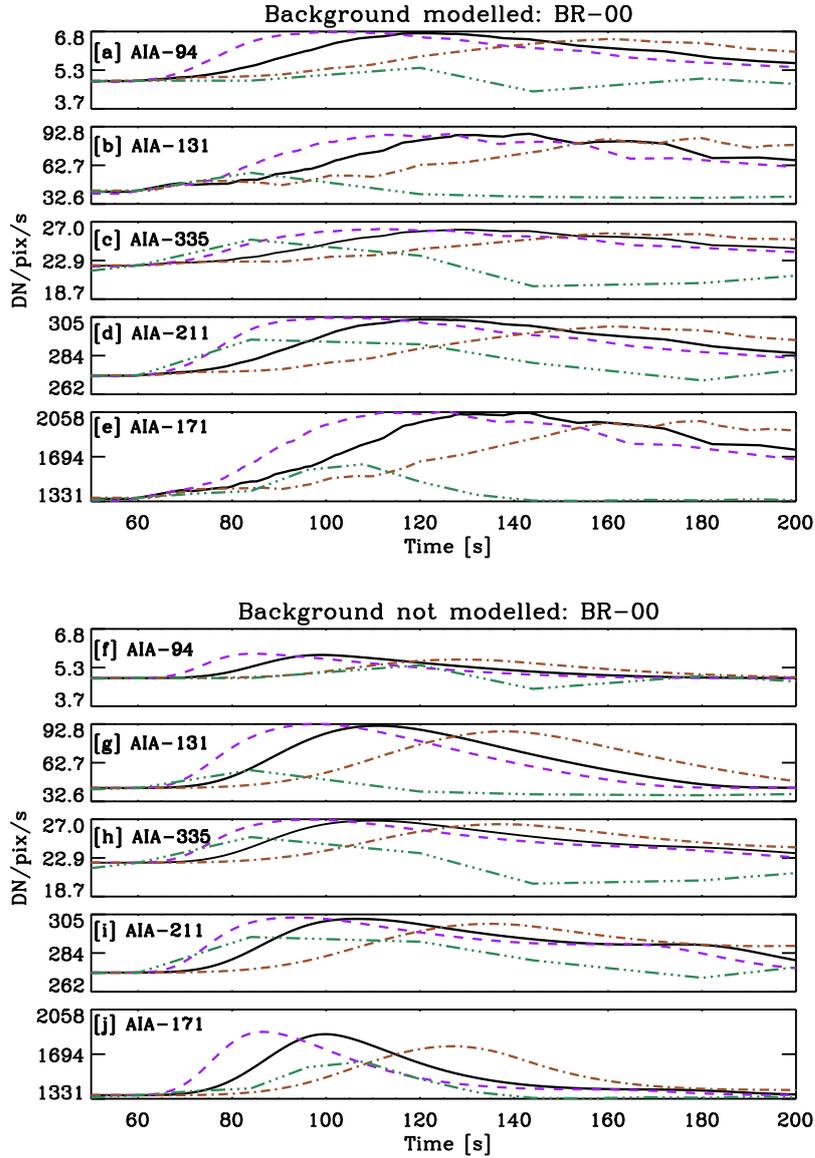

Figure 4.3: Observed and simulated light curves using input parameters set using method 1 (which models the background and the transient together; top) and method 2 (which considers the background separately from the transient; bottom) for BR-00 for AIA channels 94 (panel a[f]), 131 (panel b[g]), 335 (panel c[h]), 211 (panel d[i]) and 171 Å (panel e[j]) as labelled. The colors and line styles correspond to those in Figure 4.2. The simulated light curves obtained using method 1, where input parameters have been selected using transient + background (upper panel) and method 2, where input parameters have been selected using only transient (lowe panel) have been increased or decreased by a constant offset in each filter to match the observed background level. Note that the best durations chosen for the two methods are different, and therefore the 1/2x and 2x durations also differ.

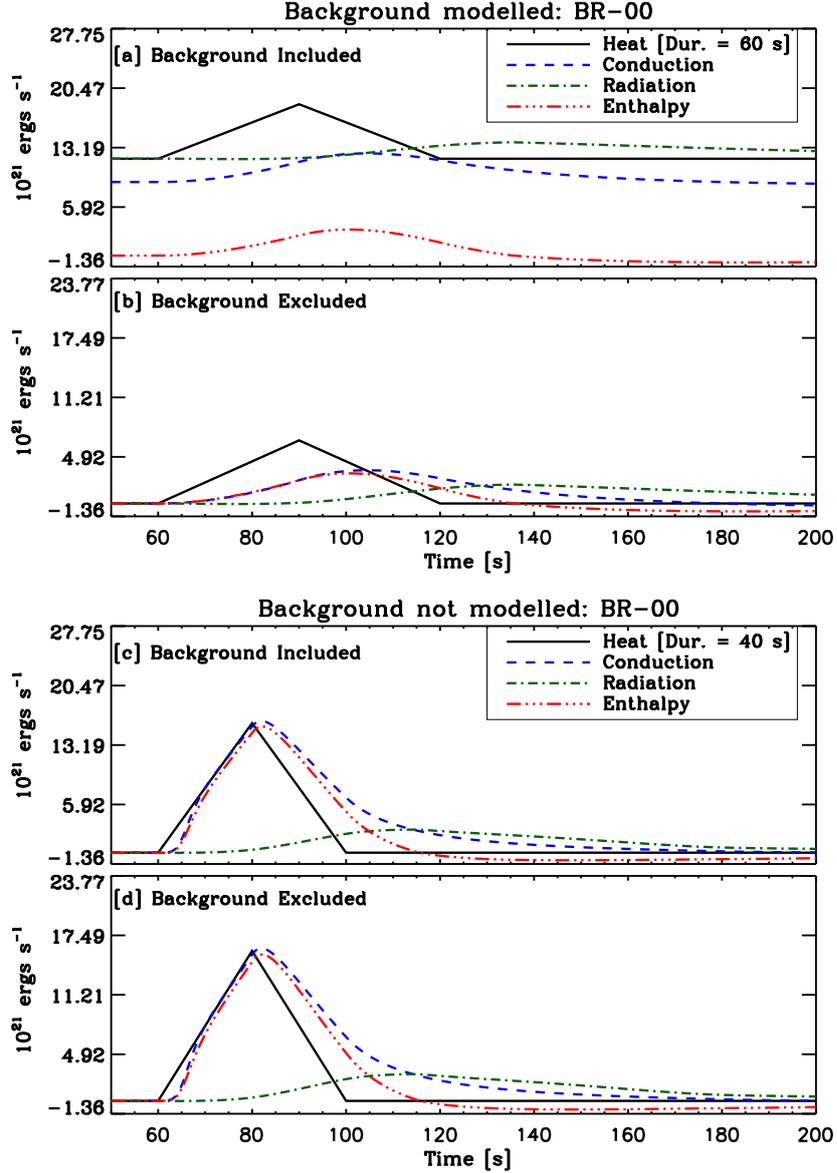

Figure 4.4: Energy loss and transfer components from EBTEL simulations designed to match BR-00. Upper two panels [a & b] correspond to method 1 (background included in the modeling), and the lower two panels [c & d] correspond to method 2 (backgound treated as separate from the transient). The heat input (solid curves), conduction (blue dashed curves), radiation loss (green dot-dashed curves), and enthalpy (red triple-dot-dashed curves) are shown. The panels [a & c] show curves that include the background, while panels [b & d] are shown with background values subtracted. Note that the radiation loss term includes contributions from both the corona and the transition region.

The energy loss and transfer terms (conduction and radiation losses, enthalpy, and heat input) for the simulation with the nominal heating duration are shown in Figure 4.4. The panels [a & c] include the contribution from the ambient background, and the panels [b & d] isolate the effects due only to the brightening event. Note that enthalpy (red triple-dot-dashed curves; equation 4.4) can be positive or negative depending on whether plasma is flowing into or out of the corona. As expected, conduction losses (blue dashed curve; equation 4.1) dominate at the beginning (until ≈70 s) of the event, and radiation losses (green dot-dashed curve; equation 4.3) dominate at later times. The enthalpy into the corona keeps pace with conduction loss in the early phase (for ≈40 s) and drops off ≈10 s before conduction does. The enthalpy reverses the sign at approximately the same time that radiation becomes the dominant loss mechanism. The total time-integrated conduction loss from the coronal loop (which is eventually radiated) is $3.0 \times 10^{23}$ ergs. The net enthalpy is positive (into corona) and is equal to $4.5 \times 10^{22}$ ergs. This is an order of magnitude less than the heating function. Notice that the radiation loss drops below the ambient background level at the beginning of the event, as does conduction loss about 120 s after the heat pulse. These represent small perturbations in the ambient coronal structure and do not affect the energetics.

## 4.5.1.2 Method 2: Background not modelled

An impulse with a triangular profile dissipating a total of $10^{23.2}$ ergs in 40 s is best suited for a loop of half-length of $0.65 \times 10^8$ cm subjected to uniform background heating of $10^{-4}$ ergs $cm^{-3}$ s$^{-1}$ (see 4.4.2.1). The plots in Figure 4.2 display the observed and simulated light curves for 193 Å (panel d), plasma temperature (panel e), and density (panel f) when the background is not modeled. Different colors belong to $t_{dur}$ (black), $\frac{1}{2} \times t_{dur}$ (purple) and $2 \times t_{dur}$ (brown) as labelled. The simulated background intensity is almost three orders of magnitude smaller than the simulated peak values and hence negligible. Note that to bring the background level of the synthetic light curves to that observed, we have added a case-specific offset (see Table 4.3). As expected, the peak values of temperature, density, and intensities correlate with the heating duration. Even though the initial density and temperature, in this case, are an order of magnitude lower than the values obtained by simulations using method 1, the peak values of temperature match in both cases, and the peak density produced by method 2 is less than that produced by method 1 by a factor $\lesssim 2$.

We have also obtained the synthetic light curves for the other channels of AIA and plotted them after adding case-specific background levels, in panels [f-j] of Figure 4.3. The colors have the same meaning as those in panels [f-j] of Figure 4.2. We have also over-plotted the observed light curves for the corresponding AIA channels for comparison.

We plot the various energy loss and transfer terms, including and excluding the background contribution in Figure 4.4 (see panels c & d). The qualitative feature of an initial conduction-dominated loss is the same as that obtained using method 1 and is shown in panels a & b of Figure 4.4. Moreover, the time at which enthalpy changes its sign is also approximately the same as where radiation loss starts dominating conduction loss, similar to method 1. However, we note that enthalpy plays a more important role in method 2 than in method 1. The time-integrated intensities obtained from synthetic light curves in all filters show a better correspondence with those obtained from the observed light curve, except for 131 and 171 Å, which shows the highest discrepancy. In this method, secondary peaks are observed in all three brightenings. This is because the temperature covers a wider range of values and the less dominant peaks in the response function play a role at those temperatures.

### 4.5.2 Modelling the Dynamics of BR-07

#### 4.5.2.1 Method 1: Background modelled

We follow the same procedure to analyze BR-07 here as we did for BR-00 in Section 4.5.1.1. A loop of half-length of $0.9 \times 10^8$ cm, with $10^{23.1}$ ergs deposited in $t_{dur} = 70$ s mimics the observed rise times, peak intensity (DN pix$^{-1}$ s$^{-1}$), and approximate duration. The background heating used for matching the base intensity (DN pix$^{-1}$ s$^{-1}$) levels in AIA 193 maintains plasma at a temperature and density of 0.92 MK and $8.7 \times 10^9$ cm$^{-3}$. The simulation is run for 60 s before a triangular heat pulse is applied for a duration of $t_{dur}=70$ s. The evolution of plasma obtained from simulations was used to predict light curves in all channels. Simulation light curves of AIA 193 intensity (DN pix$^{-1}$ s$^{-1}$), temperature and density of plasma for $t_{dur}$, $2 \times t_{dur}$ and $\frac{1}{2} \times t_{dur}$ along with observed AIA 193 light curve are shown in Figure 4.5. Simulated light curves in remaining filters (after increasing or decreasing by case-specific offsets) are shown in Figure 4.6. The energy loss and transfer terms for sim-ulations with nominal heating duration, i.e., 70 s in this case, are shown in Figure 4.7. The scheme used in the figures is the same as that used for BR-00.

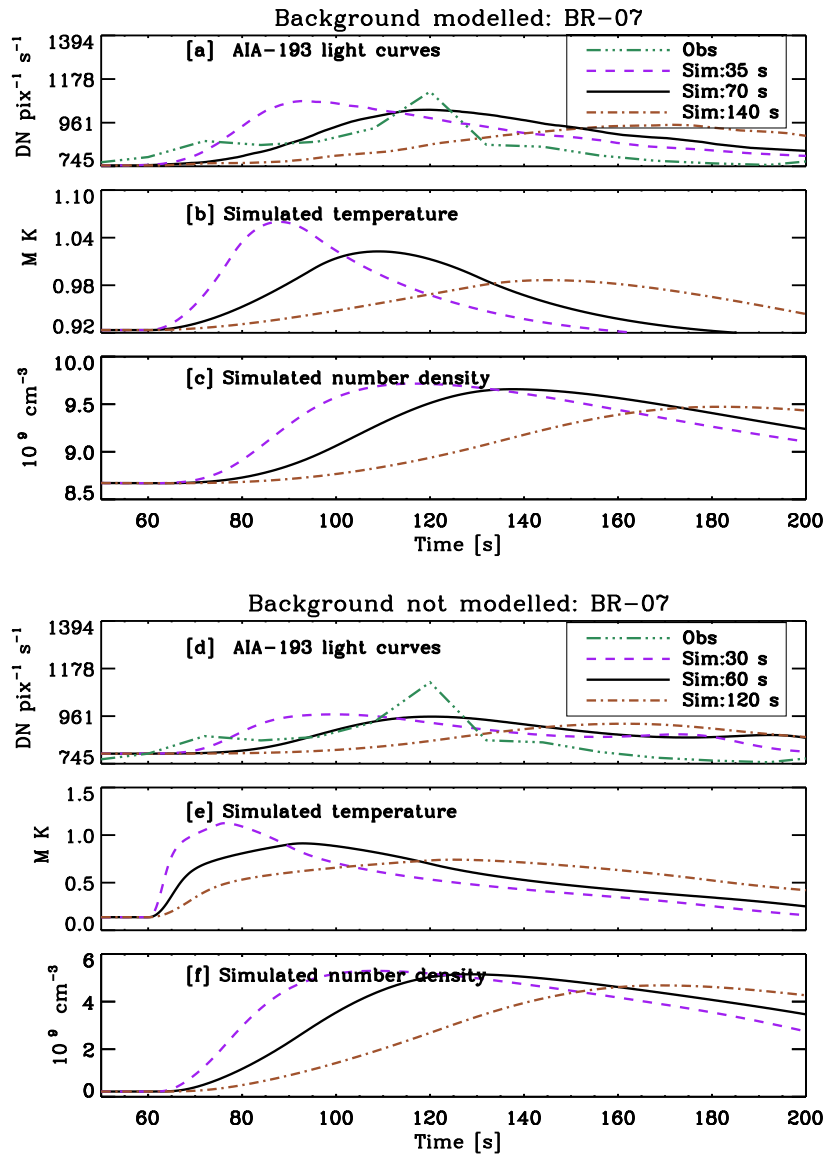

Figure 4.5: Same as in Figure 4.2 but for BR-07.

The peak intensity in all filters, temperature, and density is the largest for most impulsive heating. The simulated intensities (averaged over the lifetime of the event), match observations well for AIA 193. It differs in other filters by factors between 1.4× (AIA 94) and 5.5× (AIA 171). The event has an initial conduction-dominated cooling phase lasting for $\approx 70$ seconds, followed by radiation dominated cooling phase at later times. Enthalpy starts dropping $\approx 10$ s before conduction. The total time-integrated conduction loss from the coronal loop is $3.7 \times 10^{23}$ ergs. The net enthalpy is positive (into corona) and is equal to $5.1 \times 10^{22}$ ergs.

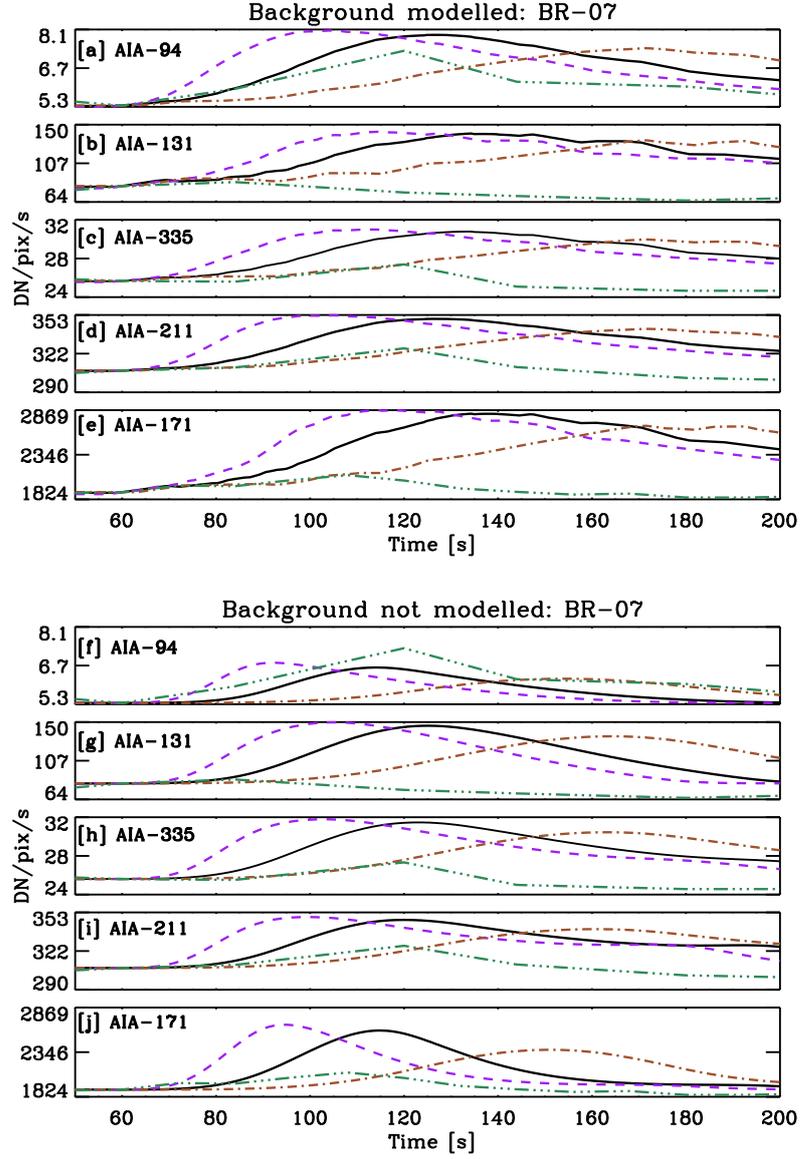

Figure 4.6: Same as Figure 4.3 but for BR-07.

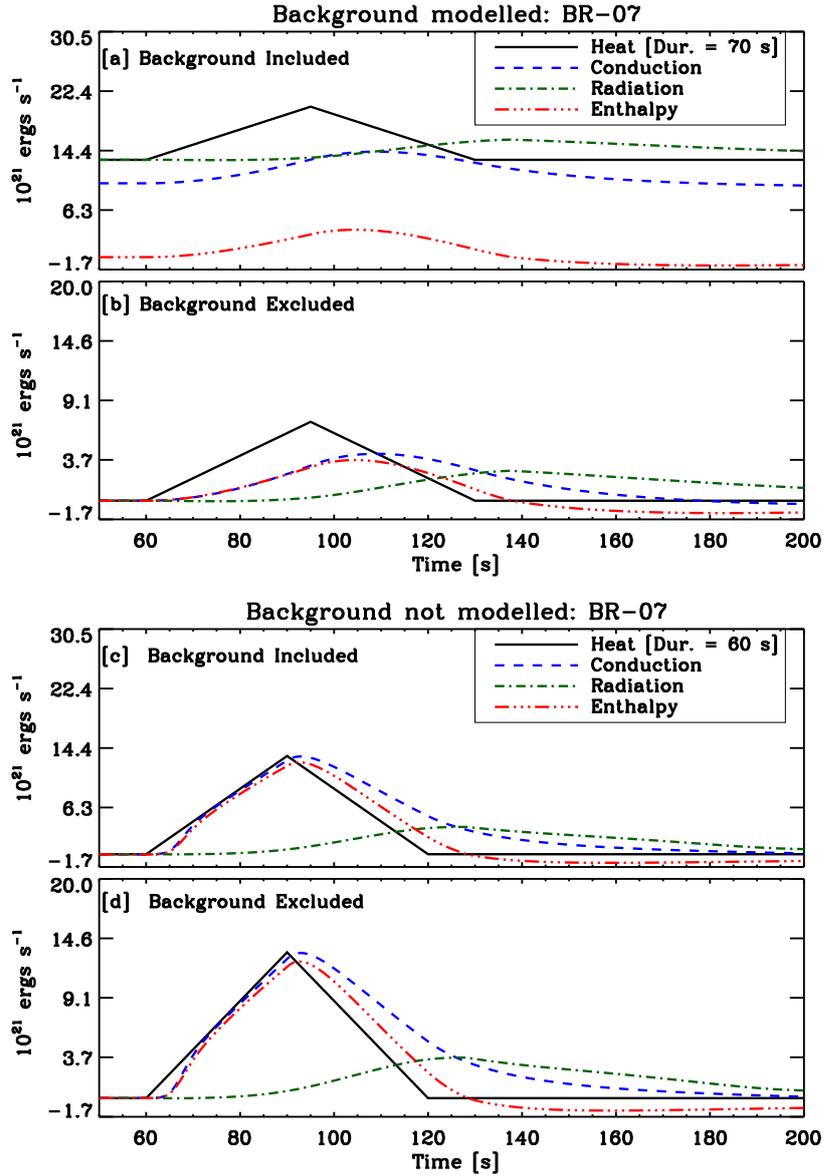

Figure 4.7: Same as in Figure 4.4 but for BR-07.

## 4.5.2.2 Method 2: Background not modelled

We follow the same procedure to analyze BR-07 as in Section 4.5.1.2. A heating event having a triangular profile, which dissipates $10^{23.3}$ *ergs* in 60 *s* is best suited for a loop of half-length of $0.65 \times 10^8$ cm subjected to uniform background heating of $10^{-4}$ *ergs* $cm^{-3}$ $s^{-1}$. A case-specific offset has been added to all the simulated light curves to make the observed and simulated pre-event intensities equal (see Figure 4.5 and 4.6). All features, including initial conduction, dominated cooling of the corona, enthalpy changing sign approximately when radiation starts dominating conduction, are qualitatively the same as that of BR-00 (see Figure 4.7). The average intensity in AIA 171 filter agrees better with the observed value.

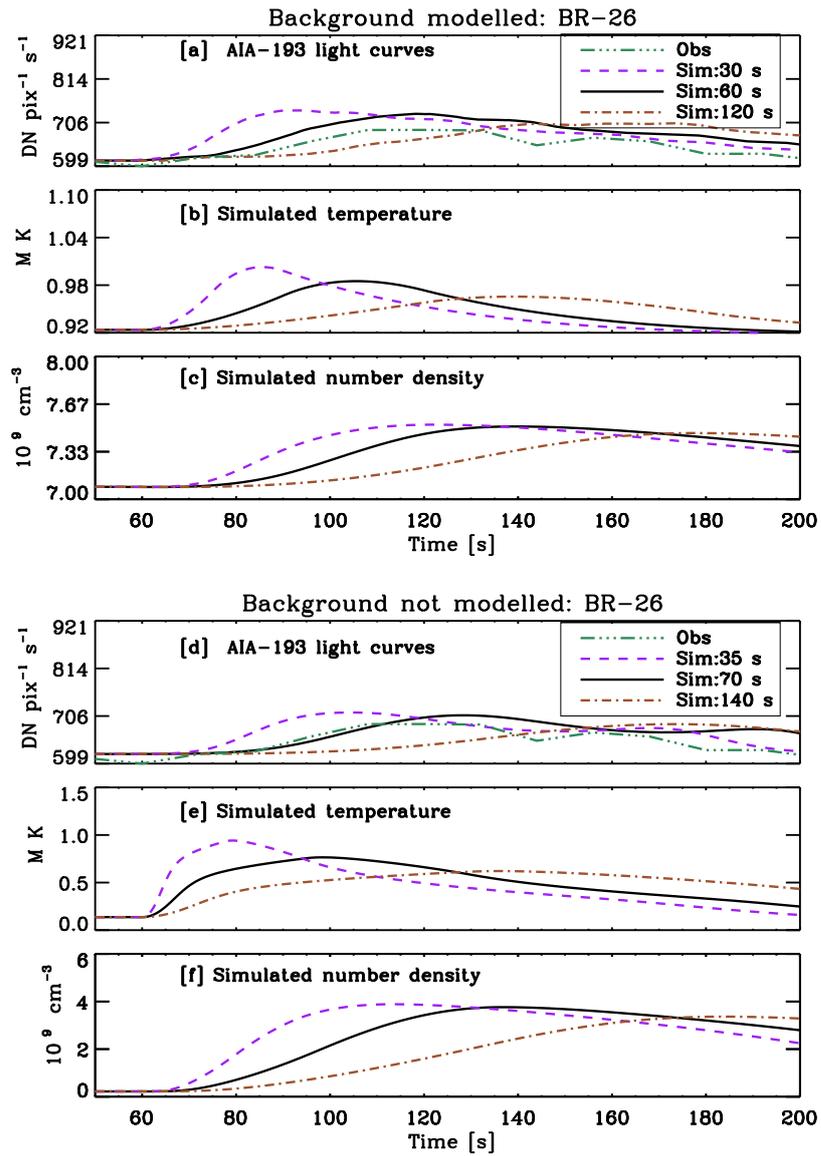

Figure 4.8: Same as Figure 4.2 but for BR-26.

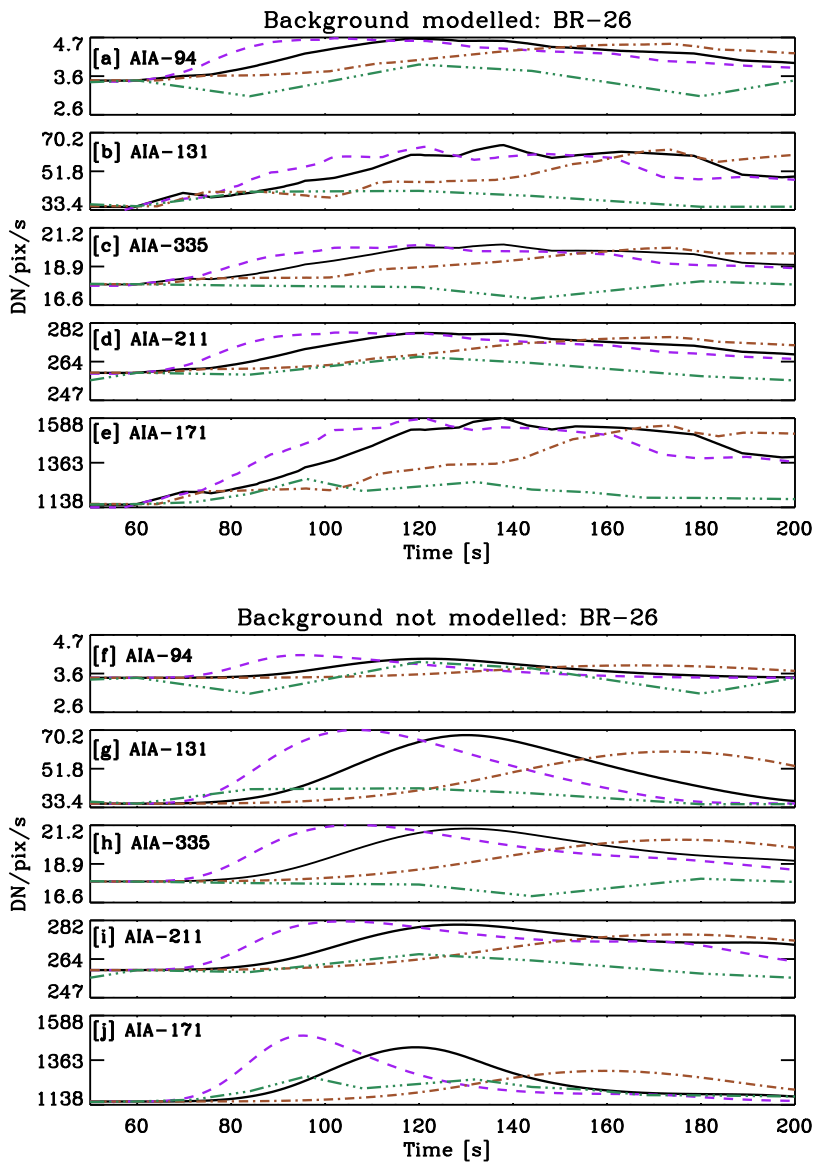

Figure 4.9: Same as Figure 4.3 but for BR-26.

### 4.5.3 Modelling the Dynamics of BR-26

#### 4.5.3.1 Method 1: Background modelled

We follow the same procedure to analyze BR-26 here as for BR-00 in Section 4.5.1.1. A loop of half-length of $1.1 \times 10^8$ cm, with $10^{22.8}$ ergs deposited in $t_{dur} = 60$ s mimics the observed light curves. The background heating used for matching the base intensity (DN pix$^{-1}$ s$^{-1}$) levels in AIA 193 Å maintains plasma at a temperature and density of 0.92 MK and $7.1 \times 10^9$ cm$^{-3}$. Simulated light curves of AIA 193 intensity (DN pix$^{-1}$ s$^{-1}$), temperature and density of plasma for $t_{dur}$, $2 \times t_{dur}$ and $\frac{1}{2} \times t_{dur}$ along with observed AIA 193 light curve are shown in Figure 4.8. Simulated light curves in remaining filters (after increasing or decreasing by case-specific offsets) are shown in Figure 4.9. The average intensities in filters apart from AIA-193 differ by factors ranging between 1.2× (AIA 94) and 3.6× (AIA 171). The energy loss and transfer terms for simulations with nominal heating duration, i.e., 60 s in this case, are shown in Figure 4.10. The scheme used in the plots is the same as that used for BR-00. Energy loss and transfer evolution trends are similar to BR-00 and BR-07. The total time-integrated conduction loss from the coronal loop is $1.9 \times 10^{23}$ ergs. The net enthalpy is positive (into corona) and is equal to $3.5 \times 10^{22}$ ergs.

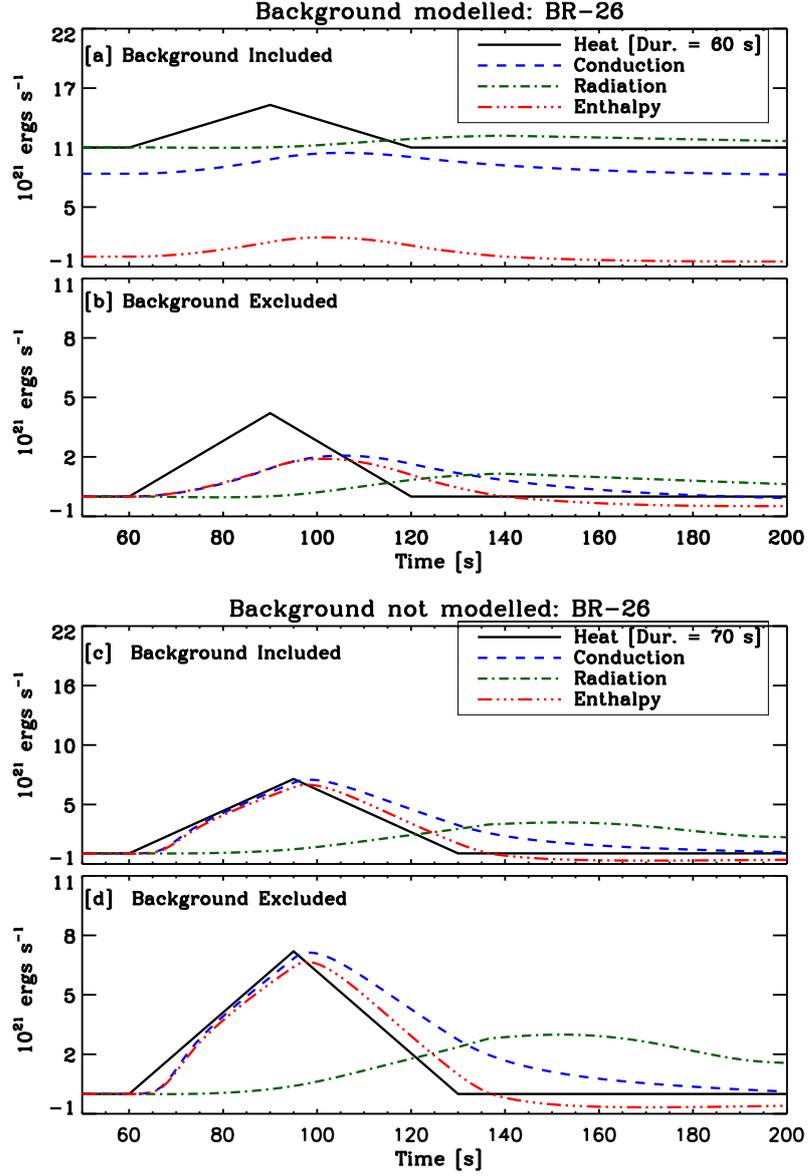

Figure 4.10: Same as Figure 4.4 but for BR-26.

### 4.5.3.2 Method 2: Background not modeled

We follow the same procedure to analyze BR-26 as detailed in Section 4.5.1.2. A heating event having a triangular profile, which dissipates $10^{23.1}$ ergs in 70 $s$ is best suited for a loop of half-length of 0.65 $\times 10^8$ cm subjected to uniform background heating of $10^{-4}$ ergs cm$^{-3}$ s$^{-1}$. A case-specific offset has been added to all the simulated light curves to make observed and simulated background intensities equal (see Figures 4.8 and 4.9). Once again, we see initial conduction-dominated cooling of the corona, enthalpy changing sign approximately when radiation starts dominating conduction, are qualitatively the same as that of BR-00 (see Figure 4.10). The average intensity in AIA-171 filter agrees better with the observed value.

### 4.5.4 Comparison of conductive flux to the radiative flux

To quantitatively assess the relative importance of conduction loss over radiation loss during different phases of the brightening, in Figure 4.11 we plot the logarithm of the ratio of absolute values of background-subtracted conduction and radiation loss rates obtained from method 1 (background modeled) for BR-00 (panel a), BR-07 (panel b) and BR-26 (panel c). The same results for input parameters selected by method 2 (background not modeled) are shown in panels d,e, and f, respectively. The solid black curves correspond to the nominal heating durations, while the dashed-blue and dashed-dotted red curves correspond to heating durations set to $\frac{1}{2}\times$ and $2\times$ the nominal duration. The cusps in the curves (e.g., those present between 80 and 100 s in the left panels and between 60 and 80 s in the right panels) are artifacts of correcting for the background, arising from the radiative loss dropping below and rising above the ambient value (see Figure 4.4, 4.7, 4.10 and discussion in Section 4.5.1). The occurrence of these spikes at different locations for the same brightenings in the two methods are due to using different input parameters in each case. For all three events, irrespective of how the input parameters are selected, we find an initial phase where conduction is the dominant cooling mechanism in the corona. This phase ends at a simulation time step of $\approx$140 s, i.e., 80 s after the heating starts.

## 4.6 Discussion and Summary

We identify loop sizes and heat inputs that mimic the intensities and durations of the observed brightenings using two methods. In the first method, we use background intensities for constraining the input parameters, and in the second method, we don't use this information. We use the simulations to study energy transfer into and out of the corona. We focus on three of the simplest brightenings identified by Subramanian et al. (2018), with single unambiguous intensity peaks. We adopt triangular heating profiles of duration 30-140 s and find that loop half lengths of $\approx 10^8$ cm and energy deposition of $\approx 10^{23}$ ergs can generate dynamical intensity profiles that mimic the observed brightenings.

We find that the average brightness of the simulated loops in the AIA 193 Å

Table 4.2: Observed and simulated average Intensities in DN pix$^{-1}$ s$^{-1}$ over AIA-193 lifetime of brightenings. The values in the parenthesis cover the range of average intensities if the duration in which energy is deposited in half and double the time duration selected as the most suitable input parameter. Observed average intensities have been taken from Subramanian et al. (2018).

| AIA Filter | Obs. | BR-00 (DN pix$^{-1}$ s$^{-1}$) Sim. | | Obs. | BR-07 (DN pix$^{-1}$ s$^{-1}$) Sim. | | Obs. | BR-26 (DN pix$^{-1}$ s$^{-1}$) Sim. | |
|---|---|---|---|---|---|---|---|---|---|
| | | Method 1 (bkg. modelled) | Method 2 (bkg. not modelled) | | Method 1 (bkg. modelled) | Method 2 (bkg. not modelled) | | Method 1 (bkg. modelled) | Method 2 (bkg. not modelled) |
| 094 | 0.50 | 0.92 [0.78-1.0] | 0.31 [0.25-0.34] | 0.89 | 1.2 [0.96-1.3] | 0.43 [0.31-0.49] | 0.40 | 0.48 [0.41-0.61] | 0.17 [0.12-0.21] |
| 131 | 7.3 | 24.0 [20.9-26.2] | 19.7 [16.9-21.45] | 9.1 | 31.7 [25.6-35.0] | 25.9 [20.4-29.3] | 4.6 | 13.9 [12.0-16.9] | 12.2 [9.3-14.0] |
| 171 | 104.3 | 376.5 [325.9-410.3] | 146.9 [124.2-155.4] | 88.6 | 489.6 [401.3-537.1] | 207.2 [153.5-228.3] | 60.1 | 214.0 [183.3-260.6] | 84.2 [53.8-97.9] |
| 193 | 83.8 | 85.6 [70.1-95.1] | 67.3 [56.8-73.9] | 101.4 | 112.2 [87.8-127.7] | 83.0 [64.6-95.4] | 39.7 | 41.5 [35.3-55.6] | 44.0 [33.1-51.0] |
| 211 | 12.0 | 15.2 [12.7-16.8] | 16.4 [14.0-17.9] | 8.7 | 19.9 [15.8-22.4] | 20.0 [15.7-22.7] | 5.2 | 7.8 [6.7-10.1] | 11.0 [8.3-12.7] |
| 335 | 2.8 | 2.0 [1.7-2.2] | 2.3 [2.0-2.5] | 1.1 | 2.6 [2.1-2.9] | 2.9 [2.3-3.2] | 0.64 | 1.1 [0.94-1.4] | 1.6 [1.2-1.8] |

Table 4.3: Background levels used to compare observed and model light curves in various filter in Figs. 4.2, 4.3, 4.5, 4.6, 4.8, 4.9. The values listed for Method 1 are offsets relative to the simulated value derived by comparison with AIA 193 light curves (see text).

| AIA Filter | BR-00 (DN pix$^{-1}$ s$^{-1}$) | | BR-07 (DN pix$^{-1}$ s$^{-1}$) | | BR-26 (DN pix$^{-1}$ s$^{-1}$) | |
|---|---|---|---|---|---|---|
| | Method 1 (bkg. modelled) | Method 2 (bkg. not modelled) | Method 1 (bkg. modelled) | Method 2 (bkg. not modelled) | Method 1 (bkg. modelled) | Method 2 (bkg. not modelled) |
| 094 | -2.1 | 4.8 | -0.95 | 5.3 | -2.8 | 3.5 |
| 131 | -176.0 | 43.0 | -116.9 | 81.7 | -163.4 | 35.2 |
| 171 | -1908.6 | 1359.8 | -1056.2 | 1908.0 | -1809.8 | 1154.5 |
| 193 | 0 | 679.0 | 0 | 790.7 | 0 | 620.6 |
| 211 | +131.4 | 272.6 | +179.5 | 307.7 | +131.3 | 259.5 |
| 335 | +2.9 | 22.3 | +7.9 | 25.6 | +0.18 | 17.8 |

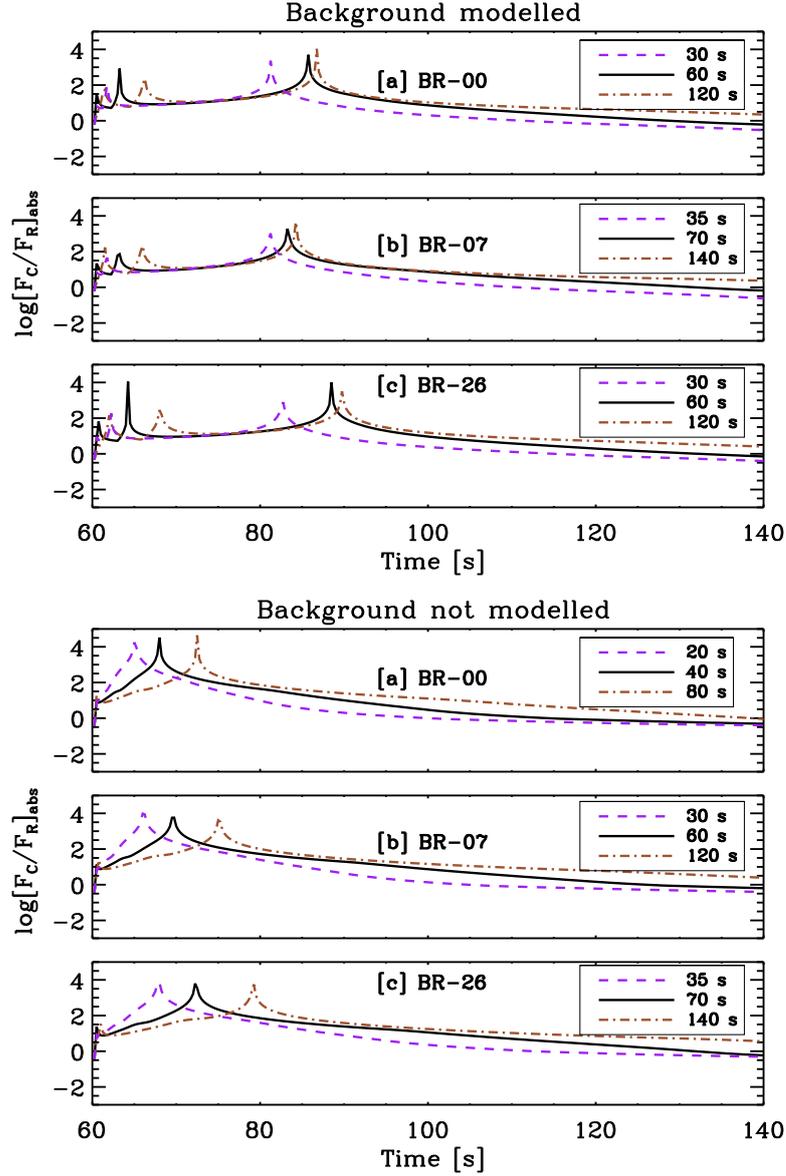

Figure 4.11: Time evolution of the ratio of absolute values of conduction and radiation loss rates for BR-00 (panel [a]), BR-07 (panel [b]) and BR-26 (panel [c]). The solid black curves are for the optimized duration while the dashed-blue and dashed-dotted red corresponds to the result obtained for half and twice the duration of the heating event as labelled.

filter matches the observations well and the match is within a factor of 5.5 in the other AIA filters if input parameters are selected by method 1 (i.e., both transient and background are modeled) and within a factor 3 if input parameters are selected by method 2 (background is not modeled). The ob-tainment of a better tally by the latter method can suggest the transient events and background being dynamically distinct. In either method, the largest discrepancies between the simulated and observed intensities, irrespective of how input parameters have been selected, arise in the AIA 171Å and AIA 131Å filters.

For all three events studied here, we find that conduction is the dominant cooling mechanism in corona in the early phase of the transients. About 80 s after the heat pulse, radiation losses begin to dominate. We observe that conduction-dominated cooling in corona during the early phase of the evolution has also been reported for flares (Cargill et al., 1995b), microflares (Gupta et al., 2018), and are also expected for nanoflares (Cargill, 1994). Our results show that transient events such as those observed in Hi-C are similar to microflares and nanoflares and are likely produced through the same underlying physical processes. We also note that our simulated plasma temperatures are lower than the observed values. And since conduction loss increases as $T^{3.5}$ while radiative loss decreases in this temperature regime, our assessment of the relative magnitude of conduction loss in the energetics of coronal plasma is an underestimate.

Suppose the background temperature were raised in the model. In that case, it necessarily requires an increase in the loop length to maintain consistency with observed intensities (see equation 4.6), resulting in loops of length $\gtrsim 8 \times 10^8$ cm and lifetimes $\gtrsim 6\times$ the observed values. If the loop lengths were fixed at $\approx 1 \times 10^8$ cm, the heating rate increases and causes the predicted model intensities to increase non-linearly, worsening the agreement between the simulated light curves and the observations.

The input parameters used for modeling might not be unique and vary depending on the filter used for reference background levels, peak values and rise time of the event. However, these variations will be small in comparison to the value of the input parameter.

Background heating rate in method 1 (background modeled), loop lengths are only functions of loop length. In the case of AIA-193 Å filter, an increase of half length of the loop from $0.9 \times 10^8$ cm to $1.1 \times 10^8$ cm changes the background light curve value from roughly 600 DN pix$^{-1}$ s$^{-1}$ to 750 DN pix$^{-1}$ s$^{-1}$. The difference in background levels of simulation and observations in each filter is similar to the difference between background levels of different brightenings (caused by a change in half length of the loop by $0.2 \times 10^8$ cm) in the same filter. Hence, depending on the filter considered for constraining half length of the loop using the match between the observed and simulated background level, the loop lengths would have been slightly different, but these differences would have been less than $10^8$ cm. Hence the conclusion of loop lengths being around $10^8$ cm is reliable. This is not an issue in method 2, where the background is not modeled, and the loop length is calculated using the spatial extent of the brightenings.

The peak value of lightcurves changes from 700 DN pix$^{-1}$ s$^{-1}$ to 1000 DN pix$^{-1}$ s$^{-1}$ in AIA-193 Å filter when the energy budget changed by multiplicative factor of $10^{0.2}$ ergs. The difference in peak value of simulation and observations in each filter is similar to the difference between peak levels of different brightenings (caused by a change in dissipated energy by a factor of $\times 10^{0.2}$ ergs) in the same filter. Hence, if other filters were used, the estimated energy budget would have differed by factors close to $10^{0.2}$ ergs. Once the loop length and energy budget have been fixed, the rise time of the event depends solely on the time duration of heating. Note that the rise time was chosen because the decay times are subject to large systematic errors due to the difficulty of identifying precisely when the model intensities become in-distinguishable from the background. Since the same heating duration gives rise time in simulations in fair agreement with observations, the filter chosen for finding the rise time is unimportant. It is also important to mention that EBTEL's DEM computation at temperatures lower than $\log[\mathrm{T(K)}] = 6.0$. Consequently, if filters like AIA-131, 171, and 335 Å are used for modeling the background of these events and constraining the lifetime of these events, the agreement between observation and simulations will worsen.

The input parameters used in this work for simulating transient brightenings mimicking AIA observations were such that the simulated Mach numbers performed using EBTEL remained subsonic at all instants. However, experimentation with a wide range of input parameters showed that the simulated Mach numbers could approach or exceed unity in many cases. This indicated EBTEL's neglect of kinetic energy in the energy conservation equation is problematic. Consequently, making EBTEL reliable in cases where kinetic energy cannot be ignored is necessary. This is the subject matter of the next chapter.

# 5

# Flows in Enthalpy-Based Thermal Evolution of Loops

*An efficient method for computing approximate but quick physics-based solutions for the dynamical evolution of coronal loops is to rely on space-integrated 0D simulations. The enthalpy-based thermal evolution of loops (EBTEL) framework is a commonly used method to study the mass and energy exchange between the corona and transition region. It solves for density, temperature, and pressure, averaged over the coronal part of the loop, the velocity at the coronal base, and the instantaneous differential emission measure distribution in the transition region. The single-fluid version of the code, EBTEL2, assumes that at all stages, the flows are subsonic. However, sometimes the solutions show the presence of supersonic flows during the impulsive phase of heat input. It is thus necessary to account for this effect. Here, we upgrade EBTEL2 to EBTEL3 by including the kinetic energy term in the energy evolution equation. This chapter, including all figures and tables, has been reproduced from Rajhans et al. 2022, "Flows in enthalpy-based thermal evolution of loops", ApJ, 924, 13 (DOI: 10.3847/1538-4357/ac3009).*

## 5.1 Introduction

One of the crucial assumptions made in EBTEL is that at all stages of the loop evolution, the flows are subsonic. Therefore, the kinetic energy term is neglected in the energy conservation equation. The assumption is found to be consistent with results from 1D field-aligned simulations performed using a wide range of loop lengths and heating functions. However, in cases where 1D simulations predict subsonic flows, EBTEL may compute Mach numbers close to or even exceeding unity. This is a consequence of neglecting the kinetic energy, which is relevant during the impulsive phase of the evolution of the loop in these cases. Mach numbers exceeding unity, in reality, would lead to shocks and disrupt flows by converting kinetic energy into heat. In such situations, a simplified 0D description becomes inadequate. Therefore, it is necessary to include the kinetic energy term in the energy conservation equation to study the dynamics of the loop.

It is important to point out that the Mach numbers generated by field-aligned codes can also reach values close to or exceeding unity. Note that while field-aligned codes such as HYDRAD have the potential to tackle shocks, it is not the case for 0D codes such as EBTEL. Nevertheless, despite the simplified nature of 0D calculations, the computed temperature, density, and pressure are in reasonable agreement with field-aligned 1D simulations performed using HYDRAD.

Here, we upgrade EBTEL2 to EBTEL3 by including the kinetic energy term in the energy conservation equation. In Section 5.2, we describe the 0D description of coronal loops, including kinetic energy in Section 5.2.1. In Section 5.2.2, we discuss the validity of approximations used in Section 5.2.1, describe their shortcomings, improve upon these approximations, and demon-strate their validity. All the changes discussed in Section 5.2.1 and 5.2.2 are incorporated in a newer version of the code, EBTEL3. In Section 5.3.1-5.3.6, we show the results obtained from EBTEL3 for various cases, covering a wide range of loop length, heating function, and Mach numbers at the coronal base. The obtained results have been compared with results obtained from EBTEL2 and HYDRAD. We also investigate the parameter space where the simulated 0D Mach numbers are unreliable and develop a heuristic way to identify such instances in Section 5.3.7. Finally, we summarize our work in Section 5.4.

## 5.2 EBTEL Framework

The 0D description of coronal loops require integrating the field-aligned hydrodynamical equations over the corona and the transition region. The details of the EBTEL code, which neglects the kinetic energy and solves the hydrodynamical equation, are discussed in detail in Chapter 3. In this chapter, we will refer to the 0D code of Cargill et al. (2012a) as EBTEL2*. Here we will derive and discuss the implications of 0D equations, including the kinetic energy in the energy equation.

### 5.2.1 0D equations without the assumption of subsonic flows

We now derive the 0D equations that avoid the assumption of subsonic flows in coronal loops. Using the field-aligned mass conservation equation (3.18) and equation of motion (3.19), we write the time derivative of kinetic energy,

$$\frac{\partial}{\partial t}\left(\frac{1}{2}n\mu v^2\right) = -\frac{1}{2}\mu v^2 \frac{\partial}{\partial s}(J - nv) - \frac{\partial}{\partial s}\left(\frac{1}{2}n\mu v^3\right) - v\frac{\partial P}{\partial s} + n\mu v g_{\parallel}. \quad (5.1)$$

*This work followed the first EBTEL paper by Klimchuk et al. (2008). Hence, named EBTEL2

Plugging equation 5.1 in energy conservation equation 3.20, we obtain

$$\frac{1}{\gamma-1}\frac{\partial P}{\partial t} = \frac{1}{2}\mu v^2 \frac{\partial}{\partial s}(J-nv) + v\frac{\partial P}{\partial s} - \frac{\partial}{\partial s}\left(\frac{\gamma}{\gamma-1}Pv + F\right) + Q - n^2\Lambda(T)\,. \tag{5.2}$$

We can include the effect of non-thermal electrons by replacing $J$ with $nv + J_{nt}$, and $F$ with $F_t + E_{nt}J_{nt}$. This gives,

$$\frac{1}{\gamma-1}\frac{\partial P}{\partial t} = \left(\frac{1}{2}\mu v^2 - E_{nt}\right)\frac{\partial}{\partial s}(J_{nt}) + v\frac{\partial P}{\partial s} - \frac{\partial}{\partial s}\left(\frac{\gamma}{\gamma-1}Pv + F_t\right) + Q - n^2\Lambda(T)\,. \tag{5.3}$$

Characteristic velocity of $10^{6-7}$ cm s$^{-1}$ in corona gives kinetic energy of an ion $(\sim \mu v^2) \approx 10^{-12} - 10^{-10}$ ergs. This is similar to average thermal energy per electron $(\sim k_B T) \approx 10^{-10}$ ergs for $T$=1 MK. For the non-thermal electrons to escape the thermal pool, their average energy should be much larger than the average thermal energy per electron $(\frac{1}{2}\mu v^2)$, and hence the above equation 5.3 can be simplified, as

$$\frac{1}{\gamma-1}\frac{\partial P}{\partial t} = v\frac{\partial P}{\partial s} - \frac{\partial}{\partial s}\left(\frac{\gamma}{\gamma-1}Pv + F\right) + Q - n^2\Lambda(T)\,. \tag{5.4}$$

Integrating equation 5.4 from the coronal base of the loop ($s=0$) to the loop apex ($s=L$) we find,

$$\frac{1}{\gamma-1}L\frac{d\bar{P}}{dt} = \int_0^L v\frac{\partial P}{\partial s}ds + \frac{\gamma}{\gamma-1}P_0v_0 + F_0 + \bar{Q}L - \bar{n}^2\Lambda(\bar{T})L\,. \tag{5.5}$$

Similarly, on integrating equation 5.4 spatially over the loop from the base of the transition region ($s=-l$) to the coronal base of the loop ($s=0$), we have

$$\frac{1}{\gamma-1}l\frac{d\bar{P}_{tr}}{dt} \approx 0 = \int_{-l}^0 v\frac{\partial P}{\partial s}ds - \frac{\gamma}{\gamma-1}P_0v_0 - F_0 - c_1\bar{n}^2\Lambda(\bar{T})L\,. \tag{5.6}$$

where $c_1$ is the ratio of total radiative losses from the transition region and corona (see equation 3.25). To solve the integral on the right-hand side of equation 5.6, we rearrange and integrate equation of motion (3.19) from $s = 0$ to $s = -l$. This gives us

$$\int_{-l}^{0} v \frac{\partial P}{\partial s} ds = -\int_{-l}^{0} \left( n\mu v \frac{\partial v}{\partial t} + n\mu v^2 \frac{\partial v}{\partial s} - n\mu v g_{||} \right) ds \approx -\frac{1}{2} n_0 \mu v_0^3 \qquad (5.7)$$

where we have assumed that the transition region is in a steady state, i.e., $\frac{\partial v}{\partial t} = 0$ and $nv = n_0 v_0$. Using estimates of $g \approx 10^4$ cm s$^{-2}$, $l \approx 10^7$ cm, and $v_0 \approx 10^{6-7}$ cm s$^{-1}$ (Klimchuk et al., 2008), we find that $n_0 \mu v_0 g_{||} l$ is negligible in comparison to $\frac{1}{2} n_0 \mu v_0^3$.

Using equations 5.7 in 5.6 we obtain

$$0 = -\frac{1}{2} n_0 \mu v_0^3 - \frac{\gamma}{\gamma - 1} P_0 v_0 - F_0 - c_1 \bar{n}^2 \Lambda(\bar{T}) L \,, \qquad (5.8)$$

which refers to the statement of energy balance in the transition region. The energy lost by radiation is the balance of the first three terms, i.e., energy fluxes of energy (kinetic, enthalpy, and heat) passing between the transition region and corona. It has been assumed that any mass and energy flow through the bottom of the transition region is negligible. Work done by or against gravity and direct heating in the thin transition region has also been neglected. However, note that in the presence of a large variation in the cross-sectional area, the transition region can thicken substantially (Cargill et al., 2021). The kinetic energy term in equation 5.8 is the only difference from the earlier versions of EBTEL. Note that equation 5.8 is cubic in $v_0$ and can be solved analytically.

Using $P_0 = 2n_0 k_B T_0$, we can re-arrange equation 5.8 in the following form

$$0 = v_0^3 + \frac{4\gamma k_B T_0}{\mu(\gamma - 1)}v_0 + \frac{2}{\mu n_0}\left(F_0 + c_1 \bar{n}^2 \Lambda(\bar{T})L\right) \quad (5.9)$$

The generic cubic equation is given by $v_0^3 + a v_0^2 + b v_0 + c = 0$, for real a, b, and c can either have three real roots, only one real root, or repeated real roots (i.e., two roots being same). Two quantities, f, and q, are defined as

$$f = b - \frac{a^2}{3} = \frac{4\gamma k_B T_0}{\mu(\gamma - 1)} \quad \& \quad q = \frac{2a^3}{27} - \frac{ab}{3} + c = \frac{2}{n_0 \mu}\left(F_0 + c_1 \bar{n}^2 \Lambda(\bar{T})L\right) \quad (5.10)$$

In the above equation, we have used the fact that the coefficient of $v_0^2$ is 0 i.e., $a = 0$.

A discriminant $(\triangle)$ can be defined as

$$\triangle = \frac{q^2}{4} + \frac{f^3}{27} = \frac{c^2}{4} + \frac{b^3}{27} \quad (5.11)$$

If $\triangle$ is positive we have only one real root, which is given by

$$v_0 = \left(-\frac{q}{2} + \sqrt{\triangle}\right)^{\frac{1}{3}} + \left(-\frac{q}{2} - \sqrt{\triangle}\right)^{\frac{1}{3}} \quad (5.12)$$

For real c, $c^2$ hence $q^2$ cannot be negative. $f^3$ is also positive. Hence we readily get the only real solution for $v_0$.

Adding equation 5.8 and equation 5.5 we get

$$\frac{1}{\gamma - 1}L\frac{d\bar{P}}{dt} = -\frac{1}{2}n_0 \mu v_0^3 + \int_0^L v\frac{\partial P}{\partial s}ds + \bar{Q}L - (1 + c_1)\bar{n}^2 \Lambda(\bar{T})L\,. \quad (5.13)$$

Field-aligned hydrodynamic simulations performed using HYDRAD for benchmarking our modified code show that

$$\int_0^L v\frac{\partial P}{\partial s}ds \approx -\frac{1}{2}n_0 \mu v_0^3 \quad (5.14)$$

holds well most of the time. It should be noted that this integral, under steady state assumption in corona, should have been

$$\int_0^L v \frac{\partial P}{\partial s} ds \approx - \int_0^L \left( n\mu v \frac{\partial v}{\partial t} + n\mu v^2 \frac{\partial v}{\partial s} - n\mu v g_{\parallel} \right) ds \approx \frac{1}{2} n_0 \mu v_0^3 \qquad (5.15)$$

if the third term was ignorable in the corona as it is in the transition region (see equation 5.7). This gives us the same magnitude as the expression we use, however, the sign is the opposite. Plugging this value in equation 5.13, we obtain

$$\frac{1}{\gamma - 1} L \frac{d\bar{P}}{dt} = -n_0 \mu v_0^3 + \bar{Q}L - (1 + c_1) \bar{n}^2 \Lambda(\bar{T}) L . \qquad (5.16)$$

The description of the system is completed by the 0D equation of mass conservation (3.27) and equation of state (3.29).

To test the results obtained by equations 3.27, 3.29, 5.8, and 5.16, collectively labeled as EBTEL2+KE, we consider an exemplar case which generates supersonic flows in EBTEL2 and subsonic flows in HYDRAD. We simulate the evolution of a loop of half-length $6.5 \times 10^9$ cm, with background heating adjusted such that the initial density and temperature are $11.05 \times 10^8$ cm$^{-3}$ and $2.51 \times 10^6$ K, respectively. The system is subjected to a symmetric triangular heating profile lasting for 200 s, with the maximum heating rate being 0.5 ergs cm$^{-3}$ s$^{-1}$ at $t = 100$ s. This corresponds to energy deposition of $3.25 \times 10^{11}$ ergs cm$^{-2}$ in 200 s.

The results of this test case are shown in Figure 5.1, with simulations from HYDRAD (solid black curves), EBTEL2 (dashed red curves), and EBTEL2+KE (dash-triple-dotted green curves). Following Cargill et al. (2012a) and Klimchuk et al. (2008), we have identified the coronal part of the loop in HYDRAD as the portion where the temperature is greater than or equal to 0.6 times the temperature at the loop top. To obtain coronal averages, pressure, density, and temperature have been averaged over this region. The lowest point of this region is taken as the coronal base, where velocity and Mach number have been evaluated. The apex quantities have been obtained by HYDRAD, averaging over the top 20% of the length of the loop.

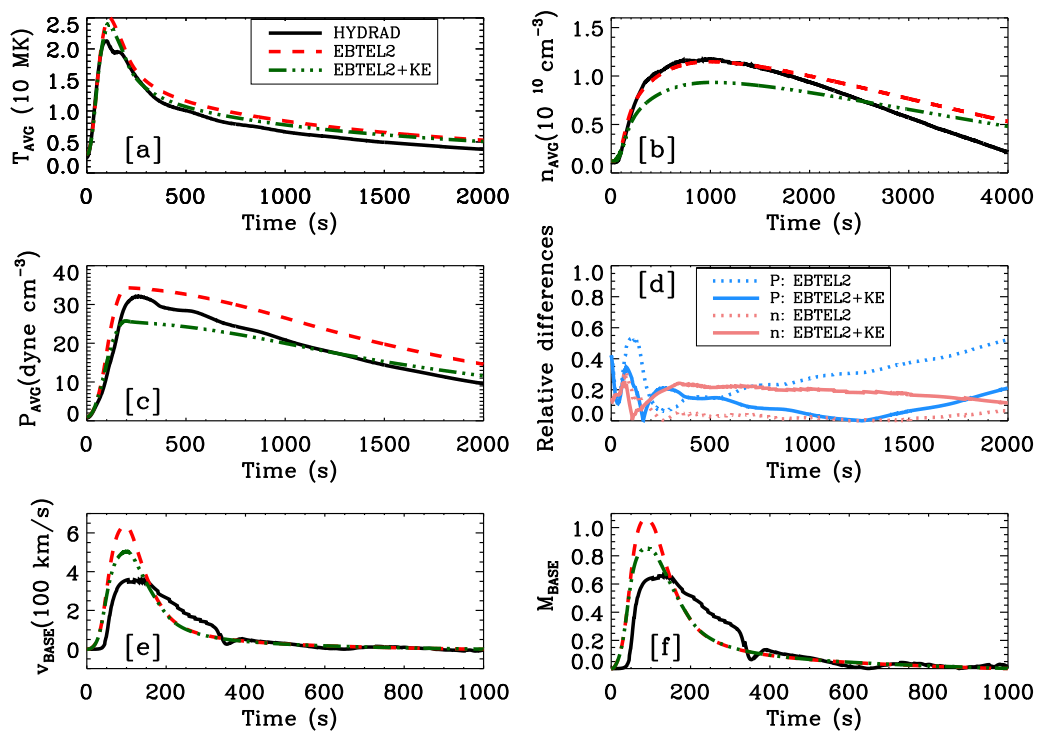

Figure 5.1: Simulation results for the exemplar case of subsonic flows due to large flare in $6.5 \times 10^9$ cm long loop. The curves demonstrate the time evolution of $\bar{T}$ (panel [a]), $\bar{n}$ (panel [b]), and $\bar{P}$ (panel [c]), the discrepancy in $\bar{n}$ (red curves) and $\bar{P}$ (blue curves) between HYDRAD and those calculated from EBTEL2 (dotted) and EBTEL2+KE (solid) (panel [d]), the velocity at the coronal base of loop ($v_0$) (panel [e]), and corresponding Mach number ($M_0$) (panel [f]).

Figure 5.1 shows the time evolution of average temperature (panel [a]), the average electron number density (panel [b]), and the average pressure (panel [c]). We also plot in panel [d] the absolute relative differences of $\bar{P}$ and $\bar{n}$ as computed from EBTEL2 or EBTEL2+KE relative to HYDRAD, which are defined as

$$\frac{\Delta \bar{P}}{\bar{P}} = \left| \frac{\bar{P}_{1D} - \bar{P}_{0D}}{\bar{P}_{1D}} \right| \quad \& \quad \frac{\Delta \bar{n}}{\bar{n}} = \left| \frac{\bar{n}_{1D} - \bar{n}_{0D}}{\bar{n}_{1D}} \right| . \qquad (5.17)$$

In panels [e] and [f], we show the time evolution of velocity and Mach number at the coronal base ($v_0$ and $M_0$), respectively.

We find that the temperatures obtained using EBTEL2 and EBTEL2+KE are nearly identical (panel [a]). However, the electron number densities obtained with EBTEL2+KE are consistently lower than those obtained with EBTEL2 (panel [b]). However, the relative error remains less than 25 % (see panel [d]). The average pressure plotted in panel [c] shows that the agreement between HYDRAD and EBTEL2+KE is better, except during a small time window close to where pressure peaks. A better agreement for velocity is also seen in HYDRAD and EBTEL2+KE (panel [e]) and Mach number (panel [f]) at the base. While HYDRAD produces subsonic flows at the base (panel [f]), EBTEL2 produces Mach numbers exceeding 1. If these are trusted, it should lead to shocks, which 0D simulations cannot tackle. EBTEL2+KE brings down Mach numbers below unity. However, we note that improvements in Mach numbers are unsatisfactory, given the small improvement in velocity and Mach numbers at the cost of deterioration in electron number density. Therefore, further improvements are needed.

## 5.2.2 Assessment and improvements of the approximations

While seeking solutions to the equations 3.27, 3.29, 5.8, and 5.16, we have made two approximations, which we now discuss in some details. The first approximation (also present in EBTEL2) is related to the ratio of pressure and electron number density at the coronal base with their average values. EBTEL2 computes the ratio of the pressure at the coronal base and average pressure by assuming the system to be in hydrostatic equilibrium i.e., $P(s) = P_0 \exp(-s/L_H(\bar{T}))$. The variation of temperature along the loop is neglected and scale height $(L_H)$ is computed using a temperature equal to the coronal average value. Cargill et al. (2012a) discuss that this is equivalent to $[\frac{P_0}{P}]_{hse} = \exp(2L \sin(\frac{\pi}{5})/\pi L_H(\bar{T}))$ for a semicircular loop, where the sub-script *hse* indicates that the loop is in hydrostatic equilibrium at a uniform temperature ($\bar{T}$). Using the constants $\frac{\bar{T}}{T_a} = c_2 = 0.9$ and $\frac{T_0}{T_a} = c_3 = 0.6$, the temperature at the coronal base ($T_0$) is 0.67 times the average temperature of loop ($\bar{T}$). Using these along with $P_0 = 2n_0 k_B T_0$ and $\bar{P} = 2\bar{n} k_B \bar{T}$, the ratio of electron number density at the coronal base to its coronal average,

$$\left[\frac{n_0}{\bar{n}}\right]_{hse} = \frac{3}{2}\exp(2L\sin(\frac{\pi}{5})/\pi L_H(\bar{T})). \tag{5.18}$$

To assess this approximation, we define the following two coefficients and compute these using HYDRAD,

$$c_4 = \left[\frac{P_0}{\bar{P}}\right]\left[\frac{P_0}{\bar{P}}\right]_{hse}^{-1} \quad \& \quad c_5 = \left[\frac{n_0}{\bar{n}}\right]\left[\frac{n_0}{\bar{n}}\right]_{hse}^{-1} \tag{5.19}$$

We show the evolution of $c_4$ and $c_5$ as the solid curves in panels [a] and [b] of Figure 5.2, respectively. To relate the pressure and electron number density at the base of the corona and their averages across the loop, EBTEL2 assumes a hydrostatic profile, with $c_4$ and $c_5$ approximated as constants, $c_{4A} = 1$ and $c_{5A} = 1$ respectively; these are shown as dashed lines in Figure 5.2. These approximations are discrepant by up to factors of 2 during the impulsive phase. Hence, we need to develop a better approximation to relate the base pressure and electron number density to the quantities computed within the EBTEL framework, $\bar{P}, \bar{n}, \bar{T}$ and $v_0$

We attribute the variation of $c_4$ and $c_5$ from their constant values of 1, mainly to two aspects: 1) since the plasma is accelerated into the loop, it requires larger pressure gradients than hydrostatic values in lower parts of the loop, 2) since the plasma velocity at the apex of the loop is zero, the flow must be decelerated, and therefore the pressure gradient in the upper parts of the loop should be smaller than hydrostatic or perhaps even in the opposite direction. From the plots, it appears that for the first 150 s, the first cause dominates, while the second dominates during the later phase till about 350 s.

To ascertain the validity of our second approximation (equation 5.14), we compute following two quantities,

$$I = \int_0^L v \frac{\partial P}{\partial s} ds - \frac{1}{2} n_0 \mu v_0^3 \qquad (\textit{without approximation})$$
$$I_A = -n_0 \mu v_0^3 \qquad (\textit{with approximation})$$

(5.20)

using HYDRAD and plot them in panel [c] of figure 5.2. The reasonable match between the two curves suggests that this approximation is satisfactory, albeit relatively poor, between 150-350 seconds. This could be attributed to the following effects. In the impulsive phase (up to 150 s), to fill plasma into the loop, there should be acceleration. Hence dominating contribution to $I$ should be from portions of the loop where force due to pressure gradients ($-\frac{\partial P}{\partial s}$) are in the same direction as the velocity of plasma ($v$). This means pressure gradients are opposite to plasma velocity at locations where a larger contribution to $I$ comes from. Since our approximation $I_A$ predicts negative values, it matches well with $I$ in this duration. However, between 150-350 seconds, the effect of pressure gradients with directions opposite to hydrostatic pressure gradients (for decelerating plasma) becomes more important (see panel [a] of figure 5.2). Since plasma is being decelerated, a larger contribution to $I$ should come from portions of the loop where pressure force ($-\frac{\partial P}{\partial s}$) is opposite to velocity ($v$). For such regions, the pressure gradient is along with velocity, but since our approximation, $I_A$ still gives negative values, it introduces errors.

To improve on the first approximation, we first develop insight into the form of pressure at the base of the corona in terms of quantities computed within the framework of 0D simulations under the assumption of steady flow and uniform temperature ($\bar{T}$) along the loop. While these conditions are not met in the impulsive phase, where kinetic energy dominates, the resulting expressions can be generalized.

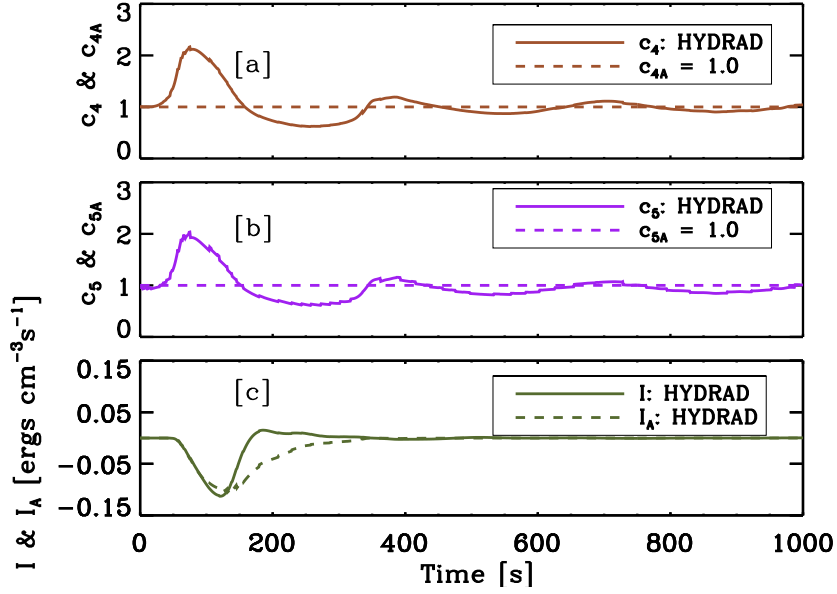

Figure 5.2: Comparing the effects of different approximations within HYDRAD for the exemplar case (see Figure 5.4). [a] The top panel (with brown curves) shows the evolution of the pressure coefficient $c_4$ (solid curve; equation 5.19) and the approximation used in EBTEL2, $c_{4A} = 1$ (dashed line). [b] The middle pane (purple curves) shows the evolution of the electron number density coefficient $c_5$ (solid curve; equation 5.19) and the approximation used in EBTEL2, $c_{5A} = 1$ (dashed line). [c] The bottom panel (olive curves) shows the evolution of energy flux $I$ and our approximation $I_A$ (solid and dashed curves respectively; equation 5.20), with both $I$ and $I_A$ are computed using HYDRAD. Note that the approximations to the pressure and electron number density coefficients $c_{4A}$ and $c_{5A}$ are discrepant by factors of up to 2, whereas the energy flux approximation $I_A$ is adequate.

Assuming a steady flow in equation 3.19 and combining that with equation 3.9, we have

$$\frac{1}{P}\frac{\partial P}{\partial s} = \frac{\mu}{2k_B \overline{T}}\left(g_{||} - \frac{1}{2}\frac{\partial v^2}{\partial s}\right) \tag{5.21}$$

Under the assumption that acceleration due to gravity and any variation in temperature can be neglected, we integrate equation 5.21 to obtain

$$\frac{P(s)}{P_0} = \exp\left[\frac{\mu}{2k_B \overline{T}}\left(sg_{||} - \frac{1}{2}(v^2 - v_0^2)\right)\right] = \exp\left[-\frac{s}{L_H}\right]\exp\left[\frac{\gamma}{2}(M_0^2 - M^2)\right] \tag{5.22}$$

where $M = \sqrt{\frac{\mu v^2}{2\gamma k_B \bar{T}}}$, and $L_H = -\frac{2k_B \bar{T}}{\mu g_{||}}$. The minus sign in $L_H$ is due to $g_{||}$ being negative. $M_0$ denotes Mach number at base $s = 0$.

We now convert the above expression into a form that can be used to place bounds on pressure at the base of the corona in terms of quantities that can be computed within the domain of 0D description of loops, viz., average pressure and Mach number at the coronal base. We consider two functions $X_1$ and $X_2$ of time and loop length such that for a particular loop of length L, the minimum and maximum values of $\exp\left[-\frac{\gamma}{2}(M_0^2 - M^2)\right]$ at a time are $\exp\left[X_1 M_0^2\right]$ and $\exp\left[X_2 M_0^2\right]$, respectively. Hence, averaging equation 5.22 over the coronal part of the loop, we obtain

$$\left[\frac{P_0}{\bar{P}}\right]_{hse} \exp\left[X_1 M_0^2\right] \leq \left[\frac{P_0}{\bar{P}}\right] \leq \left[\frac{P_0}{\bar{P}}\right]_{hse} \exp\left[X_2 M_0^2\right] \qquad (5.23)$$

where $\left[\frac{P_0}{\bar{P}}\right]_{hse}$ is the ratio of base and average pressure if the system was in hydrostatic equilibrium and isothermal with a temperature $\bar{T}$, while $\left[\frac{P_0}{\bar{P}}\right]$ is the actual value. The expressions in equation 5.23 are consistent with the fact that in the regime where Mach numbers approach 0, we will recover the hydrostatic expression.

Without any loss of generality, we can express the ratio $\left[\frac{P_0}{\bar{P}}\right]$ and $\left[\frac{n_0}{\bar{n}}\right]$

$$\left[\frac{P_0}{\bar{P}}\right] = \left[\frac{P_0}{\bar{P}}\right]_{hse} \exp[\phi(t,L)M_0^2] \quad \& \quad \left[\frac{n_0}{\bar{n}}\right] = \left[\frac{n_0}{\bar{n}}\right]_{hse} \exp[\phi(t,L)M_0^2] \quad (5.24)$$

where $\phi$ is a function of time and loop length. Moreover, for a particular loop at a time t, it is bounded by the relation $X_1 \leq \phi \leq X_2$. Even though we arrived at equation 5.24 assuming steady flow in an isothermal loop, one can always invert it to express $\phi(t,L)$ at each time step in terms of $\left[\frac{P_0}{\bar{P}}\right]$, and $\left[\frac{P_0}{\bar{P}}\right]_{hse}$, for the general case of non-steady flows in multi-thermal loops.

However, it is infeasible to derive an exact expression for $\phi(t, L)$ as it would require a field-aligned simulation in the first place to be computed. Therefore, we approximate it as a constant in time independent of loop length. We compared the results between HYDRAD and EBTEL for different values of $\phi$ while also comparing $c_4$ and $c_5$ with the new code. We compute again the approximations $c_{4A}$ and $c_{5A}$ to coefficients $c_4$ and $c_5$ using the expressions

$$c_{4A} = \exp[\frac{3}{2}M_0^2] \quad \& \quad c_{5A} = \exp[\frac{3}{2}M_0^2] \tag{5.25}$$

We find that the calculations are not sensitive to the precise value adopted and choose $\phi = \frac{3}{2}$.

We plot $c_{4A}$ and $c_{5A}$ in panels [a] and [b] of Figure 5.3 using brown dashed ($c_{4A}$) and purple dashed ($c_{5A}$) lines. For comparison, we have also plotted $c_4$ and $c_5$ computed by HYDRAD using solid lines. The plots reveal that the new approximations on $c_4$ and $c_5$ resulting in $c_{4A}$ and $c_{5A}$, respectively, show the most prominent peak observed during the impulsive phase, albeit at slightly later stage. Furthermore, other peaks occurring in $c_4$ and $c_5$ during the later phase of the evolution are also close to the values of $c_{4A}$ and $c_{5A}$ in the limit $M_0 \rightarrow 0$. However, the prominent dips observed in $c_4$ and $c_5$ between 150 and 350 sec are not captured by $c_{4A}$ and $c_{5A}$. We have analyzed these shortcomings and found that the errors due to these are rather small because of the decent match between profiles of $\bar{P}, \bar{n}, \bar{T}$ and $v_0$, computed from field-aligned and 0D simulations for different cases considered (see Section 5.3 and Figure 5.4).

We stress that the resultant expressions in equation 5.25 are crude because approximating $\phi(t, L)$ in equation 5.24 with a constant value of $\frac{3}{2}$ does not capture its complicated variation with time and loop length. However, it achieves the twin goals of capturing the most prominent feature (the first peak) of $c_4$ and $c_5$ computed from HYDRAD and improving the results of 0D simulations over a wide range of parameter space of solar coronal loops as long as the flows in HYDRAD are subsonic.

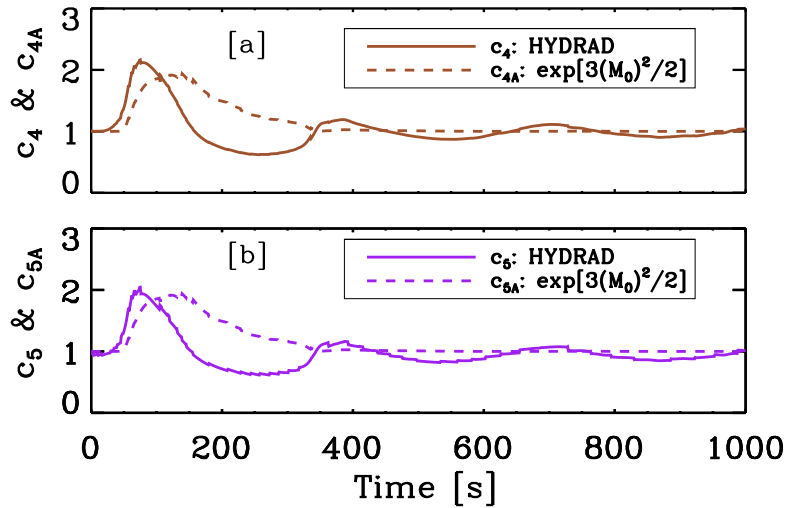

Figure 5.3: As in panels [a] and [b] of Figure 5.2, but comparing the HYDRAD pressure and electron number density coefficients $c_4$ and $c_5$ (solid curves) with approximations made as $[\exp\left(\frac{3}{2}M_0^2\right)]$ (dashed curves; see equation 5.25) where the Mach number at the base of the corona, $M_0$, is computed via HYDRAD. Note that the modifications offer a qualitative improvement in the approximations.

Due to the modifications made to the expression of the base pressure and hence base electron number density (see equation 5.25), equation 5.9 is no longer cubic and has the form

$$v_0(t)^3 + b\,v_0(t) + c\,\exp(d\,v_0(t)^2) = 0\,.$$

Therefore, to solve the above equation, we first use an adaptive time grid where the time step $\delta t$ is set to 0.1 times the smallest of the conductive, radiative, and sonic timescales at time $t$. At each time step $t$, we use the velocity at the previous time step $(v_0(t - \delta t))$ to find the analytical root of the equation cubic in $v_0(t)$, i.e.,

$$v_0(t)^3 + b\,v_0(t) + c\,\exp(d\,v_0(t - \delta t)^2) = 0\,,$$

The values of the physical quantities obtained in consecutive time steps differ by <10% due to the adaptive time stepping. Note that since the loop evolution always begins from a stationary state, i.e., at $t = 0$, velocity is zero.

Finally, using the above-described approximations, we simulate the plasma dynamics of the monolithic loop discussed in Section 5.2.1 and plot the obtained results in Figure 5.4. We denote the results obtained using the above-described approximations, including the addition of KE, with EBTEL3. For comparison, we have plotted the results obtained from HYDRAD (black-solid) and EBTEL2 (red-dashed) in Figure 5.4. The results show that the temperature (see panel [a]) produced by EBTEL2 and EBTEL3 almost overlap. However, both density and pressure are lower in EBTEL3 than EBTEL2. However, the density derived using EBTEL2 is better matched with HYDRAD than that derived with EBTEL3 (see panel [b]). This discrepancy is partially due to the errors in $c_{4A}$ and $c_{5A}$. However, pressure is better reproduced using EBTEL3 (see panel [c]). We plot the velocities and Mach numbers measured at the base of the corona in panels [e] and [f], respectively, which clearly show that EBTEL3 performs better than EBTEL2. Though EBTEL2 predicts Mach numbers exceeding unity, EBTEL3 manages to bring it down to values predicted by HYDRAD. We plot the relative errors in average pressure and electron number density in panel [d]. We find that the deterioration in density due to our approximation over EBTEL2 is less than the improvement in pressure.

## 5.2.3 Density computed by EBTEL2 and EBTEL3

The question remains, however, as to why EBTEL2 does a better job than EBTEL3 in predicting the average density when compared to HYDRAD, even when the predicted velocity at the coronal base shows a substantial discrep-ancy. This may be explained as follows. During the impulsive phase, conductive losses dominate, and radiative losses are negligible (see e.g. Rajhans et al., 2021; Subramanian et al., 2018; Cargill et al., 1995a). Therefore, the sum of enthalpy and kinetic energy flux across coronal base $\left(\frac{\gamma}{\gamma-1}P_0 v_0 + \frac{1}{2}\mu n_0 v_0^2\right)$ is approximately equal to the conduction flux $(F_0)$ across it. This approximate equality can be used in addition to the equation of state (3.9) to write the mass flux across the coronal base as

$$\mu n_0 v_0 = -\frac{F_0}{\frac{2\gamma k_B T_0}{\mu(\gamma-1)} + \frac{v_0^2}{2}} \qquad (5.26)$$

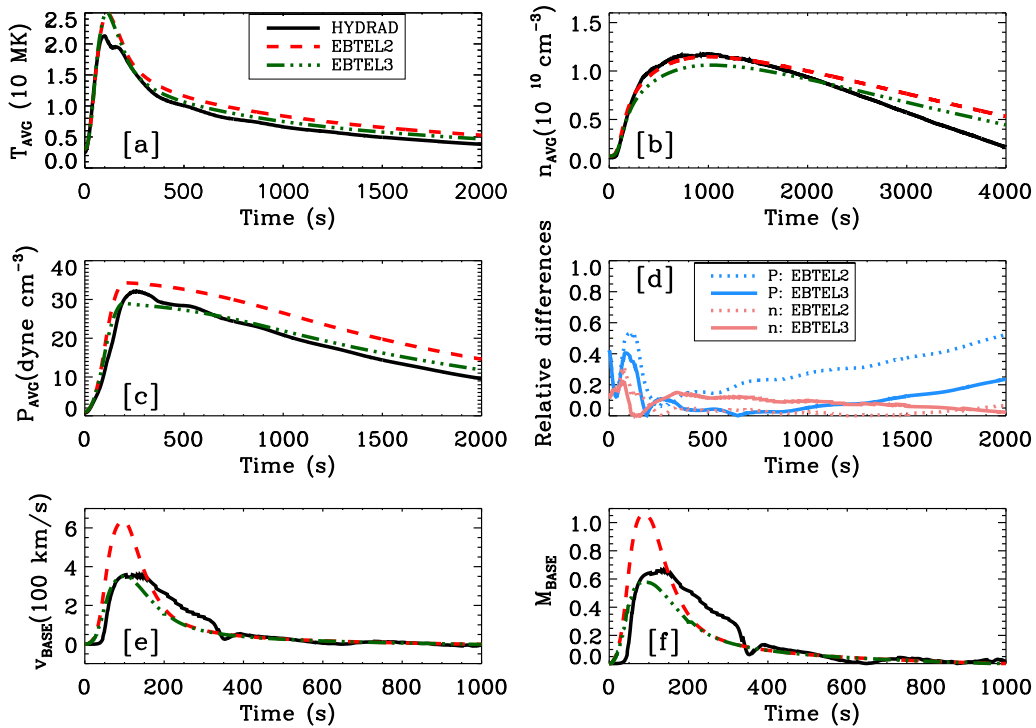

Figure 5.4: As in Figure 5.1 but with over-plotted green curves obtained with EBTEL3.

We now look at the quantities involved in this equation, namely temperature, conduction flux, and velocity. The temperature estimated by EBTEL2 and EBTEL3 is higher than those obtained using HYDRAD, which is due to the conduction flux being underestimated in EBTEL2 and EBTEL3. Hence a lower magnitude of $F_0$ in EBTEL2 and EBTEL3 than HYDRAD will tend to make a magnitude of $\mu n_0 v_0$ lower in EBTEL2 and EBTEL3 than HYDRAD. Additionally, an overestimated temperature $T_0$ in EBTEL2 and EBTEL3 will tend to make a magnitude of $\mu n_0 v_0$ lower than those obtained from HY-DRAD. However, the neglect of kinetic energy ($\propto \frac{v_0^2}{2}$) in EBTEL2 will tend to make the magnitude of mass flux higher than HYDRAD. This is not the case with EBTEL3. This last effect is the most important contributor to the discrepancy between mass flux in EBTEL2 and HYDRAD. The mass flux hence obtained from EBTEL3 agrees better with HYDRAD.

However, neither EBTEL2 nor EBTEL3 consider the dynamic change in the length of the coronal part of the loop. Both assume that the whole loop is in the corona. To compare the EBTEL2 or EBTEL3 results, the coronal portion of loops in HYDRAD is defined dynamically by measuring the length of the part of the loop with temperature $\geq 0.6T_a$. Due to this definition, the coronal portion of the loop identified by HYDRAD could be significantly smaller than the full length of the loop considered by EBTEL2 and EBTEL3 in the impulsive phase. This is consistent with the requirement of larger temperature gradients to transfer larger conduction flux to the transition region. The error caused by using the larger length of the coronal part in equation 3.27 tends to compensate for the higher mass flux in EBTEL2. As a result, $\bar{n}$ computed by EBTEL2 and HYDRAD match well. The improvement in velocity in EBTEL3 brings the mass flux computed by EBTEL3 closer to that by HYDRAD, but the error caused by taking fixed $L$ in equation 3.27 is not canceled sufficiently. This leads to a worsening of densities in EBTEL3 than EBTEL2. However, we emphasize that the worsening in densities is significantly smaller than the improvement in pressure and velocity in terms of agreement with results from HYDRAD.

## 5.3   Results

We next carry out a systematic verification covering a useful range of the parameter space of solar coronal loops. Cases 2–4 have been taken from Cargill et al. (2012a). All the heating functions have a symmetric triangular profile. The results are compared with those from EBTEL2 and HYDRAD. Table 5.1 provides the details of the test cases chosen for simulations. For the sake of completeness, we have also provided in the same table input parameters for case 1 (subsonic flows due to large flare in 6.5 $\times 10^9$ cm long loop), which was discussed in §5.2.1 and Section 5.2.2. We discuss the results from all the test cases one by one below. Note that the test cases are chosen such that they cover a wide range of loop lengths, and heating functions and the maximum velocities reached in HYDRAD at the coronal base cover the regimes of subsonic (cases 1-4), transonic case (5-6) and supersonic (case 7) flows. The maximum Mach numbers achieved at the base ($M_0$) in HYDRAD simulations are also listed in Table 5.1. We discuss the results from all the test cases one by one below.

### 5.3.1   Case 2: Subsonic flows due to small flare in $7.5 \times 10^9$ cm long loop

We simulate the plasma dynamics in a loop of half-length of $7.5 \times 10^9$ cm with background heating adjusted such that the initial electron number density and temperature are $0.62 \times 10^8$ cm$^{-3}$ and $0.85 \times 10^9$ K, respectively. We provide a symmetric triangular heating profile lasting for 500 s, with the maximum heating rate being 0.0015 ergs cm$^{-3}$ s$^{-1}$ at $t = 250$ s.

The results are plotted in Figure 5.5. As for the first case described above, there is little difference in the temperature profile (panel a), but the elec-

Table 5.1: Simulation parameters such as peak heating rate (2nd column), duration of heating (3rd column), loop half-length (4th column), initial electron number density (5th column), initial temperature (6th column), maximum Mach numbers at coronal base computed by HYDRAD (7th column), EBTEL2 (8th column) and EBTEL3 (9th column) for various test cases.

| Index | Peak heating rate [ergs cm$^{-3}$ s$^{-1}$] | Duration of heating [s] | Half Length [$10^8$ cm] | Initial $\bar{n}$ [$10^8$ cm$^{-3}$] | Initial $\bar{T}$ [$10^6$ K] | $(M_0)_{max}$ HYDRAD | $(M_0)_{max}$ EBTEL2 | $(M_0)_{max}$ EBTEL3 |
|---|---|---|---|---|---|---|---|---|
| 1 | 0.5 | 200.0 | 65.0 | 11.05 | 2.51 | 0.67 | 1.10 | 0.58 |
| 2 | $1.5\times10^{-3}$ | 500.0 | 75.0 | 0.62 | 0.85 | 0.57 | 0.88 | 0.50 |
| 3 | $1.0\times10^{-2}$ | 200.0 | 25.0 | 2.46 | 0.73 | 0.49 | 0.61 | 0.42 |
| 4 | 2.0 | 200.0 | 25.0 | 22.32 | 2.06 | 0.55 | 0.81 | 0.51 |
| 5 | $1.0\times10^{-2}$ | 200.0 | 50.0 | 0.84 | 0.71 | 0.75 | 1.32 | 0.63 |
| 6 | $1.5\times10^{-2}$ | 200.0 | 75.0 | 0.80 | 0.92 | 0.94 | 1.46 | 0.65 |
| 7 | $1.0\times10^{-3}$ | 200.0 | 60.0 | 0.13 | 0.42 | 1.15 | 1.51 | 0.65 |

tron number density (panel b) is underestimated using EBTEL3 compared to
that from EBTEL2, and the match with HYDRAD deviates. Similar to the
underestimation of electron number density, pressure is also underestimated
with EBTEL3 compared to EBTEL2 and shows better correspondence with
the results from HYDRAD. The relative errors in electron number density
and pressure obtained by EBTEL3 and EBTEL2 with that of HYDRAD are
also plotted in panel [d]. Similar to case 1, we note that the deviation in the
electron number density due to approximation in EBTEL3 is smaller than
the improvements seen in pressure.

The velocity and the Mach number estimated at the base of the corona are
lower in simulations using EBTEL3 than EBTEL2 and are in good agreement
with those obtained with HYDRAD. This is a sub-sonic case where EBTEL2
and HYDRAD produce Mach numbers at the base lower than 1.

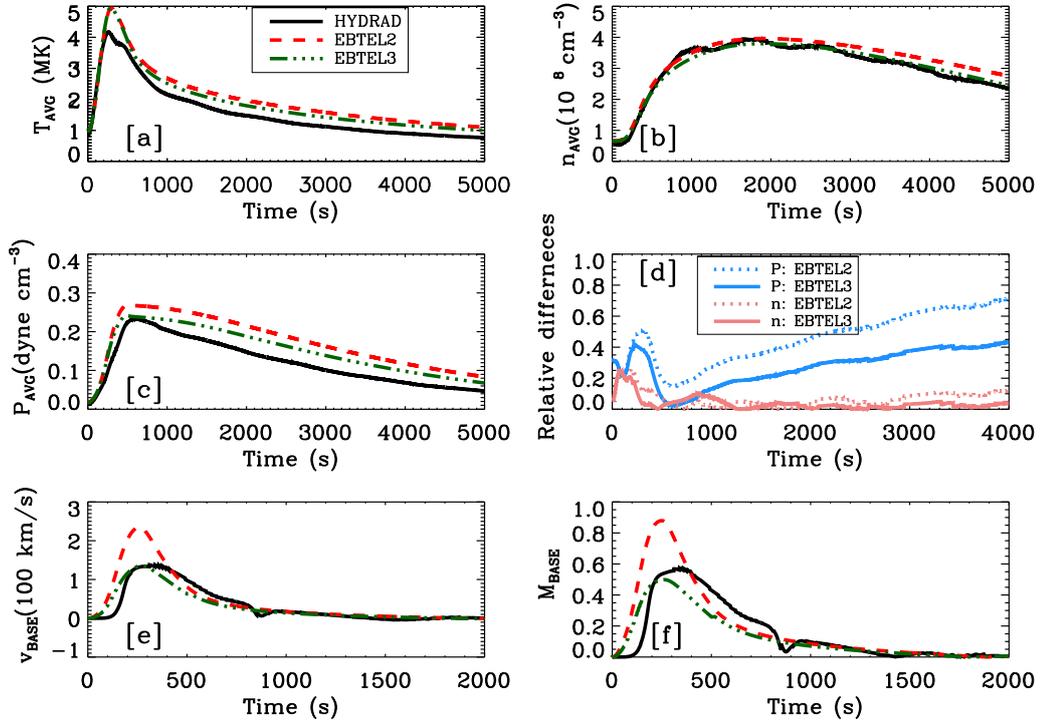

Figure 5.5: As in Figure 5.4 but for Case 2: A $7.5 \times 10^9$ cm loop receiving 2.8125 $\times 10^9$ ergs cm$^{-2}$ in 500 s.

## 5.3.2 Case 3: Subsonic flows due to small flare in $2.5 \times 10^9$ cm long loop

The third case simulates the plasma dynamics in a loop of half-length of $2.5 \times 10^9$ cm. The background heating is adjusted such that the initial electron number density and temperature are $2.46 \times 10^8$ cm$^{-3}$ and $0.73 \times 10^6$ K, respectively. We provide a symmetric triangular heating profile lasting for 200 s, with the maximum heating rate being 0.01 ergs cm$^{-3}$ s$^{-1}$ at $t = 100$ s. We plot the results in Figure 5.6. Pressure, velocity, and Mach number show significant improvements compared with EBTEL2.

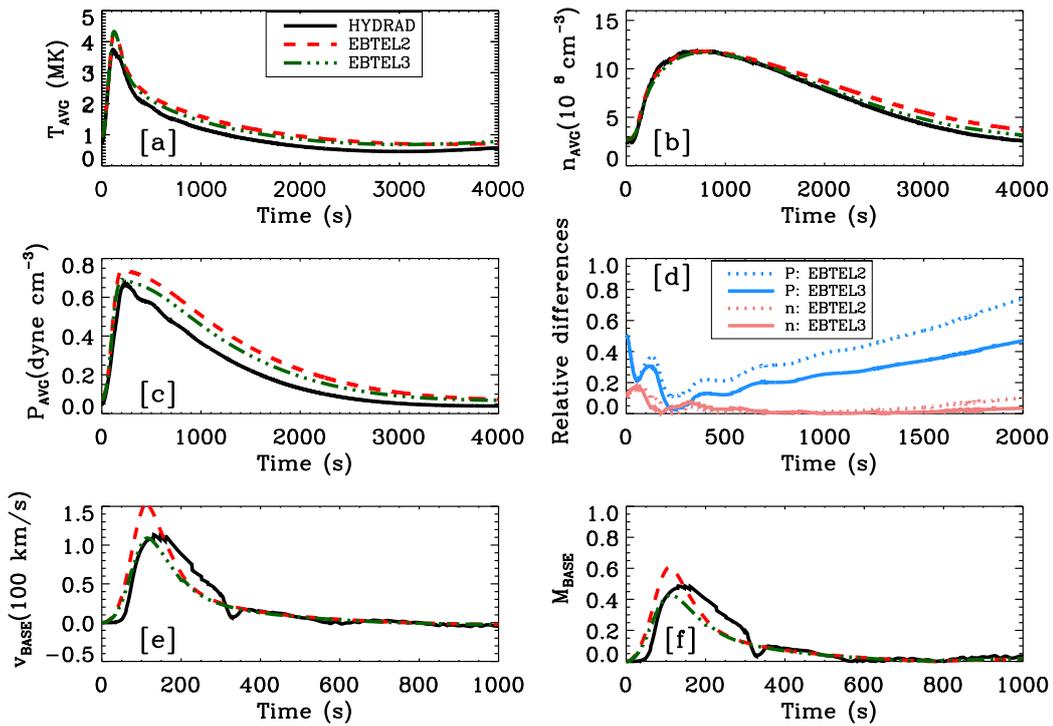

Figure 5.6: As in Figure 5.4 but for Case 3: A $2.5 \times 10^9$ cm loop receiving $2.5 \times 10^9$ ergs cm$^{-2}$ in 200 s.

### 5.3.3 Case 4: Subsonic flows due to large flare in $2.5 \times 10^9$ cm long loop

Here we take a loop of half-length of $2.5 \times 10^9$ cm with background heating adjusted such that the initial electron number density and temperature are $22.32 \times 10^8$ cm$^{-3}$ and $2.06 \times 10^6$ K, respectively. We provide a symmetric triangular heating profile lasting for 200 s, with the maximum heating rate being 2.0 ergs cm$^{-3}$ s$^{-1}$ at $t = 100$ s. The results are shown in Figure 5.7. The temperature, electron number density, and pressure show similar evolution to the other examples described above. The velocity and Mach number plots are shown in panel [e], and [f] shows remarkable improvements in EBTEL3 over EBTEL2 compared to HYDRAD.

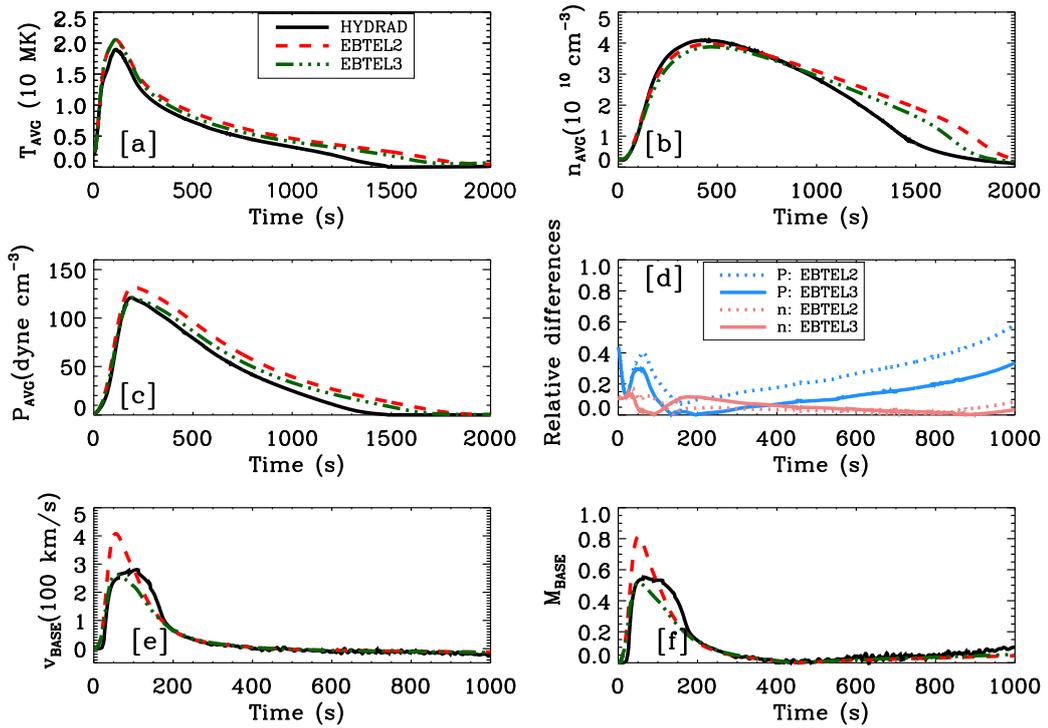

Figure 5.7: As in Figure 5.4 but for Case 4: A $2.5 \times 10^9$ cm loop receiving $5.0 \times 10^{11}$ ergs cm$^{-2}$ in 200 s.

### 5.3.4 Case 5: Transonic flows in $5.0 \times 10^9$ cm long loop

Here we consider a loop of half-length of $5.0 \times 10^9$ cm with initial electron number density and temperature are $0.84 \times 10^8$ cm$^{-3}$ and $0.71 \times 10^6$ K, respectively. The heating profile lasts for 200 s, with the maximum heating rate being 0.01 ergs cm$^{-3}$ s$^{-1}$ at $t = 100$ s. We plot the results in Figure 5.8. Panel [a] shows the temperature produced by EBTEL2 and EBTEL3 are not overestimated but are in close agreement with HYDRAD. This is because, in this case, even though velocities at the base remain subsonic throughout, velocities produced by HYDRAD at intermediate positions in the loop reach supersonic velocities, which leads to shocks and hence dissipation of kinetic energy. This adds to the temperature produced by HYDRAD. Since the 0D description cannot incorporate the physics of shocks, there is no increase in temperature due to the dissipation of kinetic energy. Panels [b]-[d] show similar electron number density and pressure trends as in previous cases. Panel [e] and [f] show the velocity and Mach number at the coronal base, respectively. The maximum Mach numbers produced at the base of the corona in HYDRAD simulations is 0.75, and those produced by EBTEL3 are $\approx$ 0.63. Mach number produced in EBTEL2 are supersonic $\approx$ 1.4. The match between Mach numbers computed using HYDRAD and EBTEL3 is worse than in previous cases.

### 5.3.5 Case 6: Transonic flows in $7.5 \times 10^9$ cm long loop

The sixth case is of a loop of half-length of $7.5 \times 10^9$ cm with initial electron number density and temperature are $0.80 \times 10^8$ cm$^{-3}$ and $0.92 \times 10^6$ K, respectively. The heating profile lasts for 200 s, with the maximum heating rate being 0.015 ergs cm$^{-3}$ s$^{-1}$ at $t = 100$ s. The results are shown in Fig-ure 5.9. HYDRAD, in this case, predicts a higher temperature than EBTEL due to the dissipation of kinetic energy by shocks in HYDRAD (see panel [a]). We observe the same pattern in the time evolution of electron number density and pressure (see panels [b]-[d]). Panel [e] and [f] show the velocity and Mach number at the coronal base of the loop. In this case, HYDRAD predicts a maximum Mach number of 0.94, while EBTEL2 and EBTEL3 predict maximum Mach numbers of 1.5 and 0.65, respectively. Though the Mach numbers produced by EBTEL3 are unreliable because of being significantly lower than HYDRAD, the agreement between velocities is better.

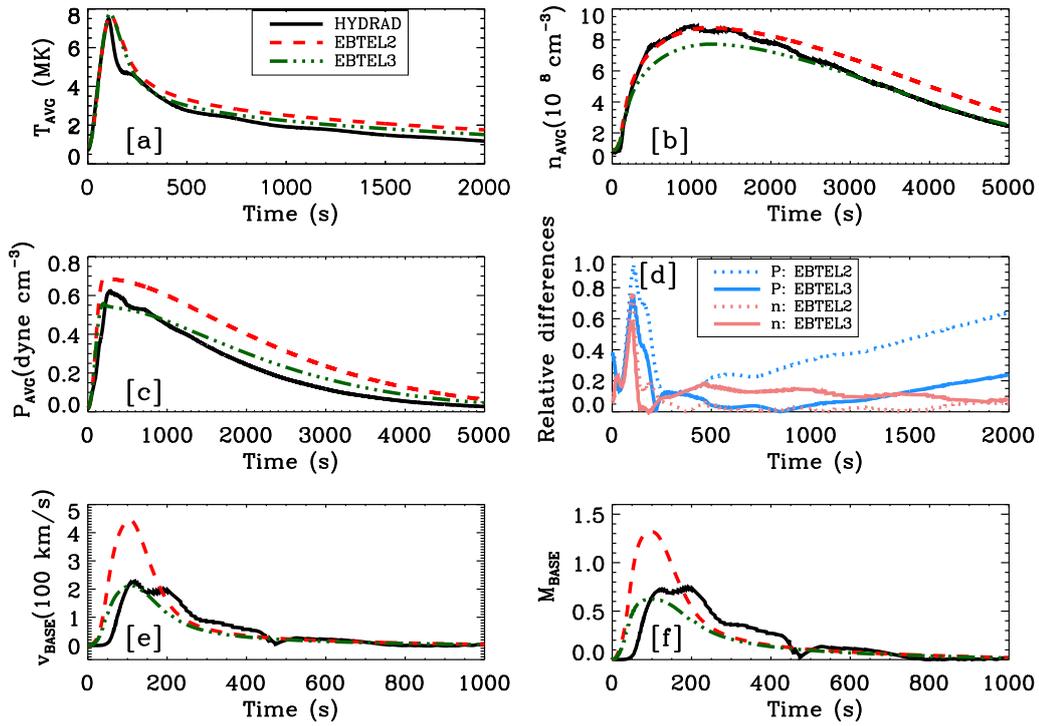

Figure 5.8: As in Figure 5.4 but for Case 5: A $5.0 \times 10^9$ cm loop receiving $5.0 \times 10^9$ *ergs cm⁻² in 200 s.*

Interestingly the agreement between velocity predicted by EBTEL3 and HYDRAD is worse in this case than in the previous cases where the maximum Mach number at the coronal base was well below 1. This worsening of velocity computed by EBTEL3 can be understood as follows. Shocks are formed from the abrupt slowing of a supersonic evaporative upflow. A discontinuous increase in density accompanies the discontinuous decrease in velocity with height. Clearly, $\frac{n_0}{\bar{n}}$ must be higher without the shock than with it. Consequently equation 5.25 gives higher values of $n_0$ in EBTEL3. The overestimation of $n_0$ in equation 5.8 leads to a lower value of $v_0$.

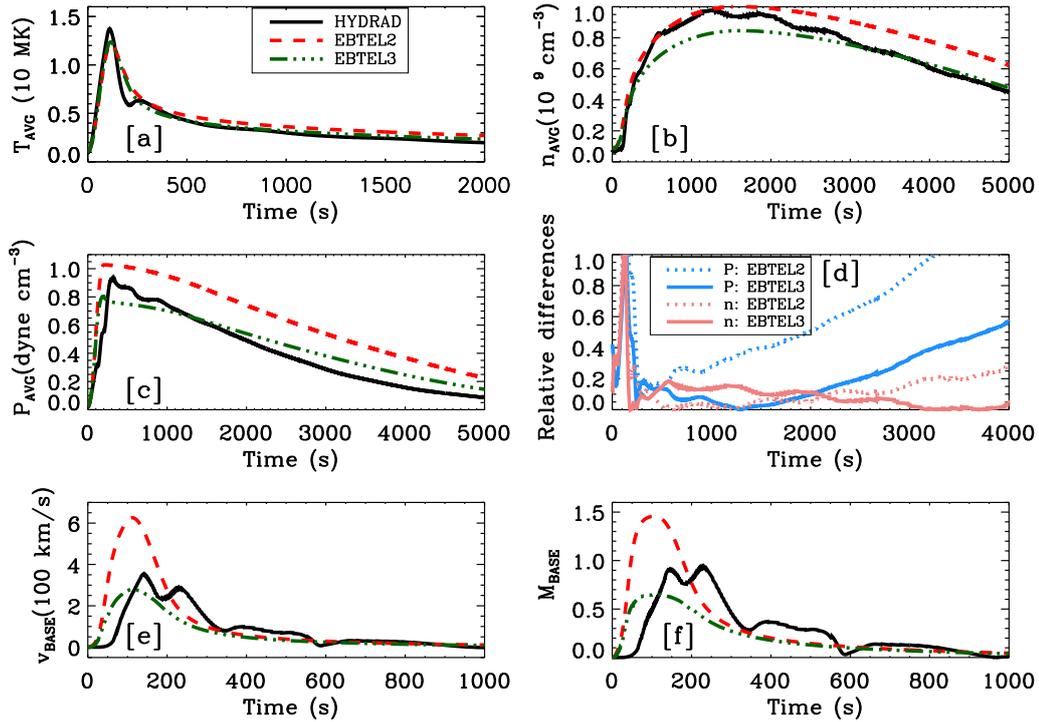

Figure 5.9: As in Figure 5.4 but for Case 6: A $7.5 \times 10^9$ cm loop receiving 1.125 $\times 10^{10}$ ergs cm$^{-2}$ in 200 s.

### 5.3.6 Case 7: Supersonic flows in 6.0 ×10⁹ cm long loop

The seventh case is of a loop of half-length of $6.0 \times 10^9$ cm with background adjusted such that the initial electron number density and temperature are $0.13 \times 10^8$ cm$^{-3}$ and $0.4 \times 10^6$ K, respectively. It receives a symmetric triangular heating profile lasting for 200 s, with the maximum heating rate being 0.001 ergs cm$^{-3}$ s$^{-1}$ at $t = 100$ s. The results are plotted in Figure 5.9.

In this case, maximum Mach numbers reached by HYDRAD exceed unity

(1.15) at the base. The maximum Mach numbers produced by EBTEL2 and
EBTEL3 are 1.51 and 0.65, respectively, and both are unreliable because
0D simulations cannot handle shocks, which are expected to occur in reality.
Despite this, we see that temperature (panel [a]), density (panel [b]), and
pressure (panel [c]) computed from 0D simulations are reliable. As discussed
in Section 5.3.5, HYDRAD computes higher temperatures than EBTEL2 and
EBTEL3 (see panel [a] of Figure 5.10). Panel [e] shows that EBTEL3 un-
derestimates velocities at the coronal base. Even though the peak velocities
computed by HYDRAD match better with EBTEL3 than EBTEL2, the Mach
numbers computed by EBTEL3 are subsonic and hence cannot be trusted.

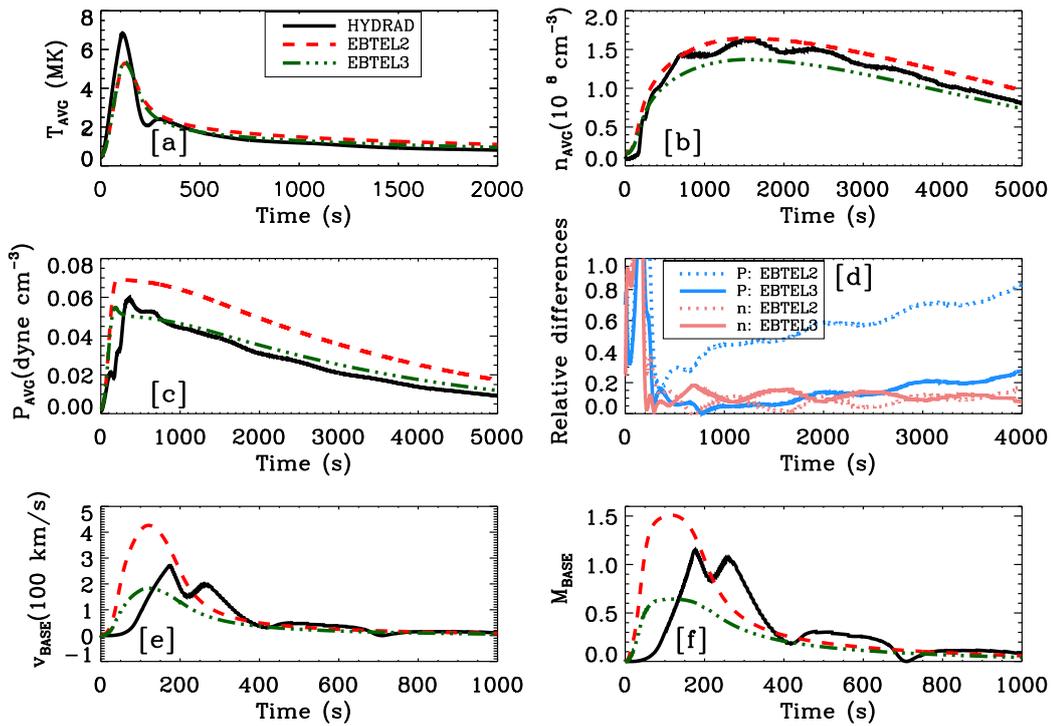

Figure 5.10: As in Figure 5.4 but for Case 7: A 6 ×10⁹ cm loop receiving $6.0 \times 10^8$
ergs cm$^{-2}$ in 200 s.

### 5.3.7 Prediction of onset of shocks

From the simulations performed with EBTEL2, EBTEL3, and HYDRAD for various input parameters, we find reasonable agreement between $\bar{P}, \bar{n}$, and $\bar{T}$ obtained from HYDRAD and EBTEL3. The agreement for $v_0$ and $M_0$ is better when the maximum $M_0$ produced in HYDRAD is subsonic but starts worsening when $M_0$ starts approaching or exceeding unity. Despite this, we find that both EBTEL2 and EBTEL3 generate values of $\bar{P}, \bar{n}$, and $\bar{T}$ within a factor 2× that from HYDRAD. While EBTEL2 provides better $\bar{n}$, EBTEL3 provides better $\bar{P}$. Additionally, EBTEL3 does better for velocities. Note, however, that the match between $M_0$ computed from HYDRAD and EBTEL3 is not acceptable in the last two cases, where the maximum $M_0$ produced in HYDRAD is $\gtrsim 1$, but EBTEL3 produces subsonic flows (see Table. 5.1). This is because such $M_0$ in HYDRAD would lead to shocks, which a simple 0D description of loops cannot model. Consequently, this may lead to the erroneous conclusion that flows in the loop are subsonic. However, we note that even in these cases, 0D is still sufficient to study the electron number density, temperature, and pressure as they are not affected as shown in Figures 5.9 and 5.10. This is consistent with qualitative arguments made in Cargill et al. (2012b).

In light of the above arguments, it becomes necessary to be able to use results from 0D simulations for predicting regimes where the maximum $M_0$ produced in field-aligned simulations is $\gtrsim 1$. In such cases, detailed field-aligned hydrodynamical simulations are better suited to compute reliable Mach numbers.

Based on the EBTEL3 simulations, we find a simple criterion that can be employed to predict this regime. A measurement of the ratio of full width at half maximum (FWHM) of the Mach number profile predicted by EBTEL3 and that of the heating functions may be used as a diagnostic; when this ratio is $\gtrsim 2$ flows in HYDRAD approach or exceed supersonic velocities. Since the heating functions have symmetric triangular profiles, their FWHM is half the total duration. $M_0$ profiles, however, have no concrete shape. Hence we resort to finding it numerically by implementing the definition of FWHM. We tabulate these ratios in Table 5.2 for each case and the maximum Mach number predicted by HYDRAD. A shock should produce a local increase in average temperature in the corona, i.e., a local peak at some time apart from the peak, corresponding to maximum direct heating. The presence of such local peaks in $\bar{T}$ and sonic-supersonic flows at the base in HYDRAD is seen in cases where the ratio of FWHMs of $M_0$ and the heating function is $\gtrsim 2$. In such cases, there is a large departure from $M_0$ predicted using EBTEL3 results. Hence, if this ratio is $\gtrsim 2$, the Mach numbers obtained with EBTEL3 should be treated with caution.

## 5.4 Discussion and Summary

Understanding the complete plasma dynamics in coronal loops is important for understanding the energetics of the corona. Under the assumption of an absence of cross-field conduction, these loops can be modeled reasonably well using field-aligned hydrodynamical simulations, viz., HYDRAD. However, field-aligned simulations are computationally expensive for estimating the evolution of loops in more realistic scenarios where thousands of elemental strands are present, and multiple heating events are involved. Even for cases of a monolithic loop, obtaining quick and approximately accurate estimates of loop evolution over a wide range of parameters, 0D simulations like EBTEL provide a useful alternative.

Table 5.2: Ratio of FWHM of profiles of Mach number at base and heating function

| Case | Maximum value of $M_0$ in HYDRAD | FWHM of $M_0$ profile in EBTEL2 ($r_1$) | FWHM of input heating profile ($r_2$) | $r_1/r_2$ |
|---|---|---|---|---|
| 1 | 0.67 | 165 | 100 | 1.65 |
| 2 | 0.57 | 396 | 250 | 1.58 |
| 3 | 0.49 | 155 | 100 | 1.55 |
| 4 | 0.55 | 120 | 100 | 1.20 |
| 5 | 0.75 | 193 | 100 | 1.93 |
| 6 | 0.94 | 227 | 100 | 2.27 |
| 7 | 1.15 | 259 | 100 | 2.59 |

Cargill et al. (2012a) and Klimchuk et al. (2008) had assumed subsonic flow exists at all stages, neglecting the kinetic energy term in the energy equation. While this assumption holds good during most of the evolution of the loop, it fails during the impulsive phase of some of the heating events. In this paper, we have relaxed the assumption of subsonic flows by not neglecting the kinetic energy term in the energy equation.

To solve the equations, we have made the following two assumptions:

$$(i) \ \left[\frac{P_0}{\bar{P}}\right] = \left[\frac{P_0}{\bar{P}}\right]_{hse} \left[\exp\left(\frac{3}{2}M_0^2\right)\right] \implies \left[\frac{n_0}{\bar{n}}\right] = \left[\frac{n_0}{\bar{n}}\right]_{hse} \left[\exp\left(\frac{3}{2}M_0^2\right)\right]$$

$$(ii) \ \int_0^L v\frac{\partial P}{\partial s}ds \approx -\frac{1}{2}n_0\mu v_0^3$$

The results obtained concerning the plasma dynamics in different kinds of loops by EBTEL3 show significant improvements in average pressure in the corona and predicted velocities at the coronal base, compared with the results obtained with field-aligned simulations using HYDRAD. Though the electron number density produced by EBTEL2l are less accurate than those produced by EBTEL2, the discrepancy remains less than 20%. The improvement in pressure estimates by EBTEL2 is larger than the deterioration in electron number density estimates. However, the main improvement of EBTEL3 over EBTEL2 is that of velocities, which match better with HYDRAD results. Additionally, EBTEL3 guarantees $M_0$ to remain subsonic if $M_0$ produced by HYDRAD remain subsonic.

Furthermore, we have developed a simple heuristic to check whether field-aligned simulations produce subsonic flows without performing field-aligned simulations. This is useful in deciding if the Mach numbers computed by 0D simulations can be trusted because 0D simulations are not designed to tackle supersonic plasma flows, which lead to complicated situations like shocks. For this, we look at the ratio of FWHM of profiles of $M_0$ and heating profiles.

If the ratio is larger than 2, the maximum Mach numbers produced at the base of the corona are close to or even larger than 1. Nevertheless, even in the cases where HYDRAD predicts trans-sonic and supersonic flows, and the Mach numbers derived by EBTEL2 and EBTEL3 cannot be trusted, we find the coronal averages ($\bar{T}, \bar{P}$ and $\bar{n}$) calculated by EBTEL3 to be in good agreement with HYDRAD.

It is important to note, however, that while EBTEL may be suitable for getting rough estimates of the evolution of coronal loops, a major drawback is that any spatial variation in quantities across the loop cannot be investigated. Hence in situations where spatial variation across loops needs to be studied, 1D simulations are more suited. The major advantage of EBTEL is speed. While using 1D codes with an adaptive grid can take a few minutes, 0D codes study the same in less than a second. Hence, in situations where approximate and quick physics-based responses to an ensemble of loops to different energetic events are needed, EBTEL is a viable alternative.

Adding kinetic energy to the EBTEL framework makes it more suitable for tackling impulsive heating scenarios where kinetic energy cannot be ignored. The more impulsive an event, the larger are temperatures reached by the loop. While AIA observations can be used to constrain the thermal evolution of plasma components at temperatures less than 5 MK, X-ray measurements by FOXSI can be used to constrain the thermal evolution of plasma at temperatures larger than 5 MK. Consequently, combined observations from AIA and FOXSI can be used for observing plasma at a wider range of temperatures. These observations can be modeled better using EBTEL3. In the next chapter, we study events observed in EUV and X-rays using multi-stranded simulations performed using EBTEL3.

# 6

# Simulations of AR studied by FOXSI and AIA: single power-law distribution of events

*Solar coronal heating, which maintains the corona at $\gtrsim 10^6$ K, can be attributed to uniform steady background heating and transients like impulsive flaring events. Here we explore the possibility that the steady heating is attributable to a collection of small impulsive events with a single power-law distribution of different flare energies. We perform 0D hydrodynamical simulations of a multi-stranded system of loops, mimicking EUV data obtained from Atmospheric Imaging Assembly (AIA) and X-ray data Focusing Optics X-ray Solar Imager (FOXSI-2) for an isolated loop complex. We parameterize the system with the slope of power-law distribution, minimum and maximum energy dissipated in a single event, constrained by total energy from AIA and FOXSI-2 and strand radius from Hi-C. Preliminary results indicate that the observed light curves can be best explained by a power-law with a negative slope ($-\alpha$) in the range $1.6 \leq -\alpha \leq 1.8$ and maximum and minimum energies differing by more than 7–8 orders of magnitudes. We discuss the implications of these results and possible extensions.*

## 6.1 Introduction

Ever since Hudson (1991) suggested that heating events of different energy follow a power-law distribution, numerous attempts have been made to quantify this power-law distribution. See Section 1.7 for a detailed discussion on the various estimates of slopes. Power-law distribution of different energetic events may indicate self-organized criticality (Lu and Hamilton, 1991). This would imply that events with different energy follow the same physics.

In this work, we consider the possibility of simulating transient events observed by FOXSI-2 and AIA by a single power-law distribution of heating events with different energies. Hydrodynamical simulations of coronal loops generally impose a uniform background heating for the corona to always be at temperatures exceeding $0.5 \times 10^6$ K (Priest, 2014; Klimchuk, 2006). Here, we do not impose any background on our hydrodynamical simulations and generate different energetic events from the same power-law distribution.

The remaining chapter is structured as follows. Section 6.2 discusses the FOXSI-2 (X-ray) and AIA/SDO (EUV) observations of the event, which we have modeled. Section 6.3 provides details of our simulation setup. Section 6.4 describes the analysis performed on simulated data. We present results and future scope in Section 6.5.

## 6.2 Observations and Data

### 6.2.1 Observations from FOXSI-2

FOXSI-2 provided hard X-ray (HXR) observations of two sub-A class microflares. The flight was designed to coordinate with XRT/Hinode and AIA/SDO.

This gave spatially and temporally coaligned observations of the region of interest in soft X-rays and EUV. These observations were used for constraining DEM for these microflares. Athiray et al. (2020) found the DEM of these microflares to be peaking at $3 \times 10^6$ K. The temperature of multi-thermal plasma observed extends beyond $10^7$ K. Since emission measure at temperatures higher than $5 \times 10^6$ K was less than $10^{26}$ cm$^{-5}$, HXR measurements are extremely helpful in understanding such components of plasma. Using these estimates of DEM, the total radiative energy for the microflares in AR12230 and AR12234 were $5.0 \times 10^{28}$ and $1.6 \times 10^{28}$ ergs, respectively. Figure 6.1 (left panel) shows the X-ray image of AR 12230 and AR 12234 taken from FOXSI-2. These images have been made using photons in the energy band of 5–8 keV. The middle and right panels show the EUV and soft X-ray images taken using AIA/SDO and XRT/Hinode, respectively.

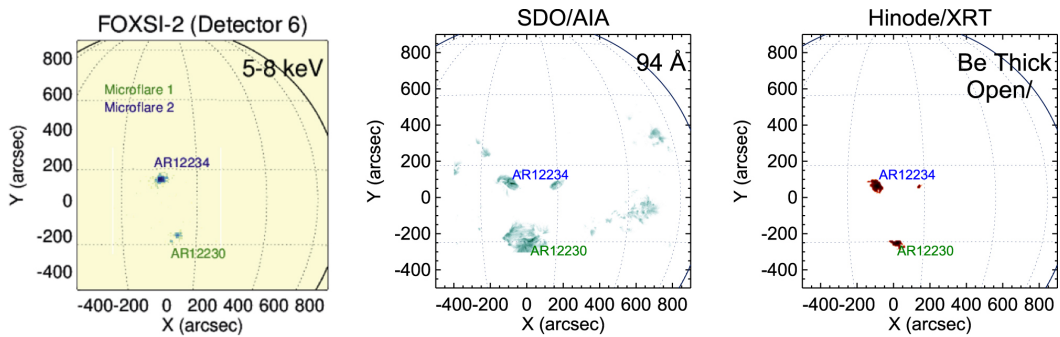

Figure 6.1: A coordinated image of microflares producing active region using FOXSI-2 (left), AIA-94 (middle), and XRT (right). Image courtesy: Athiray et al. (2020)

The AR12230 was observed in its flaring phase by all three instruments. Therefore, it provides an excellent opportunity to study the physics of the multi-thermal nature of plasma in microflares. Consequently, in the remaining parts of this chapter, we will only discuss the microflare observed in AR12230.

FOXSI-2 had six targets, each of which observed this active region at a different time. The left panel of Figure 6.2 shows the HXR image recorded by Target C of FOXSI-2. The combined observations from different targets give the count rates in different energy bands. The right panel of Figure 6.2 shows the count rates in the energy range of 5–8 keV.

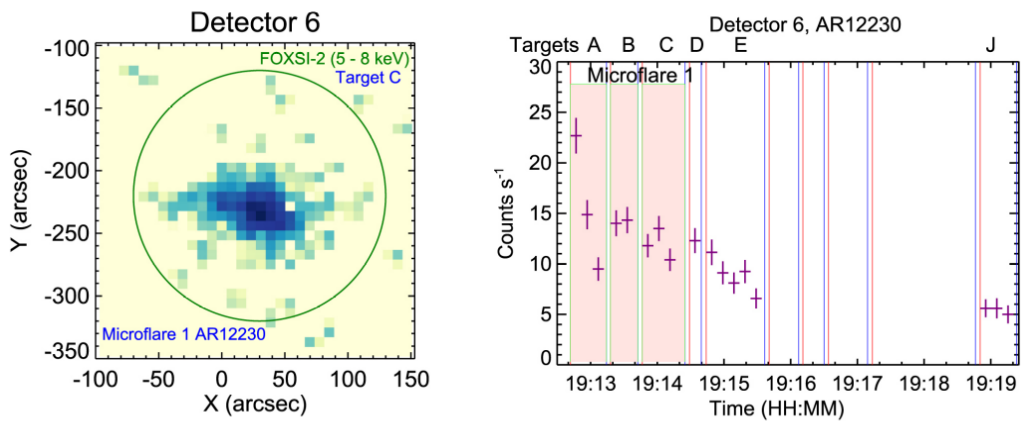

Figure 6.2: [Left] Hard X-ray image of the microflare observed in AR12230 made using Target C of FOXSI-2. [Right] Observed count rates in the 5–8 keV band for the same event. Image courtesy: Athiray et al. (2020)

## 6.2.2 Observations from AIA/SDO

Figure 6.3 shows EUV images of the active region AR12230 in six filters (94, 131, 171, 193, 211, and 335 Å) around 7 minutes before FOXSI-2 started taking observations. The images demonstrate that a lot of emissions come from outside the isolated loop complex that we are interested in modeling. Consequently, we need to select contributions from those pixels that show statistically significant enhancement in intensities. For this purpose, we have used the Automark package developed by Wong et al. (2016); Xu et al. (2021). We provide a series of images in six EUV filters as input to the Automark algorithm, which detects abrupt changes in the emission distribution over the filters centered on different wavelengths. Figure 6.4 shows the average intensity from pixels, which detect the event in 5 $\sigma$ confidence intervals. The section of the light curves between the two solid vertical lines is the time window during which FOXSI-2 provided observation for 100 s. The AIA light curves within this window are what we aim to match with synthetic light curves.

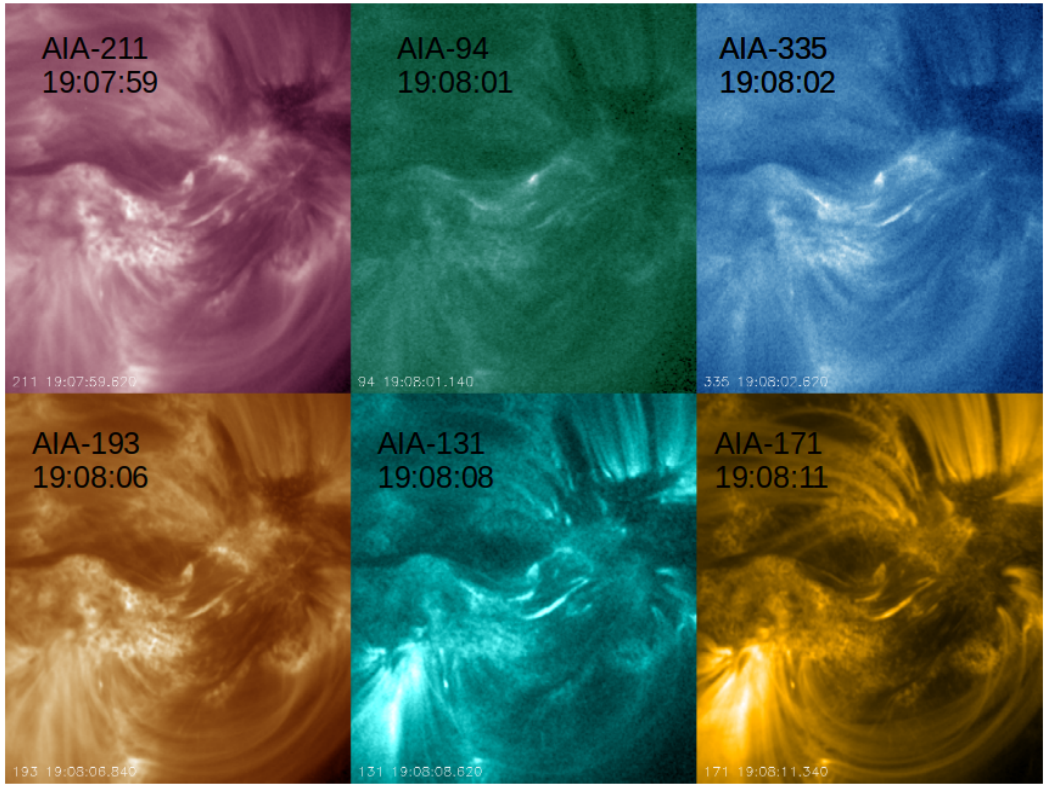

Figure 6.3: . Image courtesy: AIA images of the region of interest in 6 EUV channels (94, 131, 171, 193, 211, and 335 Å) were taken around 19:05, i.e., roughly 7 minutes before FOXSI-2 started taking observations.

## 6.3   Simulation setup

This section details the multi-stranded 0D hydrodynamical simulations performed using the EBTEL code. We provide details of the geometry and the number of strands that compose the loop in Section 6.3.1. The generation of the heating function is discussed in Section 6.3.2. The details of EBTEL simulations are discussed in section  6.3.3

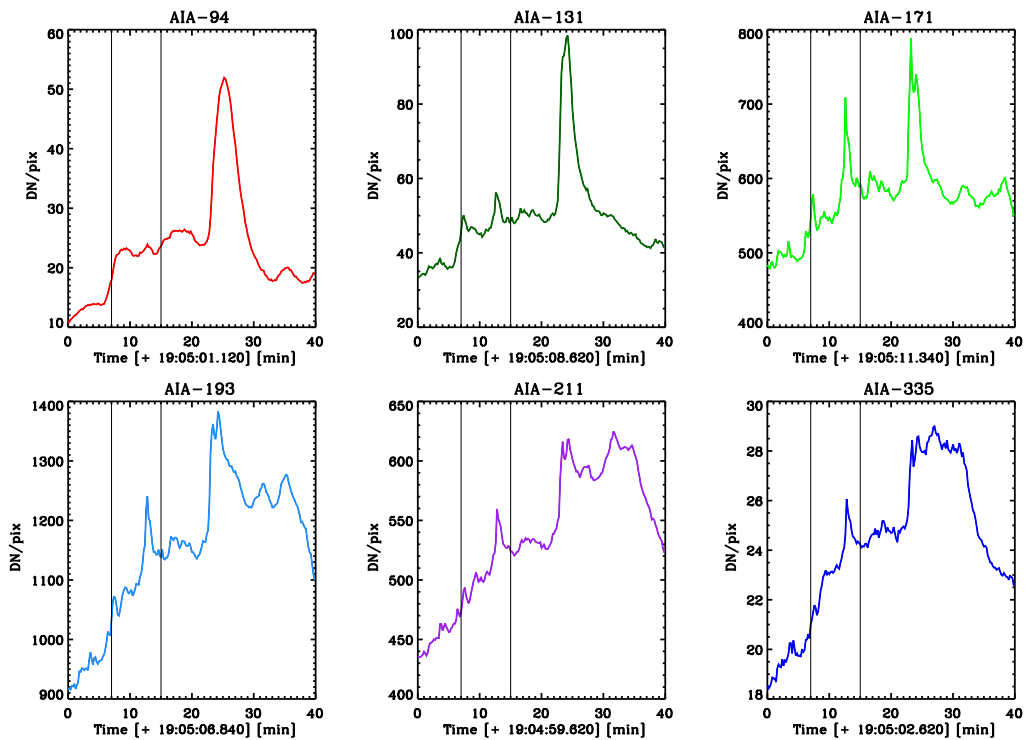

Figure 6.4: The light curves in six EUV filters (94, 131, 171, 193, 211, and 335 Å) averaged over pixels which detected the event within 5 $\sigma$ confidence interval. The vertical black lines show the time period during which FOXSI-2 provided observations.

## 6.3.1 Number and geometry of strands

We consider AIA-94 Å images to approximate the distance between the footprints. It turns out to be 35 arcseconds correspondingly. Under the assumption of a semicircular loop between these foot points, the half-length of the loop turns out to be 1.5 $^9$ cm. Changing the loop length by a factor of 2 or $\frac{1}{2}$ did not change the results significantly. Additionally, since our main aim is to study the distribution of heating events, we have fixed the half length of the loop to be $10^9$ cm. All strands and the composite loop are assumed to be semicircular, and their lengths are equal for simplicity.

Williams et al. (2020) studied the width of 49 coronal "strands" observed by Hi-C. The radius of these strands varied between 100–500 km. Consequently, the corresponding footprint area ($\pi R_s^2$) of these strands turns out to be 0.46–1.15 AIA pixels. This corresponds to an area of $3.7 \times 10^{14}$ cm$^2$ – $2.3 \times 10^{15}$ cm$^2$.

We consider the limiting case where the semicircular loop projects a length equal to the diameter of the loop $2R_L$ on the plane containing footpoints. Hence the projected area on the plane becomes $A = 4\,R_L^2$. The maximum number of non-overlapping strands is given by

$$N = \frac{\pi R_L^2}{\pi R_s^2} = \frac{\pi A}{4\pi R_s^2} \qquad (6.1)$$

The number of AIA pixels in which the event is detected over the $5\sigma$ confidence interval is $\sim 500$ pixels. Hence for $A = 500$ pixels and $\pi R_s^2$ varying from 0.46–1.15 pixels, the number of strands varies from 400–800 strands. However, note that several approximations are made while arriving at these estimates, which might differ from the real loop geometry. Hence, for simplicity, we have taken the number of strands to 1000.

### 6.3.2 Generation of heating function

Athiray et al. (2020) estimated radiative energy to be of the order of $10^{28}$ ergs over the period of 100 s for which FOXSI-2 provided observations. We aim to generate different representations of the event such that in each of them, $10^{28}$ ergs is the energy budget over a duration of 100 seconds. Hence, we consider total energy dissipation ($E_{Bgt} =$)$10^{30}$ ergs over $10^4$ s in the multistranded loop. We assume that all the heating events have a symmetric triangular time profile that lasts for $t_{dur} = 100$ s. The time profile of volumetric heating rate ($H(t)$) in ergs cm$^{-3}$ s$^{-1}$ can be written for each event as

$$H(t) = \frac{H_p}{2} \left( \frac{t - t_i}{t_p - t_i} \left[ \theta(t - t_i) - \theta(t - t_p) \right] + \frac{t_f - t}{t_f - t_p} \left[ \theta(t_f - t) - \theta(t_p - t) \right] \right)$$

$$(6.2)$$

where, $H_p$ is the peak volumetric heating rate in ergs cm$^{-3}$ s$^{-1}$, and $t$ denotes the time. $t_i, t_p$, and $t_f$ are the starting, peaking, and ending time, respectively, for the event and $t_f - t_p = t_p - t_i = 50$ s. $\theta(x)$ is the sign function. It is +1, -1, or 0 for $x$ being positive, negative, or zero.

In this work, we are interested in generating all heating events from the same power-law distribution without imposing an arbitrary external background heating. For a given $E_{Bgt}$ and geometry of strands, the heating function can be specified using three parameters: (i) the maximum energy which can be dissipated in a single event ($E_{max}$) (ii) the minimum energy which can be dissipated in a single event ($E_{min}$) and (iii) slope of power-law distribution of heating events ($\alpha$). The number of heating events ($N$) required for the case of $[E_{min}, E_{max}, \alpha]$ can be determined using the relations

$$E_{Bgt} = \int_{E_{min}}^{E_{max}} AE^{1+\alpha} dE$$

$$(6.3)$$

and

$$N = \int_{E_{min}}^{E_{max}} AE^{\alpha} dE$$

$$(6.4)$$

where A is the normalization constant, and $N$ has been rounded.

We generate $N$ random numbers uniformly distributed between 0 and $10^7$. These correspond to the time at which any event peaks. We then generate the N random numbers following a power-law distribution of slope $\alpha$. These are the energy dissipated in each of the events. We can write the peak volumetric heating rate ($H_p$) for an event with energy ($E$) as

$$H_p = \frac{E}{t_{dur} L \pi R_s^2} \tag{6.5}$$

Performing this for all the events occurring at a random time gives a time series of the heating function. Since we aim to perform multistranded simulations, we dissipate $10^{30}$ ergs over $10^7$ seconds and then divide the time series into 1000 parts. Each strand is then allotted a time series of $10^4$ second. This way, we maintain randomness in the total energy allotted to each strand. We need to ensure some initial heating for having coronal density and temperature as initial conditions in simulations. To address this issue, we have introduced two small events of energy $10^{21}$ ergs in the first 100 seconds. It forms a negligible contribution to the total energy budget.

### 6.3.3 EBTEL simulations

The heating function obtained above is provided as input along with the semi-loop length for performing 0D simulations using EBTEL. Each strand evolves independently in response to its train of heating events. We only consider the parts of simulations where a clear background corona has been established to remove any dependence on initial conditions. In all our cases, this happens within the first 1000 seconds. Consequently, we only study the next 8000 seconds of the evolution of all strands. We used the EBTEL3 code (discussed in Chapter 5) for this work. It includes kinetic energy terms in the energy conservation equation and has an adaptive time grid. It returns the instantaneous DEM in the corona and transition region of all the strands in temperature bins of $\delta \log(T[K]) = 0.01$. The total contribution (corona+TR) from all strands is added using the additivity of DEMs. The total DEM is used for generating synthetic time series of EUV intensities in AIA and X-ray count rates in 5-8 keV energy bands of FOXSI-2.

## 6.4 Analysis

This section describes the range of input parameters used in our studies. These are $E_{min}$, $E_{max}$, $\alpha$, and $R_s$. We have fixed $E_{min}$ in the following manner. We assume that the minimum average temperature of the corona is $10^{5.5}$ K. The $H_P$, which can be associated with an average coronal temperature ($\bar{T}$) of around $10^{5.5}$ is given by

$$H_P = \frac{2}{7}\kappa_0 \frac{T_A^{\frac{7}{2}}}{L^2} \tag{6.6}$$

where $T_A = \frac{1}{0.9}\bar{T}$ (Klimchuk et al., 2008). Using $\bar{T} = 10^{5.5}$, we get $H_P = 10^{-5}$. For a strand with radius of $10^7$ cm and semi-length of $10^9$ cm, $E_{min}$ becomes $10^{20}$ ergs. We experimented with $E_{min}$ of $10^{19}$ and $10^{21}$ ergs. However, the difference was negligible; hence, we have fixed $E_{min}$ to be $10^{20}$ ergs in all the simulations. This matches the energy estimates of the smallest nanoflares in active region provided by Katsukawa and Tsuneta (2001).

We have performed multistrand simulations for 10,000 seconds for all possible combinations of $E_{max}$, $\alpha$, and $R_S$, where $E_{max}$ varied between $10^{24}$–$10^{25}$ ergs ($\delta log[E_{max}(ergs)] = 10$), -$\alpha$ ranged between 1.6–2.2 ($\delta\alpha = 0.1$), and $R_S$ varied between $1 \times 10^7$–$5 \times 10^7$ cm. Table 6.1 shows the list of all the variable and fixed input parameters explored in this work.

EBTEL simulates all the synthetic emissions to be coming from one pixel. Additionally, we need to take into account the filling factor as well. To compare these simulated results with observations, we need to divide the observed AIA/SDO intensities by a factor that considers the filling factor and the number of pixels contributing to the emission. We accomplish this task in the following way.

Table 6.1: List of input parameters

| Parameter | Value(s) |
|-----------|----------|
| Radius of strand ($R_s$) | [1, 2, 3, 4, 5] $\times 10^7$ cm |
| Maximum energy in single event($E_{max}$) | [$10^{24}, 10^{25}, 10^{26}, 10^{27}, 10^{28}, 10^{29}$] ergs |
| Slope of power-law distribution($\alpha$) | -[1.6, 1.7, 1.8, 1.9, 2.0, 2.1, 2.2] |
| Energy budget ($E_{Bgt}$ over $10^4$ s) | $10^{30}$ ergs |
| Maximum energy in single event($E_{min}$) | $10^{21}$ ergs |
| Duration of single event ($t_{dur}$) | 100 s |
| Number of strands (N) | 1000 |

We first divide the synthetic lightcurves generated for 9000 s into 21 parts, each of 420 s (equivalent to the observation time of FOXSI-2). For each segment, we compute a multiplicative factor $J_{p,c}$ for each of these 21 representations (indexed by $p$) and a particular case, i.e., combination of $E_{max}, \alpha$, and $R_s$ (indexed by $c$) by using the observed average intensity in 94 Å filter $\bar{I}_{obs}$ and the average synthetic intensity in 94 Å filter in the $p^{th}$ representation ($\bar{I}_{sim:p,c}$). $J_p$ is given by

$$J_p = \frac{\bar{I}_{obs}}{\bar{I}_{sim:p,c}} \quad (6.7)$$

This multiplicative factor is used for all the AIA filters and FOXSI-2 energy bins.

## 6.5 Results and Outlook

We describe the results for some of the cases explored. We take the case of $Rs$ = 3$\times 10^7$ cm (modal value of $R_s$ in Williams et al. (2020)) and log[$E_{max}$(ergs)] = 29 and discuss the results for four values of $\alpha$ viz, -1.6, -1.8, -2.0 and -2.2.

### 6.5.1  Case 1: $\alpha = -1.6$

For the slope of $-1.6$, we have the highest contribution from large energy events as compared to the case for any other slope. The modal value of observed and simulated lightcurves in the AIA-94 Å filter is the same because of being used for adjusting the filling factor. The modal value of observed and simulated light curves in other AIA filters are slightly shifted with respect to each other. The shift is more prominent in FOXSI-2 energy bands. Simulated curves tend to give higher FOXSI-2 counts than observations. Figure 6.5 shows the histograms of observed (blue) and simulated (red) time series of AIA/SDO intensities and FOXSI-2 energy bands. The histograms have been normalized with a peak value of unity.

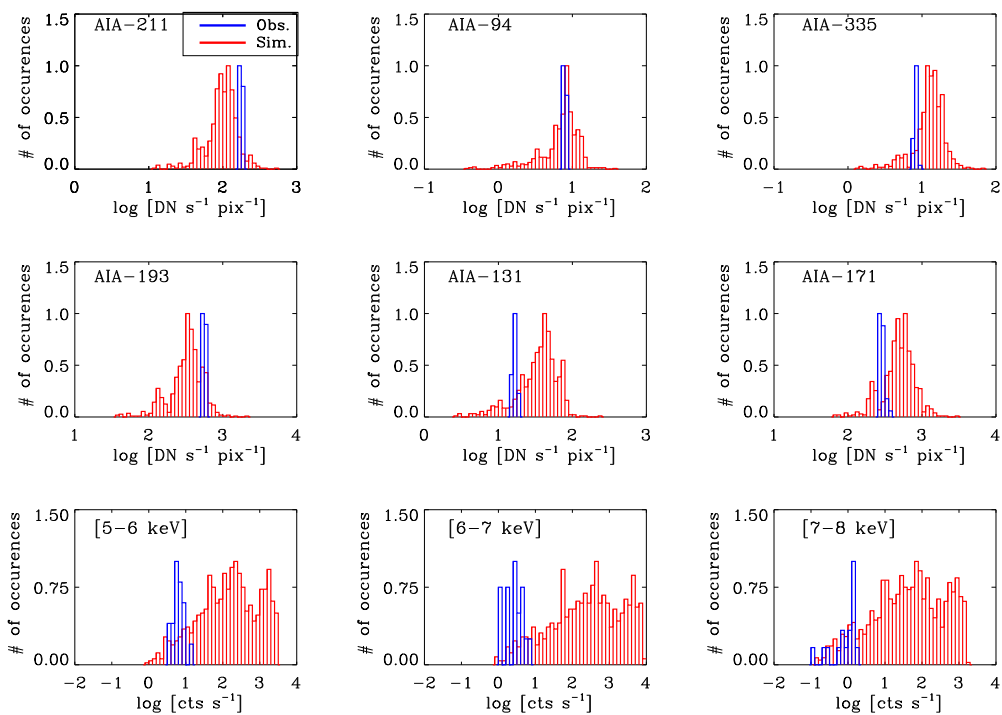

Figure 6.5: Histograms of observed and simulated lightcurves of AIA and FOXSI-2 for $E_{max} = 10^{29}$ ergs, $R_s = 3 \times 10^7$ cm and $\alpha = -1.6$. Note the area under observed and simulated light curves are different because of different durations.

## 6.5.2   Case 2: $\alpha = -1.8$

Figure 6.6 shows the results for the same parameters in Figure 6.5 except for a slope of -1.8. The difference between modal values of observed and simulated FOXSI-2 counts has been further reduced. However, the corresponding shift in AIA-171 has started worsening. This is because the larger the slope, the more events have lower energy and hence lower temperature. This results in higher intensities in synthetic lightcurves in AIA-171 Å.

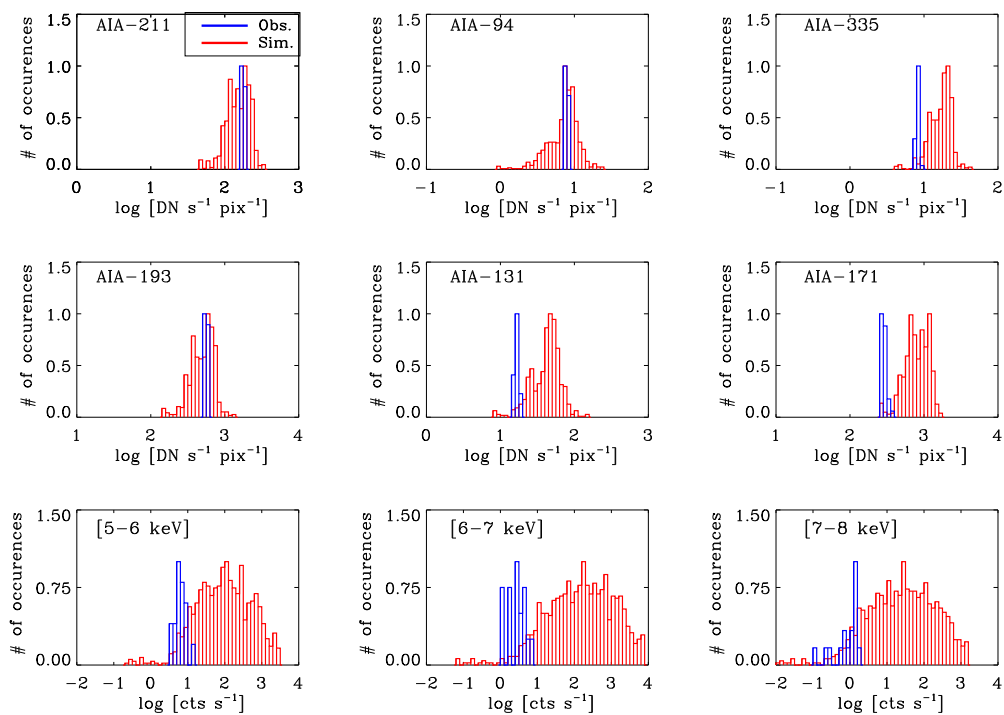

Figure 6.6: Histograms of observed and simulated lightcurves of AIA and FOXSI-2 for $E_{max} = 10^{29}$ ergs, $R_s = 3 \times 10^7$ cm and $\alpha$ = -1.8.

## 6.5.3   Case 3: $\alpha = -2.0$

Figure 6.7 shows the results for the same parameters in Figure 6.5 but for a slope of -2.0. As the match between observed and simulated modal values of FOXSI-2 counts has significantly improved, the corresponding match in all the AIA filters has deteriorated.

### 6.5.4  Case 4: $\alpha = -2.2$

Figure 6.7 shows the results for the same parameters in Figure 6.5 but for a slope of -2.2. Now the match between modal values of simulated and observed FOXSI-2 counts has also started deteriorating along with the corresponding match for the AIA intensities.

We see that slopes (-$\alpha$) $\leq 2.0$ match better with observations of the event observed in AR12230 by FOXSI-2 and AIA. We have performed the simulations for all the cases.

The next step is to perform statistical analysis to rank these cases in terms of the match with observations from FOXSI-2 and AIA. We also aim to validate these statistical tests using machine learning methods.

So far, we have performed field-aligned simulations for either monolithic or multistranded coronal loops. These simulations result in field-aligned flows leading to chromospheric evaporation and subsequent condensation. Such flows can be observed as Doppler shifts in different emission lines and should show center to limb variation. Additionally, they must vanish on the limb. Redshifts dominate the Doppler measurements in the transition region. Measurement of Doppler shifts in the transition region lines like Si-IV show significant center to limb variation, but there are sizeable flows at the limbs. Furthermore, field-aligned simulations predict Doppler shifts in Si-IV lines to be an order of magnitude smaller than observations. The next chapter discusses this problem and its relation to transition region heating.

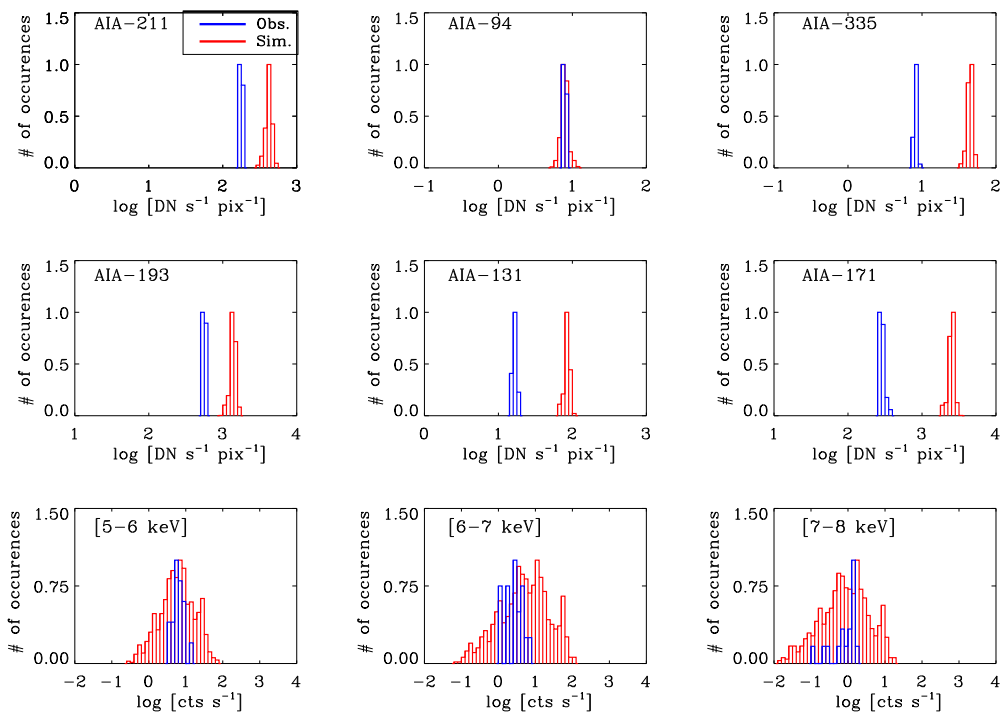

Figure 6.7: Histograms of observed and simulated lightcurves of AIA and FOXSI-2 for $E_{max} = 10^{29}$ ergs, $R_s = 3 \times 10^7$ cm and $\alpha$ = -2.0.

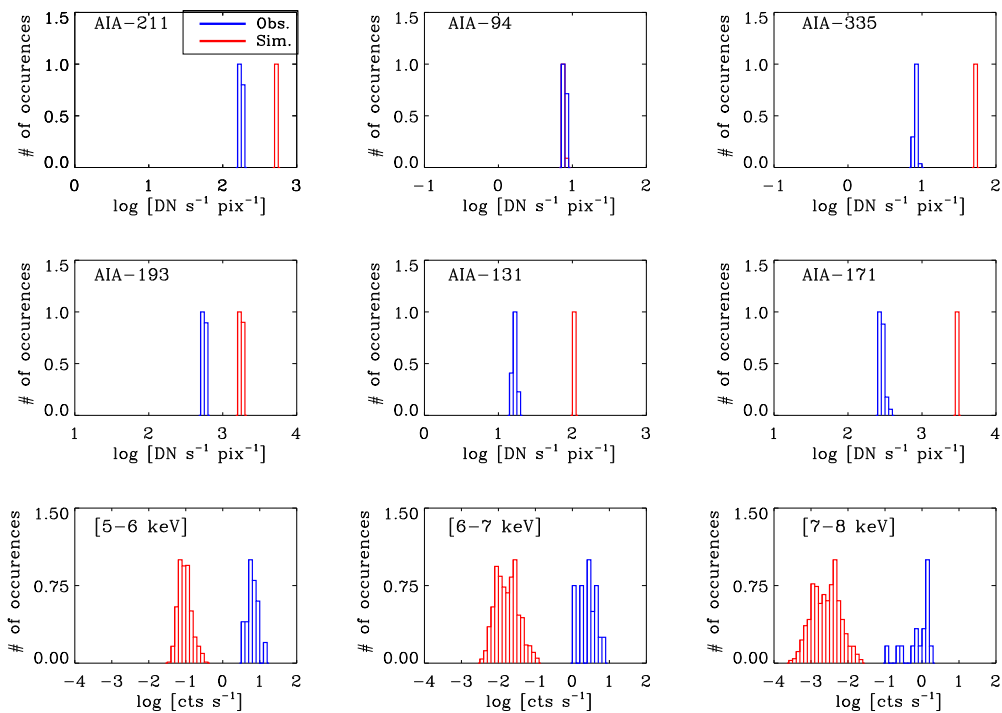

Figure 6.8: Histograms of observed and simulated lightcurves of AIA and FOXSI-2 for $E_{max} = 10^{29}$ ergs, $R_s = 3 \times 10^7$ cm and $\alpha$ = -2.2.

# 7

# Center to limb variation of Doppler shifts in active regions of solar transition region

*Redshifts dominate the Doppler measurements in the transition region. It has been proposed that these downflows indicate the cooling and draining of plasma in coronal loops following chromospheric evaporation due to impulsive coronal heating. However, such flows would show center-to-limb variation (CLV) of measured Doppler shifts that vanishes entirely at the limb. Ghosh et al. (2019) tracked an active region and measured the Doppler shift as it evolved and showed that although there were hints of CLV, it did not become zero at the limb. Moreover, the measured flow speed was an order of magnitude larger than predicted by field-aligned simulations. They suggested that these flows are primarily due to type II spicules and proposed the idea of a chromospheric wall associated with classical type-I spicules that diminish the CLV of the Doppler shifts. However, this study lacked the coverage of longitude range. Here we study the CLV of Doppler shifts in Si IV lines for 50 active regions covering almost all the longitude range on the solar disk. We find that, though there is a CLV in the data, there are sizable flows at the limb, and the measurements have substantial scatter. This study supports the idea of flows related to type II spicules in the presence of a chromospheric wall.*

## 7.1 Introduction

The heating of the solar atmosphere continues to be an extraordinarily challenging problem. Though magnetic fields are known to be responsible, the exact mechanism for energy dissipation and the transport of mass and energy across different layers of the atmosphere is elusive. It is believed that any mechanism that can explain the solar atmosphere's heating should be impulsive (see for review Klimchuk, 2006). Impulsive heating results in the evaporation of chromospheric plasma along the loops into the corona, followed by condensation. Hence, studying flows in different layers of the solar atmosphere sheds valuable insights on the heating and possible ways the solar atmosphere may be coupled.

Observations show that the transition region has ubiquitous presence of redshifts (downflows) in the active regions as well as quiet Sun and coronal holes. Early observations from Orbiting Solar Observatory (OSO-8; Bruner, 1977a), the Naval Research Laboratory (NRL) normal incidence spectrograph onboard Skylab (S082-B), NRL High-Resolution Telescope and Spectrograph (HRTS; Bartoe and Brueckner, 1975), and the Ultra-Violet Spectrometer and Polarimeter (UVSP; Woodgate et al., 1980) onboard the Solar Maximum Mission (SMM; Simnett, 1981) show downflows in the range of 5–20 km s$^{-1}$ in ultraviolet spectral emission lines from bright regions in the chromosphere and the transition region (also see Lemaire et al., 1978; Brueckner et al., 1980; Gebbie et al., 1980; Lites, 1980; Brueckner, 1981; Athay et al., 1982, 1983; Dere, 1982; Rottman et al., 1982; Brekke, 1993; Achour et al., 1995). Moreover, transition region downflows in the range of 80–100 km s$^{-1}$ have also been reported in small regions within active regions. However, due to their rare occurrence these are considered to be associated with transients (Nicolas et al., 1982; Dere et al., 1984).

Studies with similar scientific goals have also been performed using observations from Solar Ultraviolet Measurements of Emitted Radiation (SUMER; Wilhelm et al., 1995), the Coronal Diagnostic Spectrometer (CDS; Harrison et al., 1995) onboard SOlar and Heliospheric Observatory (SOHO; Domingo et al., 1995a), EUV Imaging Spectrometer (EIS; Culhane et al., 2007) onboard Hinode (Kosugi et al., 2007b), and Interface Region Imaging Spectrograph (IRIS; De Pontieu et al., 2014a). Observations from SUMER show downflows in the active regions to be ranging from $\sim 0$ km s$^{-1}$ at log[T(K)] = 4.3 to $\sim 15$ km s$^{-1}$ at log[T(K)] = 5.0. At log[T(K)] = 5.8 blueshifts $\sim 8$ km s$^{-1}$ are observed. Further studies on plasma flows were conducted using observations from EIS (see, e.g., Del Zanna, 2008; Brooks and Warren, 2009; Tripathi et al., 2009, 2012; Dadashi et al., 2011; Gupta et al., 2015; Ghosh et al., 2017) in warm loops as well as moss regions (transition region counterpart of hot loops). Persistent downflows were reported across the range of temperature EIS observed, i.e., log, T=5.8 to 6.3. However, a limitation of these studies was using spectral lines formed in the lower transition region (O IV, O V, and Mg V) for wavelength calibration because these lines were very weak in the observations (Young et al., 2007) and, moreover, did not provide absolute calibration.

At first glance, it was thought that these downflows were due to impulsive heating occurring in the solar corona (see for review Klimchuk, 2006; Reale, 2014). Under this scenario, the redshift is due to field-aligned downflows of condensing plasma that was pushed up in the coronal loops due to chromospheric evaporation. However, such flows should show center-to-limb variation (CLV) and vanish as one approaches the limb. Feldman et al. (1982) used data from NRL onboard Skylab (S082-B) for tracking two active re-gions as they traversed across the solar disk for studying the Doppler shifts in the temperature range log T = 4.7–5.0. They found the downflows to be in the range of 4–17 km s$^{-1}$. However, there was almost no CLV observed. Moreover, the redshifts extended out to the limb. Klimchuk (1987, 1989) used UVSP data and found similar results in measuring Doppler shifts relative to the average over the full raster.

The IRIS instrument provides regular spectroscopic observations of transition region in the Si IV line, with an accuracy of about 1 km s$^{-1}$. Moreover, the presence of multiple spectral lines due to neutral and single-ionized ions provides the best opportunity to measure and characterize flows in the transition region. Ghosh et al. (2019); Ghosh et al. (2021) studied the Doppler shift and non-thermal velocities in the Si IV line and their CLV for a single active region as it traversed the central meridian. They found that flows in active regions were redshifted by 5–10 km s$^{-1}$ with moderate CLV.

The observation of persistent downflows was explained by Antiochos (1984) as a signature of field-aligned flows due to condensation. Moreover, to explain the absence of CLV and non-diminishing flows at the limb, Antiochos (1984) introduced the idea of a chromospheric well, which is formed due to the enhanced localized pressure created by impulsive events. Under this scenario, the absence of CLV naturally arises due to projection effects. However, there are several drawbacks to this scenario. Under the scenario of impulsive heating, field-aligned hydrodynamical simulations show downflows with much lower amplitude than those observed at similar temperatures. For example, the downflows in the Fe VIII line formed at an approximate temperature of 0.4 MK is ∼0.9 km s$^{-1}$ (see, e.g., López Fuentes and Klimchuk, 2018). Considering the constant mass flux along a given flux tube, the magnitude of the downflows in Si IV, the speed of downflows should be less than 0.1 km s$^{-1}$, which is about two orders of magnitude that observed speeds in the lower transition regions. To mitigate the above discrepancy, Ghosh et al. (2019) suggested the downflows observed in the transition regions are very likely related to the downflow of type-II spicules and proposed the idea of a *chromospheric wall* formed by cold spicules heated to a temperature of about $10^4$ K in the vicinity of hot spicules, which get heated to $10^5$ K. They argued that the optical depth of surrounding cold spicules is close to but less than unity, hence, allowing some center to limb variation in Si IV line.

We note that Ghosh et al. (2019) performed the Doppler measurements for a single active region while it crossed the central meridian. Therefore, the longitude coverage is limited. In this work, we extend the analysis of Ghosh et al. (2019) to 50 active regions observed at different times and locations of the disk. This provides a statistically large sample and better coverage of the longitude. The rest of the chapter is structured in the following manner. In §7.2 we describe the data from different instruments used in this study. In §7.3 we describe the various procedures involved in analyzing data from different instruments, *viz* (i) wavelength calibration, (ii) coalignment of data from AIA-1600, HMI, and IRIS, (iii) identification of strong-field regions within the active regions, and (iv) computation of Doppler shifts in these regions and associated radius vector. We discuss the results for all active regions and their CLV in §7.4. In §7.5 we summarize and conclude.

## 7.2   Observations and Data

To study the Doppler shifts, we have used IRIS observations. IRIS provides spectra and images with spatial resolutions varying between 0.33" and 0.4" and a cadence of up to 20 s (spectral cadence) and 10 s (image cadence). The field of view (FOV) can extend to 175"×175". The spectra obtained allow us to resolve velocities upto 1 km s$^{-1}$.

IRIS records a pair of Si IV lines at 1393.78 Å and 1402.77 Å, with peak formation temperature $10^{4.9}$ K. Under the optically thin conditions, the line at 1393.78 Å is a factor of two stronger than that at 1402.77 Å (Dere et al. (1996); Landi et al. (2013); see however, Gontikakis and Vial (2018); Tripathi et al. (2020)). Hence, following Ghosh et al. (2019), we use the line at 1393.78 Å for our study.

We have also used observations from Atmospheric Imaging Assembly (AIA; Lemen et al., 2012b) in 1600 Å filter for co-alignment purposes because the peak formation temperature in this filter is log[T()K] = 5.0 which is close to peak formation temperature of Si IV lines (log[T()K] = 4.9). Our aim is to study the Doppler shifts in the two major polarities of the ARs. Hence to identify the two polarities, we have used the line of sight (LOS) magnetograms obtained from Helioseismic and Magnetic Imager (HMI; Schou et al., 2012a,c).

To study the CLV of the Doppler shift, we have selected 50 active regions, listed in Table 7.1, observed at various locations covering the full range of longitudes. Figure7.1 displays the location of all the active regions over AIA 1600 Å image taken on Jul 8, 2014. The yellow-colored box represents the exemplar case that is described in detail.

## 7.3 Data Analysis and results

To measure the absolute Doppler shift, we need to perform wavelength calibration. Also, since HMI and IRIS observe the Sun from two different vantage points, a proper coalignment needs to be ensured. For this purpose, we co-align IRIS observations with those obtained using AIA 1600Å passband. Since

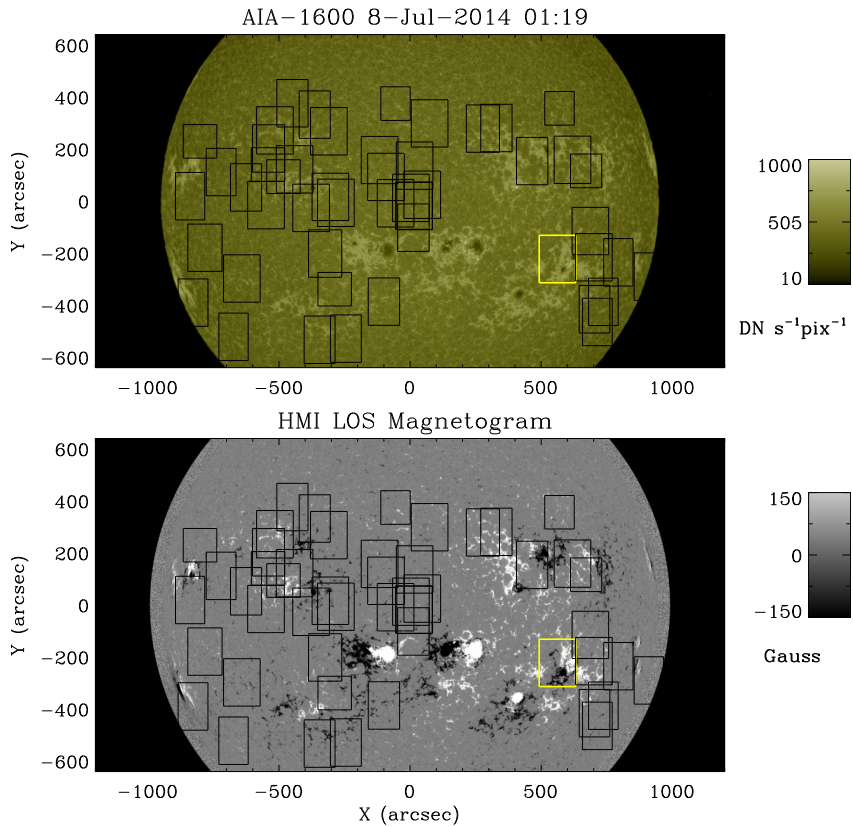

Figure 7.1: Top panel: AIA 1600 Å image taken on 8-Jul-2014 overplotted with the field of view (FOV) of all the IRIS rasters studied in the paper. Bottom panel: the corresponding HMI LOS magnetograms superposed with the FOV of IRIS rasters. The yellow-colored box in both panels shows the exemplar case discussed in detail.

AIA and HMI are both onboard the Solar Dynamics Observatory (SDO), the magnetograms can be readily coaligned with that of AIA. Once we obtain a calibrated Doppler map in Si IV and coaligned magnetograms, we identify the pixels associated with strong field areas of the active region and deduce the average Doppler shift. Here, we discuss the above-mentioned procedure in detail for an exemplar case of active region AR 12104. IRIS provided observations of this region from 23:35 UT on $7^{th}$ of July 2014 to 03:05 UT on $8^{th}$ of July 2014. The spatial extent of the corresponding IRIS raster extended from 490 to 630 arcseconds along the x–axis and –310" to –130" along the y–axis. The position of the raster for the exemplar case is shown with a yellow box in Figure 7.1.

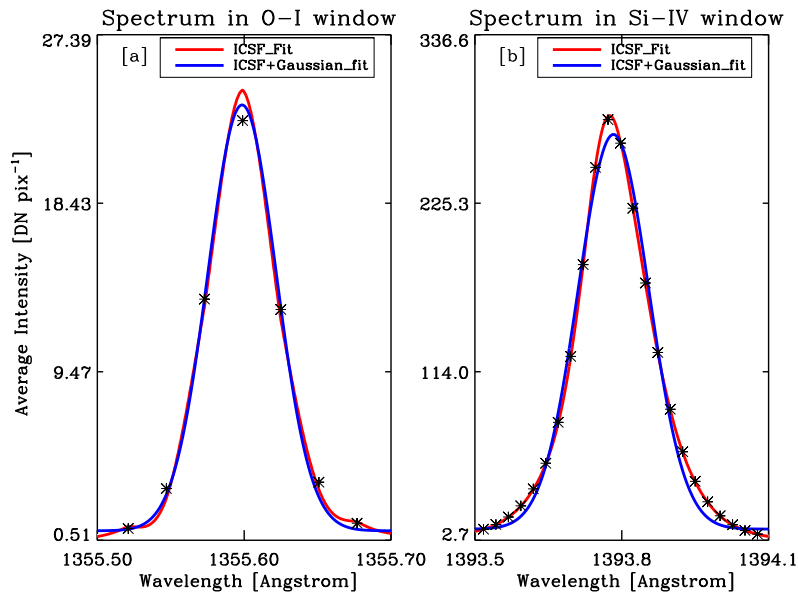

Figure 7.2: Spectrum obtained from IRIS in O I (panel a) and Si IV (panel b) line windows. These spectra have been averaged over the full raster for the exemplar case. The black asterisks show the data obtained from IRIS level2 fits files. Red curves denote the ICSF fitted spectrum. The blue curves show the Gaussian fit to the ICSF spectrum.

### 7.3.1 Wavelength Calibration

Wavelength calibration involves identifying average Doppler shifts in emission lines coming from neutral or singly ionized atoms, which are expected to be approximately at rest (Hassler et al., 1991). Such neutral or singly ionized atoms are present in the photosphere or chromosphere. There are multiple lines such as Fe II, O I, and S I present in IRIS spectral windows. Following Ghosh et al. (2019), we have used O1 (1355.6 Å) line for performing wavelength calibration.

In the absence of any preferred flow direction (i.e., average Doppler velocity over the whole raster being zero), the spectrum should peak at the rest wavelength of the line. This should be the case in the ideal scenario because we expect atoms emitting these lines to be at rest. Any deviation in the peak of the spectrum from the rest wavelength should be due to instrumental effects, which need to be corrected.

The average spectrum from a system of atoms at a finite non-zero temperature is Gaussian. However, directly fitting the obtained spectrum with a Gaussian profile has limitations. The spectrum obtained by an instrument gives the average energy recorded in different wavelength bins, not the energy associated with the center of each bin. Even though in the first approximation, the energy in the bin is associated with the central wavelength of the bin, it is valid only if the spectral line profile in the bin is linear. This certainly cannot be expected to be the case always. Consequently, to increase our accuracy in finding the line center, we have applied the method of Intensity Conserving Spline Fitting (ICSF) to the spectra using the ICSF procedure (Klimchuk et al., 2016). It preserves the total intensity in each spectral bin and performs a spline fitting to account for the line profile variation within the wavelength bin. Finally, the spectrum obtained after the application of ICSF with a Gaussian using eisautofit routine in Solarsoft (Freeland and Handy, 1998).

In Figure 7.2, we plot the spectrum obtained in O I (panel a) and Si IV (panel b) lines averaged over the full raster. The black asterisks denote the original spectrum obtained from IRIS level2 fits files. The red curve represents the spectrum obtained after applying the ICSF spectrum, and the blue curve is the final Gaussian fit to the ICSF spectrum. The reference wavelength for Si IV spectra is calibrated depending on the rest wavelength being larger or smaller than the observed peak wavelength. For the exemplar case, we find the wavelength at which the O I line peaks is 1355.5987 Å, which is larger than the lab measurements of the rest wavelength which is at 1355.5980 Å as obtained from Sandlin et al. (1986). Since this difference in wavelength corresponds to 0.15 km s$^{-1}$, which is less than 1 km s$^{-1}$, the difference between observed and lab measurement of O I wavelength is insignificant.

## 7.3.2 Co-alignment of observations from IRIS, HMI, and AIA

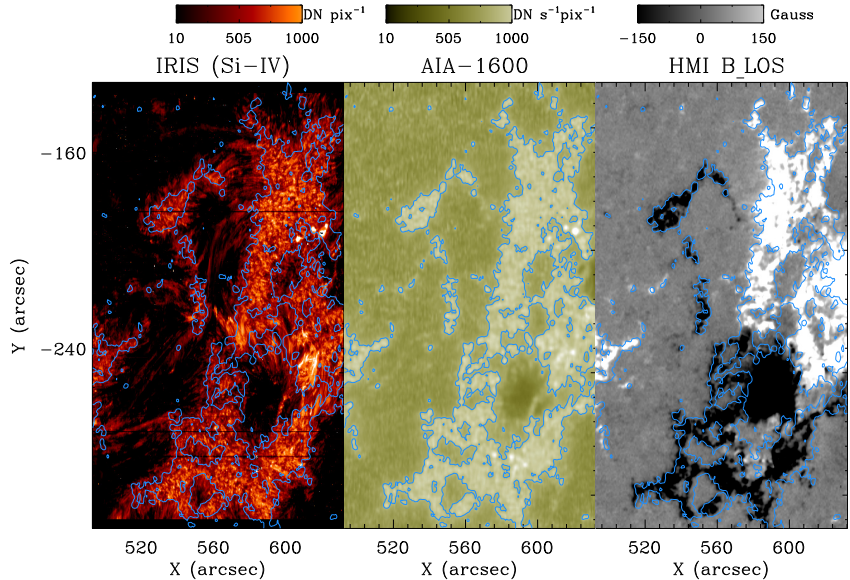

Figure 7.3: Intensity maps of emission in Si IV line (left), artificial rasters of AIA-1600 (middle), and HMI LOS magnetogram (right). Contours of 250 DN s$^{-1}$ pix$^{-1}$ in AIA-1600 Å filter (peak formation temperature of log[T(K)] = 5.0) are overplotted.

While HMI and AIA, being payloads on the same spacecraft, observe from a similar vantage point. But, IRIS observes from a different vantage point. Moreover, one needs to consider that observations may be recorded with different roll angles and roll centers of the telescope. HMI and AIA have different pixel sizes. The images taken around the same instant may have different roll angles but have the same roll center. Furthermore, the pixel size of AIA is 0.6 arc seconds, while the pixel size of HMI is 0.5 arc seconds. Using the above information, we have coaligned AIA and HMI images. On the other hand, the pixel size of IRIS spectra is 0.167 arc seconds. For our analysis, we consider AIA observations taken at 1600 Å as this is closest in temperature to that recorded by IRIS in the Si IV line.

For the purpose of coalignment of IRIS and HMI, we first take all the AIA 1600 Å images and LOS magnetograms of the region of interest during the entire duration of the IRIS raster. All the AIA images and LOS magnetograms in datacubes have been corrected for solar rotation with respect to the first AIA 1600 Å image as the reference. We then create artificial AIA-1600 and HMI LOS magnetogram rasters corresponding to IRIS rasters to ensure proper coalignment.

Figure 7.3 (left panel) displays the intensity map obtained in Si IV. The middle and right panel displays the AIA 1600 Å image and the LOS magnetogram obtained by artificial rastering. The over-plotted contours correspond to 250 DN s$^{-1}$ pix$^{-1}$ in AIA 1600 Å images. The excellent correspondence between the AIA contours on the IRIS image and the magnetogram suggests a near-perfect coalignment of the data.

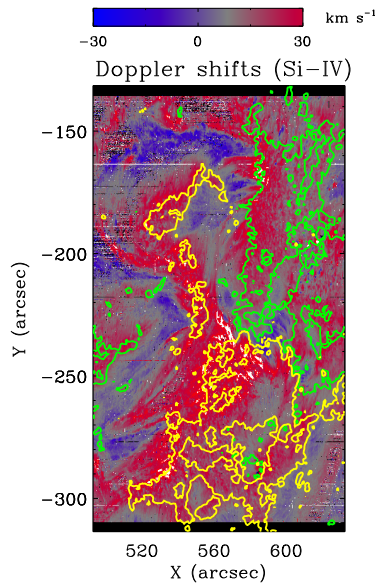

Figure 7.4: Velocity maps in Si IV line. The green and yellow contours in the right panel are of +50 and -50 Gauss, respectively.

### 7.3.3   Identification of active region and computing average Doppler shifts

After coaligning the data from different instruments and ensuring that we select the same structures from different data, we identify the strong field areas inside the active regions. Klimchuk (1987) identified the pixels in which the magnitude of the magnetic field exceeded 100 G. Ghosh et al. (2019) on the other hand, used an absolute magnitude of 50 G for the same purpose. Ghosh et al. (2019) noted that the precise value is unimportant because the magnetic field strength decays rapidly outside the strong field regions. Consequently, the contours of magnetic fields of ± 100 G or ± 50 G are not very different. Following Ghosh et al. (2019), here we have used contours of ±50 G to identify the strong field regions.

We plot the velocity maps obtained in Si4 in Figure 7.4. The over-plotted yellow and green contours correspond to ± 50 Gauss, respectively, obtained from the magnetograms shown in Figure 7.3.c. We find that the average Doppler shift in the strong field regions is 7.80 km s$^{-1}$ (downflows).

### 7.3.4   Errors

We estimate the uncertainties following the procedure discussed in the Appendix in Ghosh et al. (2019). We describe the procedure briefly in this section. Two types of errors have been discussed namely random and systematic.

(i) Random error: The first component of random error is the standard error in velocities which is defined as $E_1 = \frac{\sigma_v}{\sqrt{N}}$, where $\sigma_v$ is the standard deviation in velocity in the pixels identified as strong field regions and $N$ is the total number of such pixels. The second component of random error is related to the photon shot noise. Additionally, the central wavelength found by Gaussian fitting of the OI window of the spectrum varies from pixel to pixel. This introduces an error given by $E_2 = \frac{\sigma_\lambda}{\sqrt{N}}$. Note that $\sigma_\lambda$ is the standard deviation in $\lambda_{cent} - \lambda_{lab}$, where $\lambda_{cent}$ is the central wavelength derived by fitting the OI window of the spectrum, and $\lambda_{lab}$ is the lab wavelength of OI line. Since the two errors $E_1$ and $E_2$ are independent of each other the cumulative random error can be written as $E_R = \sqrt{E_1^2 + E_2^2}$. As can be clearly seen these errors vary between the rasters.

Systematic errors: The true rest wavelength of the O-I line with respect to which Si-IV lines have been calibrated is subject to a systematic error of 3 mÅ. Hence all estimates of velocities computed from Doppler shifts have a systematic error of $E_3$ =0.66 km s$^{-1}$. Sandlin et al. (1986) computed this error by using HRTS observations across the limb in conjunction with the rest wavelength of O-I in the laboratory. Additionally, dispersion of about 0.1 pixels for spectral resolution of 26 mÅ in FUV IRIS introduces a systematic error in wavelength by an amount $\Delta\lambda = 2.6$ mÅ. Since the rest wavelength of Si-IV line is $\lambda_0 = 1393.795$ Å, this introduces a systematic error of $E_4 = \frac{\Delta\lambda}{\lambda_0}c$ =0.56km s$^{-1}$. Since the two systematic errors are independent of each other the cumulative systematic error becomes $E_S = \sqrt{E_3^2 + E_4^2}$.

While the cumulative random error for this active region is ± 0.10 km s$^{-1}$, the total systematic error, which is the same for all the regions, is 0.87 km s$^{-1}$.

## 7.3.5 Radius vector

We need to compute the radius vector of observed active regions to study the CLV of Doppler shifts. The radius vector is defined as the fraction of the distance of the feature from the disk center and the radius of the solar disk ($R_\odot$) (Klimchuk, 1987). In this convention, radius vector zero corresponds to the disk center whereas the positive(negative) radius vector represents longitudes to the west(east) of the central meridian.

We compute the radius vector of a given IRIS raster using its central pixel. If the central position of IRIS raster is [x,y], the radius vector is computed as

$$\frac{\sqrt{x^2 + y^2}}{R_{Sun}}$$

We multiply it by $\pm 1$ for the west(east) limb. For the exemplar case under consideration, the radius vector is 0.63.

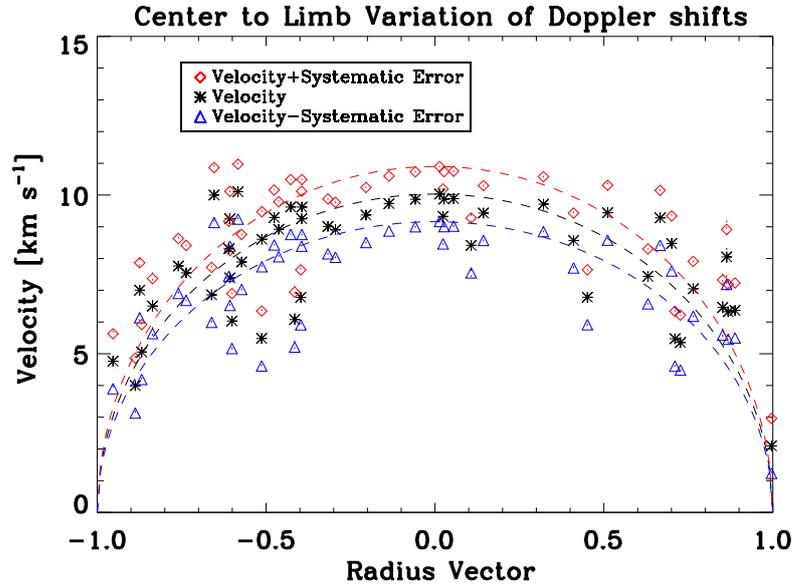

Figure 7.5: Measured Doppler shifts in the strong field regions of the active region as a function of radius vector shown with black asterisks. The red diamond and blue triangle show the values obtained after adding the systematic error $\delta v_{sys}$ to the average Doppler shift. The dotted black curve shows the variation of Doppler velocity expected from the hypothetical vertical flow. The over-plotted blue and red dashed lines show $v_d - \delta v_{sys}$ (blue), and $v_d + \delta v_{sys}$ (red). The random errors range from 0.01 km $s^{-1}$ to 0.1 km $s^{-1}$

## 7.4   Center to Limb variation of Doppler shifts

We now conduct exactly the same analysis discussed above for the other 49 active regions listed in Table. 7.1. The name of the analyzed iris level2 fits files, and the x & y coordinates of the center of the IRIS raster are given in Table 7.1. We compute the average Doppler shifts in the strong field regions of all these active regions. Figure 7.5 plots the Doppler shifts as a function of the radius vector. The black stars show the average Doppler shift in the strong field regions.

To ascertain the variation due to systematic errors, we also plot the sum and difference of average Doppler shifts and systematic errors with red diamonds and blue triangles, respectively. As stated earlier, the systematic error is 0.87 km s$^{-1}$. The diamonds, stars, and triangles also contain error bars related to random errors. However, since their magnitude is too small ($\sim$ 0.05 km), they are barely noticeable.

To study the departure of measured Doppler velocities from those that will be produced due to hypothetical vertical flows ($v_{vertical}$), we have computed the expected Doppler shifts ($v_{LOS}$) using the relation

$$v_{LOS} = v_{Vertical}\sqrt{\left[1 - \left(\frac{r}{R}\right)^2\right]}$$

,

where $\frac{r}{R}$ is the radius vector, and $v_{vertical}$ is the velocity corresponding to the active region closest to the disk center. In our sample, the active region closest to the disk center is located at a radius vector of 0.01. The average Doppler shift in its strong field regions is 8.28 km s$^{-1}$. Considering the proximity of this active region to the disk center, we take $v_{vertical}$ to be 8.28 km s$^{-1}$.

The plots suggest that the Doppler shift shows modest CLV, similar to the observation of Ghosh et al. (2019) based on a single active region. However, the plots show a large scatter at mid-longitude regions. Moreover, there are sizable Doppler shifts towards the limb.

## 7.5 Summary and Discussion

In this paper, we study the Doppler shifts and its CLV in the transition region of active regions in Si IV lines. For this purpose, we have used the observations recorded by IRIS. For the purpose of co-alignment and to identify the strong field regions in active regions, we have used the observations from AIA and HMI, both onboard SDO.

Similar to the results obtained by Feldman et al. (1976); Klimchuk (1987, 1989); Ghosh et al. (2019), we find that in transition regions, active regions are predominantly red-shifted with velocities ranging between 4–10 km s$^{-1}$. Moreover, the Doppler shifts show modest CLV, as was also reported by Ghosh et al. (2019). Note that the results obtained by Ghosh et al. (2019) were based on the tracking of a single active region AR 12641 as it crossed from the center towards the limb. Here, we have studied 50 active regions located at different locations across the solar disk. Previous studies aimed at studying CLV of transition region lines (see, e.g. Pecker et al., 1988) lacked the presence of neutral lines for performing wavelength calibration. This work provides a comprehensive measurement of Doppler shifts in the transition region of active regions and their CLV in the literature.

Our results support the conclusion that the predominant redshifts observed in the transition region are not due to the field-aligned downflows and may be due to downflows related to type II spicules, which are obscured due to cooler materials corresponding to classical type I spicules. Therefore, we suggest that the scenario proposed by Ghosh et al. (2019) for explaining large transition region downflows in active regions (larger than field-aligned downflows) with modest CLV can be considered as a generic result and not limited to the detailed geometry and magnetic field structure of a particularly active region. Further detailed observations combined with numerical simulations are required to fully comprehend transition region downflows.

Table 7.1: List of active regions studied

| Index | Fits file[a] | $X_{Cen}$ | $Y_{Cen}$ | Index | Fits file | $X_{Cen}$ | $Y_{Cen}$ |
|---|---|---|---|---|---|---|---|
| 00 | 20140215_163205_3800258296 | 687.6 | -114.9 | 25 | 20200801_011722_3620108077 | 737.3 | -386.9 |
| 01 | 20140701_195503_3820258196 | 17.47 | 136.4 | 26 | 20200801_143730_3620108077 | -363.4 | 332.3 |
| 02 | 20140707_233530_3824263396 | 562.4 | -222.5 | 27 | 20200814_054633_3620108077 | -673.1 | -521.4 |
| 03 | 20140708_192613_3824263396 | 701.0 | -215.7 | 28 | 20170303_021419_3620106076 | -55.55 | 374.7 |
| 04 | 20140815_070803_3800258196 | -440.1 | 122.3 | 29 | 20180901_175648_3620108077 | -838.8 | 19.14 |
| 05 | 20141001_224938_3800009396 | -91.69 | 93.06 | 30 | 20170306_072447_3620106076 | 569.9 | 356.5 |
| 06 | 20150207_041007_3800256196 | 14.62 | -17.50 | 31 | 20150207_041007_3800256196 | 17.58 | -18.79 |
| 07 | 20150214_150407_3820256096 | 74.63 | 298.7 | 32 | 20150915_062928_3893010094 | 750.7 | -184.8 |
| 08 | 20150222_154645_3800110096 | -377.3 | -25.98 | 33 | 20141130_070200_3893010094 | -482.7 | 94.13 |
| 09 | 20150223_233348_3800110096 | -55.18 | -8.437 | 34 | 20160115_120419_3630008076 | 671.4 | 116.3 |
| 10 | 20180601_030145_3620108077 | 329.2 | 281.1 | 35 | 20141003_044846_3893260094 | -287.3 | -338.4 |
| 11 | 20190309_044823_3620010077 | 278.0 | 278.7 | 36 | 20140701_164900_3820258196 | -116.1 | 157.9 |
| 12 | 20200315_235329_3620108077 | 714.4 | -464.4 | 37 | 20140916_044847_3893010094 | -541.5 | 140.7 |
| 13 | 20171210_021825_3630108077 | 819.8 | 117.1 | 38 | 20140702_003429_3820259296 | -641.3 | -297.4 |
| 14 | 20160227_074513_3620258078 | -282.8 | -6.481 | 39 | 20141106_024328_3893010094 | -800.9 | 229.6 |
| 15 | 20160228_153411_3620258078 | 3.246 | 12.62 | 40 | 20140129_200158_3880010095 | -782.1 | -179.0 |
| 16 | 20160310_163211_3620258078 | 619.4 | 158.8 | 41 | 20171217_031351_3610108077 | 12.8 | -101.7 |
| 17 | 20160329_061311_3600108078 | -303.8 | 16.38 | 42 | 20191222_045119_3690108077 | -448.6 | 376.2 |
| 18 | 20160330_190439_3600108078 | 46.64 | 24.57 | 43 | 20200105_000818_3690108077 | -344.8 | -532.6 |
| 19 | 20160412_020911_3600108078 | -515.3 | 271.8 | 44 | 20201205_164951_3610108077 | 910.8 | -289.8 |
| 20 | 20160413_014409_3600108078 | -309.6 | 268.4 | 45 | 20201206_090413_3610108077 | -99.40 | -386.1 |
| 21 | 20210923_130908_3620108077 | -244.8 | -528.4 | 46 | 20200501_145553_3620108077 | 794.4 | -235.2 |
| 22 | 20201118_153452_3690108077 | -827.5 | -389.9 | 47 | 20191122_101544_3690108077 | 703.4 | -415.6 |
| 23 | 20190605_051739_3620108077 | -720.3 | 111.8 | 48 | 20190507_121550_3620110077 | -539.9 | 202.4 |
| 24 | 20180205_113343_3610108077 | -625.2 | 52.62 | 49 | 20190322_105259_3620108077 | 465.6 | 154.0 |

[a] The tabulated filename excludes the common part 'iris_l2_' on the left and '_raster_t000_r00000.fits' on right.

# 8

# Summary, conclusions, & outlook

In this thesis, we addressed three scientific problems. The first problem was to understand the energetics of small transient events using the EBTEL code which is based on the 0D description of coronal loops. The second problem was to understand the distribution of events of different energy by generating transients and backgrounds from a single power-law distribution. We realized that the version of EBTEL used in the first project neglected the kinetic energy term throughout. However this rendered EBTEL unreliable for more impulsive events. Consequently, we first added kinetic energy term in EBTEL. We used the modified EBTEL for our scientific goal of simulating transients and the background from a single power law. The results obtained from these projects supported the heating events to be impulsive. The third problem was to test the viability of impulsive heating in the transition region of active regions. Below we summarize the main results obtained in this thesis.

1. We performed 0D hydrodynamical simulations using EBTEL to study the energetics of small transient events observed by Hi-C and AIA/SDO. These brightenings were simulated by dissipating $\sim 10^{23}$ ergs over a time period of $\sim 50$ s in loops of half-length $\sim 10^6$ cm. We demonstrate that conduction is the dominant cooling mechanism in the corona for these brightenings. This is a feature shared by other impulsive events such as flares, microflares, etc., suggesting the origin of these transient events can be impulsive.

2. We upgraded EBTEL to include kinetic energy terms in the 0D hydrodynamical equations, so as to facilitate the modeling of large-scale impulsive events. This leads to velocities in upgraded EBTEL matching much better with 1D field-aligned simulations than the previous version of EBTEL, especially when the flows are subsonic. Furthermore, we included an adaptive time grid in the IDL version of the code, which made it roughly ten times faster.

3. We performed multi-stranded simulation events observed in AR12230 in X-rays by FOXSI-2 and in EUV by AIA/SDO. We did not impose any steady background heating in our simulations. Instead, we generated different heating events from the same power-law distribution. Our simulations show that FOXSI and AIA observations cannot be explained by slopes $\gtrsim 2$.

4. We addressed the viability of impulsive events as the heating mechanism in the transition region of active regions. Since such a mechanism should show CLV, we studied the Doppler shift of 50 active regions in the transition region and its dependence on the radius vector. The Doppler shifts in Si-IV lines remained significantly less than those predicted by field-aligned simulations. We found modest CLV. These together provided further support to the idea of a chromospheric wall, i.e., warm type II spicules surrounded by type I spicules.

The results obtained in this thesis suggest that the physics of small-scale transient brightenings are akin to those for microflares and flares, albeit happening at very small length scales. Moreover, for the first time, we have demonstrated the possibility of modeling the background omnipresent emission in the solar corona along with identifiable transients with a single power law, thereby suggesting a uniform mechanism for heating the solar corona in quiet Sun, active regions, and flare. The results obtained in this thesis also support the idea of field-aligned downflows due to impulsive heating being insufficient at explaining the heating of the transition region. It adds to the evidence of a chromospheric wall formed by type I spicules surrounding transient type II spicules.

The results obtained in this thesis shed light on the energetics of the solar atmosphere. Additionally, it opens up a number of challenging opportunities for further study that may help comprehend the dynamic coupling of the magnetized solar atmosphere. We now discuss the future directions we will be working on as parts of ongoing projects or as follow-ups of these projects.

1. We will perform statistical analysis to rank all the cases $[E_{max}, \alpha, R_s]$ for which we have performed multi-stranded simulations in terms of their match with observations of events in AR12230. We will validate these results against the optimal parameters obtained from machine learning methods.

2. We will perform a similar set of simulations as a part of a separate project to study the quiet Sun's emission.

3. We will match the emission measure distributions obtained from such simulations with observed emission measure distribution from coronal active and quiet regions.

4. We will introduce waiting times in our multistranded simulations. It will be a new parameter in our simulations. We will study the systematic effect of its variation in a separate study.

5. We will study the center-to-limb variation of Doppler shifts in Si-IV lines using 1D simulations of the loop at different projections and locations across the disk.

6. We will study the energetics of spicules using multiwavelength data and understand their role in transition region heating.